\documentclass[twocolumn]{aastex7}

\usepackage{hyperref,amsmath, bm, empheq, mathrsfs,  cancel, multirow, xcolor,array}
\usepackage{natbib}
\usepackage{textcase}
\usepackage{bigstrut}

\usepackage{placeins}
\allowdisplaybreaks

\makeatletter

\usepackage{etoolbox}
\patchcmd{\thebibliography}{\clearpage}{}{}{}
\patchcmd{\thebibliography}{\newpage}{}{}{}
\AtBeginDocument{
  \setlength{\bibsep}{0pt plus 0.3ex}
}

\makeatother

\allowdisplaybreaks
\begin{document}

% cite aliases
\defcitealias{Clark1973}{C73}
\defcitealias{Spiegel1992}{SZ92}
\defcitealias{Gough1998}{GM98}
\defcitealias{Matilsky2022}{M22}
\defcitealias{Matilsky2024}{M24}
\defcitealias{Matilsky2025a}{M25}
\defcitealias{Korre2024}{KF24}
\defcitealias{ForgcsDajka2001}{FDP01}

% title
\title{A Dynamo Confinement Scenario for the Solar Tachocline and its Implications for Spin-down in the Radiative Spreading Regime }

% authors and affiliations
\correspondingauthor{Loren I. Matilsky}
\author[0000-0001-9001-6118]{Loren I. Matilsky}\thanks{U.S. National Science Foundation\\ Astronomy and Astrophysics Postdoctoral Fellow}	
\affiliation{Department of Applied Mathematics,
Baskin School of Engineering,
University of California, 
Santa Cruz, CA 96064-1077, USA}
\email{lmatilsk@ucsc.edu}

\author[0000-0002-0963-4881]{Lydia Korre}
\affiliation{Department of Applied Mathematics,
University of Colorado,
Boulder, CO 80309-0526, USA}	
\email{lydia.korre@colorado.edu}

\author[0000-0003-4350-5183]{Nicholas H. Brummell}
\affiliation{Department of Applied Mathematics,
Baskin School of Engineering,
University of California,
Santa Cruz, CA 96064-1077, USA}	
\email{brummell@ucsc.edu}

% get standard macros
% MY MACROS
% Universalized: 12/13/2024
% test macro to make sure everything is in sync

\providecommand{\testmacro}{\text{macros.tex v. 2025-01-02 6pm}}
% MY EMAIL
\providecommand{\myemail}{loren.matilsky@gmail.com}

% scientific notation
\providecommand{\sn}[2]{#1\times10^{#2}}
\providecommand{\sncomp}[2]{{#1}e{#2}}

% COMMON SUBSCRIPTS
\providecommand{\cz}{_{\rm{cz}}}
\providecommand{\rz}{_{\rm{rz}}}
\providecommand{\czt}{_{{\rm cz},t}}
\providecommand{\rzt}{_{{\rm rz},t}}
\providecommand{\fulll}{_{\rm{full}}}
\providecommand{\dimm}{_{\rm{dim}}}
\providecommand{\rad}{_{\rm{r}}}
\providecommand{\nrad}{_{\rm{nr}}}
\providecommand{\loc}{_{\rm{loc}}}

% MATH OPERATORS
\providecommand{\sgn}[1]{{\text{sgn}}(#1)}

\providecommand{\pderiv}[2]{\dfrac{\partial#1}{\partial#2}}
\providecommand{\matderiv}[1]{\frac{D#1}{Dt}}
\providecommand{\pderivline}[2]{\partial#1/\partial#2}
\providecommand{\parenfrac}[2]{\left(\frac{#1}{#2}\right)}
\providecommand{\brackfrac}[2]{\left[\frac{#1}{#2}\right]}
\providecommand{\bracefrac}[2]{\left\{\frac{#1}{#2}\right\}}
\providecommand{\sign}[1]{{\text{sign}}(#1)}

% averages
\providecommand{\av}[1]{\left\langle#1\right\rangle}
\providecommand{\abs}[1]{\left\lvert#1\right\rvert}
\providecommand{\magnitudeof}[1]{\left[#1\right]}
\providecommand{\avsph}[1]{\left\langle#1\right\rangle_{\rm{sph}}}
\providecommand{\avspht}[1]{\left\langle#1\right\rangle_{ {\rm sph}, t}}
\providecommand{\avt}[1]{\left\langle#1\right\rangle_{t}}
\providecommand{\avtpar}[1]{\left(#1\right)_{t}}
\providecommand{\avphi}[1]{\left\langle#1\right\rangle_{\phi}}
\providecommand{\avphit}[1]{\left\langle#1\right\rangle_{\phi,t}}
\providecommand{\avvol}[1]{\left\langle#1\right\rangle_{\rm{v}}}

\providecommand{\avcz}[1]{\left\langle#1\right\rangle\cz}
\providecommand{\avrz}[1]{\left\langle#1\right\rangle\rz}
\providecommand{\avfull}[1]{\left\langle#1\right\rangle\fulll}
\providecommand{\avczt}[1]{\left\langle#1\right\rangle\czt}
\providecommand{\avrzt}[1]{\left\langle#1\right\rangle\rzt}
\providecommand{\avfullt}[1]{\left\langle#1\right\rangle_{{\rm full}, t}}

% alternative average notation
\providecommand{\avalt}[1]{\langle#1\rangle}
\providecommand{\avaltsph}{\overline}
\providecommand{\avaltspht}[1]{\left(\overline{#1}\right)_{t}}

\providecommand{\dbprime}{^{\prime\prime}}

\providecommand{\define}{\coloneqq}
\providecommand{\definealt}{\equiv}
%\providecommand{\define}{\equiv}

% TEXT OPERATORS
\providecommand{\five}{\ \ \ \ \ }
\providecommand{\ten}{\five\five}
\providecommand{\twenty}{\ten\ten}
\providecommand{\orr}{\text{or}\five }
\providecommand{\andd}{\text{and}\five }
\providecommand{\where}{\text{where}\five }
\providecommand{\with}{\text{with}\five }

% VECTOR SHORCUTS

% operators
\providecommand{\curl}{\nabla\times}
\providecommand{\Div}{\nabla\cdot}
\providecommand{\lap}{\nabla^2}
\providecommand{\dotgrad}{\cdot\nabla}
\providecommand{\ugrad}{\bm{u}\dotgrad}
\providecommand{\vgrad}{\bm{v}\dotgrad}

% unit vectors
\providecommand{\e}{\hat{\bm{e}}}
\providecommand{\erad}{\e_r}
\providecommand{\etheta}{\e_\theta}
\providecommand{\ephi}{\e_\phi}
\providecommand{\elambda}{\e_\lambda}
\providecommand{\ez}{\e_z}
\providecommand{\exi}{\e_\xi}
\providecommand{\eeta}{\e_\eta}
\providecommand{\epol}{\e_{\rm{pol}}}
\providecommand{\emer}{\e_{\rm{mer}}}
\providecommand{\ihat}{\hat{\bm{i}}}
\providecommand{\jhat}{\hat{\bm{j}}}
\providecommand{\khat}{\hat{\bm{k}}}
\providecommand{\eone}{\e_1}
\providecommand{\etwo}{\e_2}
\providecommand{\ethree}{\e_3}

% FLUX ALIASES
\providecommand{\flux}{{\bm{F}}}

\providecommand{\fluxrad}{\flux_{\rm{r}}}
\providecommand{\fluxnrad}{\flux_{\rm{nr}}}
\providecommand{\fluxcond}{\flux_{\rm{c}}}
\providecommand{\fluxenth}{\flux_{\rm{e}}}

\providecommand{\fluxscalarrad}{F_{\rm{r}}}
\providecommand{\fluxscalarnrad}{F_{\rm{nr}}}
\providecommand{\fluxscalarcond}{F_{\rm{c}}}
\providecommand{\fluxscalarenth}{F_{\rm{e}}}

% differential rotation stuff
\providecommand{\Omzero}{\Omega_0}
\providecommand{\twoOmzero}{2\Omega_0}
\providecommand{\Omzerovec}{\bm{\Omega}_0}
\providecommand{\Omsun}{\Omega_\odot}
\providecommand{\Omrz}{\Omega\rz}
\providecommand{\Omcz}{\Omega\cz}
\providecommand{\bruntsun}{N_\odot}
\providecommand{\dOm}{\Delta\Omega}
\providecommand{\dOmsixty}{\dOm_{\rm 60}}
\providecommand{\dOmcz}{\Delta\Omega\cz}
\providecommand{\dOmrz}{\Delta\Omega\rz}
\providecommand{\dOmczt}{\Delta\Omega\czt}
\providecommand{\dOmrzt}{\Delta\Omega\rzt}
\providecommand{\dOmloc}{\Delta\Omega\loc}
\providecommand{\Deltaloc}{\Delta\loc}

% REFERENCE STATE AND THERMO. VARIABLES

% may want to append this
\providecommand{\ofr}{(r)}
\providecommand{\rprime}{{r^{\prime}}}
\providecommand{\ofrprime}{(\rprime)}

% reference-state constants
\providecommand{\cv}{c_{\rm{v}}}
\providecommand{\cp}{c_{\rm{p}}}
\providecommand{\cvcap}{C_{\rm{v}}}
\providecommand{\cpcap}{C_{\rm{p}}}
\providecommand{\cs}{c_{\rm s}}
\providecommand{\gasconst}{\mathcal{R}}
\providecommand{\gammaone}{\Gamma_1}

% total thermal variables (may wish to switch this stuff around later--I've always hated this notation of "no subscripts" = perturbation
\providecommand{\tot}{_{\rm{tot}}}
\providecommand{\rhotot}{\rho\tot}
\providecommand{\tmptot}{T\tot}
\providecommand{\prstot}{P\tot}
\providecommand{\stot}{S\tot}
\providecommand{\dsdrtot}{\frac{dS\tot}{dr}}
\providecommand{\dsdrtotline}{dS\tot/dr}

% Delta S
\providecommand{\Dentr}{\Delta s}

% mean state
\providecommand{\rhoover}{\overline{\rho}}
\providecommand{\tmpover}{\overline{T}}
\providecommand{\prsover}{\overline{P}}
\providecommand{\entrover}{\overline{s}}
\providecommand{\inteover}{\overline{u}}
\providecommand{\enthover}{\overline{h}}
\providecommand{\heatover}{\overline{Q}}
\providecommand{\heatradover}{\overline{Q}_{\rm r}}
\providecommand{\coolover}{\overline{C}}
\providecommand{\bruntsqover}{\overline{N^2}}
\providecommand{\bruntover}{\overline{N}}
\providecommand{\gravover}{\overline{g}}
\providecommand{\nuover}{\overline{\nu}}
\providecommand{\kappaover}{\overline{\kappa}}
\providecommand{\etaover}{\overline{\eta}}
\providecommand{\muover}{\overline{\mu}}
\providecommand{\deltaover}{\overline{\delta}}
\providecommand{\cpover}{\overline{\cp}}
\providecommand{\cvover}{\overline{\cv}}
\providecommand{\csover}{\overline{\cs}}
\providecommand{\cssqover}{\overline{\cs^2}}
\providecommand{\hrhoover}{\overline{H_\rho}}
\providecommand{\Dentrover}{\Delta\entrover}
\providecommand{\Dentroverf}{\Delta\entrover_{\rm f}}

\providecommand{\fluxradover}{\overline{\flux}_{\rm{r}}}
\providecommand{\fluxscalarradover}{\overline{F}_{\rm{r}}}
\providecommand{\fluxnradover}{\overline{\flux}_{\rm{nr}}}
\providecommand{\fluxscalarnradover}{\overline{F}_{\rm{nr}}}
\providecommand{\kradover}{\overline{\kappa}_{\rm rad}}

% ``typical" values of mean state (a subscript)
% mean state
\providecommand{\rhoa}{\rho_a}
\providecommand{\rhocz}{\rho\cz}
\providecommand{\rhorz}{\rho\rz}
\providecommand{\tmpa}{T_a}
\providecommand{\tmpcz}{T\cz}
\providecommand{\tmprz}{T\rz}
\providecommand{\cpcz}{c_{\rm p,cz}}
\providecommand{\cprz}{c_{\rm p,rz}}
\providecommand{\prsa}{p_a}
\providecommand{\prscz}{p\cz}
\providecommand{\prsrz}{p\rz}
\providecommand{\entra}{s_a}
\providecommand{\intea}{u_a}
\providecommand{\entha}{h_a}
\providecommand{\heata}{Q_a}
\providecommand{\fluxa}{F_a}
\providecommand{\hrhoa}{H_{\rho a}}
\providecommand{\hprsa}{H_{{\rm p} a}}
\providecommand{\heatrada}{\overline{Q}_{{\rm r}a}}
\providecommand{\coola}{C_a}
\providecommand{\bruntsqa}{N^2_a}
\providecommand{\bruntsqrz}{N^2\rz}
\providecommand{\bruntrz}{N\rz}
\providecommand{\grava}{g_a}
\providecommand{\gravcz}{g\cz}
\providecommand{\gravrz}{g\rz}
\providecommand{\nua}{\nu_a}
\providecommand{\nut}{\nu_{\rm t}}
\providecommand{\nurz}{\nu\rz}
\providecommand{\nucz}{\nu\cz}
\providecommand{\kappaa}{\kappa_a}
\providecommand{\kappat}{\kappa_{\rm t}}
\providecommand{\kapparz}{\kappa\rz}
\providecommand{\kappacz}{\kappa\cz}

\providecommand{\etaa}{\eta_a}
\providecommand{\etarz}{\eta\rz}
\providecommand{\etacz}{\eta\cz}
\providecommand{\mua}{\mu_a}
\providecommand{\deltaa}{\delta_a}
\providecommand{\cpa}{c_{{\rm p}a}}
\providecommand{\cva}{c_{{\rm v}a}}
\providecommand{\csa}{c_{{\rm s}a}}
\providecommand{\cssqa}{(\cs^2)_a}
\providecommand{\fluxrada}{F_{{\rm r}a}}
\providecommand{\fluxnrada}{F_{{\rm{nr}}a}}
\providecommand{\fluxnradcz}{F_{{\rm{nr,cz}}}}
\providecommand{\krada}{\kappa_{{\rm rad}a}}
\providecommand{\sigmaa}{\sigma_a}

% mean--state derivatives
\providecommand{\dlnrhoover}{\frac{d\ln\rhoover}{dr}}
\providecommand{\dlntmpover}{\frac{d\ln\tmpover}{dr}}
\providecommand{\dlnprsover}{\frac{d\ln\prsover}{dr}}
\providecommand{\dsdrover}{\frac{d\entrover}{dr}}
\providecommand{\dsdroverline}{d\entrover/dr}
\providecommand{\dlnrhooverline}{d\ln\rhoover/dr}
\providecommand{\dlntmpoverline}{d\ln\tmpover/dr}
\providecommand{\dlnprsoverline}{d\ln\prsover/dr}

% reference state
\providecommand{\rhotilde}{\tilde{\rho}}
\providecommand{\tmptilde}{\tilde{T}}
\providecommand{\rhottilde}{\rhotilde\tmptilde}

\providecommand{\prstilde}{\tilde{p}}
\providecommand{\entrtilde}{\tilde{s}}
\providecommand{\intetilde}{\tilde{u}}
\providecommand{\enthtilde}{\tilde{h}}
\providecommand{\heattilde}{\tilde{Q}}
\providecommand{\heatradtilde}{\tilde{Q}_{\rm r}}
\providecommand{\dentr}{\delta_s}
\providecommand{\dheat}{\delta_Q}
\providecommand{\drad}{\delta_{\rm r}}
\providecommand{\rheat}{r_Q}
\providecommand{\rentr}{r_s}
\providecommand{\lumtilde}{\tilde{L}}
\providecommand{\heatra}{\tilde{Q}_{\texttt{Ra}}}
\providecommand{\cooltilde}{\tilde{C}}
\providecommand{\brunttilde}{\widetilde{N}}
\providecommand{\brunttildesq}{\widetilde{N}^2}
\providecommand{\bruntsqtilde}{\widetilde{N^2}}
\providecommand{\bruntsqtildedim}{\widetilde{N^{*2}}}
\providecommand{\gravtilde}{\tilde{g}}
\providecommand{\nutilde}{\tilde{\nu}}
\providecommand{\kappatilde}{\tilde{\kappa}}
\providecommand{\etatilde}{\tilde{\eta}}
\providecommand{\mutilde}{\tilde{\mu}}
\providecommand{\deltatilde}{\tilde{\delta}}
\providecommand{\cptilde}{\tilde{c}_{\rm p}}
\providecommand{\cvtilde}{\tilde{c}_{\rm v}}
\providecommand{\cptildedim}{\tilde{c}_{\rm p}^*}
\providecommand{\cvtildedim}{\tilde{c}_{\rm v}^*}
\providecommand{\cssqtilde}{\tilde{\cs^2}}
\providecommand{\Dentrtilde}{\Delta\entrtilde}
\providecommand{\Dentrtildef}{(\Delta\entrtilde)_{\rm f}}

\providecommand{\fluxtilde}{\tilde{\flux}}
\providecommand{\fluxscalartilde}{\tilde{F}}
\providecommand{\fluxradtilde}{\tilde{\flux}_{\rm{r}}}
\providecommand{\fluxscalarradtilde}{\tilde{F}_{\rm{r}}}
\providecommand{\fluxnradtilde}{\tilde{\flux}_{\rm{nr}}}
\providecommand{\fluxscalarnradtilde}{\tilde{F}_{\rm{nr}}}
\providecommand{\fluxcondtilde}{\tilde{\flux}_{\rm{c}}}
\providecommand{\fluxscalarcondtilde}{\tilde{F}_{\rm{c}}}

% reference-state derivatives
\providecommand{\dlnrhotilde}{\frac{d\ln\rhotilde}{dr}}
\providecommand{\dlntmptilde}{\frac{d\ln\tmptilde}{dr}}
\providecommand{\dlnprstilde}{\frac{d\ln\prstilde}{dr}}
\providecommand{\dsdrtilde}{\frac{d\entrtilde}{dr}}
\providecommand{\dlnrhotildeline}{d\ln\rhotilde/dr}
\providecommand{\dlntmptildeline}{d\ln\tmptilde/dr}
\providecommand{\dlnprstildeline}{d\ln\prstilde/dr}
\providecommand{\dsdrtildeline}{d\entrtilde/dr}

% some other ancillary reference state quantities (sort of)
\providecommand{\grav}{g}
\providecommand{\vecg}{\bm{g}}
\providecommand{\geff}{g_{\rm{eff}}}
\providecommand{\vecgeff}{\bm{g}_{\rm{eff}}}
\providecommand{\heat}{Q}
\providecommand{\buoyfreq}{N}
\providecommand{\brunt}{N}
\providecommand{\bruntsq}{N^2}
\providecommand{\hrho}{H_\rho}
\providecommand{\hprs}{H_{\rm{p}}}
\providecommand{\hrhotilde}{\widetilde{H_\rho}}
\providecommand{\hprstilde}{\widetilde{H_{\rm p}}}

% grad ad and grad rad
\providecommand{\gradrad}{\nabla_{\rm r}}
\providecommand{\gradad}{\nabla_{\rm ad}}

% perturbations from reference state
\providecommand{\rhoprime}{{\rho^\prime}}
\providecommand{\tmpprime}{{T^\prime}}
\providecommand{\prsprime}{{p^\prime}}
\providecommand{\entrprime}{{s^\prime}}
\providecommand{\inteprime}{{u^\prime}}
\providecommand{\heatprime}{{Q^\prime}}
\providecommand{\enthprime}{{h^\prime}}
\providecommand{\fradprime}{\bm{F}^\prime_{\rm rad}}
\providecommand{\kradprime}{\kappa^\prime_{\rm rad}}
\providecommand{\fcondprime}{\bm{F}^\prime_{\rm cond}}

% horizontally symmetric perturbations
\providecommand{\rhohat}{\hat{\rho}}
\providecommand{\tmphat}{\hat{T}}
\providecommand{\prshat}{\hat{p}}
\providecommand{\entrhat}{\hat{s}}
\providecommand{\intehat}{\hat{u}}
\providecommand{\enthhat}{\hat{h}}

% perturbations from mean
\providecommand{\rhoone}{\rho_1}
\providecommand{\tmpone}{T_1}
\providecommand{\prsone}{p_1}
\providecommand{\entrone}{s_1}
\providecommand{\inteone}{u_1}
\providecommand{\enthone}{h_1}

\providecommand{\pomega}{\varpi}

% VECTOR FIELDS
\providecommand{\vecu}{\bm{u}}
\providecommand{\vecv}{\bm{v}}
\providecommand{\veca}{\bm{A}}
\providecommand{\vecb}{\bm{B}}
\providecommand{\vecom}{\bm{\omega}}
\providecommand{\vecj}{\bm{\mathcal{J}}}

\providecommand{\upol}{\vecu_{\rm{pol}}}
\providecommand{\bpol}{\vecb_{\rm{pol}}}
\providecommand{\umer}{\vecu_{\rm{m}}}
\providecommand{\bmer}{\vecb_{\rm{m}}}

\providecommand{\urad}{{u_r}}
\providecommand{\utheta}{{u_\theta}}
\providecommand{\uphi}{{u_\phi}}
\renewcommand{\uphi}{{u_\phi}}

\providecommand{\ulambda}{{u_\lambda}}
\providecommand{\uz}{{u_z}}

\providecommand{\rhoumer}{\av{\rhotilde\umer}}
\providecommand{\rhourad}{\av{\rhotilde\urad}}
\providecommand{\rhoutheta}{\av{\rhotilde\utheta}}
\providecommand{\rhoulambda}{\av{\rhotilde\ulambda}}
\providecommand{\rhouz}{\av{\rhotilde\uz}}
\providecommand{\rhoomphi}{\av{\rhotilde\omphi}}

\providecommand{\omrad}{\omega_r}
\providecommand{\omtheta}{\omega_\theta}
\providecommand{\omphi}{\omega_\phi}
\providecommand{\omlambda}{\omega_\lambda}
\providecommand{\omz}{\omega_z}

\providecommand{\brad}{B_r}
\providecommand{\btheta}{B_\theta}
\providecommand{\bphi}{B_\phi}
\providecommand{\blambda}{B_\lambda}
\providecommand{\bz}{B_z}

\providecommand{\jrad}{\mathcal{J}_r}
\providecommand{\jtheta}{\mathcal{J}_\theta}
\providecommand{\jphi}{\mathcal{J}_\phi}
\providecommand{\jlambda}{\mathcal{J}_\lambda}
\providecommand{\jz}{\mathcal{J}_z}

% primed versions
\providecommand{\vecuprime}{\bm{u}^\prime}
\providecommand{\vecvprime}{\bm{v}^\prime}
\providecommand{\vecuhat}{\hat{\bm{u}}}
\providecommand{\vecvhat}{\hat{\bm{v}}}
\providecommand{\vecbprime}{\bm{B}^\prime}
\providecommand{\vecomprime}{\bm{\omega}^\prime}
\providecommand{\vecjprime}{\bm{\mathcal{J}}^\prime}
\providecommand{\vecuover}{\overline{\bm{u}}}
\providecommand{\vecvover}{\overline{\bm{v}}}
\providecommand{\wprime}{{w^\prime}}
\providecommand{\what}{{\hat{w}}}
\providecommand{\wover}{\overline{w}}
\providecommand{\vecbover}{\overline{\bm{B}}}
\providecommand{\vecomover}{\overline{\bm{\omega}}}
\providecommand{\vecjover}{\overline{\bm{\mathcal{J}}}}

\providecommand{\upolprime}{\vecu_{\rm{pol}}^\prime}
\providecommand{\bpolprime}{\vecb_{\rm{pol}}^\prime}
\providecommand{\umerprime}{\vecu_{\rm{m}}^\prime}
\providecommand{\bmerprime}{\vecb_{\rm{m}}^\prime}

\providecommand{\uradprime}{u_r^\prime}
\providecommand{\uthetaprime}{u_\theta^\prime}
\providecommand{\uphiprime}{u_\phi^\prime}
\providecommand{\ulambdaprime}{u_\lambda^\prime}
\providecommand{\uzprime}{u_z^\prime}

\providecommand{\avurad}{\av{u_r}}
\providecommand{\avutheta}{\av{u_\theta}}
\providecommand{\avuphi}{\av{u_\phi}}
\providecommand{\avulambda}{\av{u_\lambda}}
\providecommand{\avuz}{\av{u_z}}
\providecommand{\aventrhat}{\av{\entrhat}}

\providecommand{\omradprime}{\omega_r^\prime}
\providecommand{\omthetaprime}{\omega_\theta^\prime}
\providecommand{\omphiprime}{\omega_\phi^\prime}
\providecommand{\omlambdaprime}{\omega_\lambda^\prime}
\providecommand{\omzprime}{\omega_z^\prime}

\providecommand{\bradprime}{B_r^\prime}
\providecommand{\bthetaprime}{B_\theta^\prime}
\providecommand{\bphiprime}{B_\phi^\prime}
\providecommand{\blambdaprime}{B_\lambda^\prime}
\providecommand{\bzprime}{B_z^\prime}

\providecommand{\jradprime}{\mathcal{J}_r^\prime}
\providecommand{\jthetaprime}{\mathcal{J}_\theta^\prime}
\providecommand{\jphiprime}{\mathcal{J}_\phi^\prime}
\providecommand{\jlambdaprime}{\mathcal{J}_\lambda^\prime}
\providecommand{\jzprime}{\mathcal{J}_z^\prime}

% spherical coordinates
\providecommand{\cost}{\cos\theta}
\providecommand{\sint}{\sin\theta}
\providecommand{\cott}{\cot\theta}
\providecommand{\rsint}{r\sint}
\providecommand{\orsint}{\frac{1}{\rsint}}
\providecommand{\orsintline}{(1/\rsint)}
\providecommand{\rt}{r\theta}

% angular momentum
\providecommand{\amom}{\mathcal{L}}

% SIMULATION GEOMETRY
%s subscripts
\providecommand{\minn}{_{\rm{min}}}
\providecommand{\maxx}{_{\rm{max}}}
\providecommand{\inn}{_{\rm{in}}}
\providecommand{\out}{_{\rm{out}}}
\providecommand{\bott}{_{\rm{bot}}}
\providecommand{\midd}{_{\rm{mid}}}
\providecommand{\topp}{_{\rm{top}}}
\providecommand{\bcz}{_{\rm{bcz}}}
\providecommand{\ov}{_{\rm{ov}}}
\providecommand{\rms}{_{\rm{rms}}}
\providecommand{\const}{_{\rm{const}}}

\providecommand{\nr}{N_r}
\providecommand{\nt}{N_\theta}
\providecommand{\np}{N_\phi}
\providecommand{\nmax}{{n_{\rm{max}}}}
\providecommand{\lmax}{{\ell_{\rm{max}}}}

% SOLAR AND STELLAR VARIABLES
\providecommand{\lsun}{L_\odot}

\providecommand{\msun}{M_\odot}
\providecommand{\rstar}{R_*}
\providecommand{\lstar}{L_*}
\providecommand{\mstar}{M_*}

\providecommand{\rearth}{R_\oplus}
\providecommand{\omearth}{\Omega_\oplus}
\providecommand{\mearth}{M_\oplus}

% TORQUE DEFINITIONS
\providecommand{\taurs}{\tau_{\rm{rs}}}
\providecommand{\taumc}{\tau_{\rm{mc}}}
\providecommand{\tauv}{\tau_{\rm{v}}}
\providecommand{\taurad}{\tau_{\rm{rad}}}
\providecommand{\taums}{\tau_{\rm{ms}}}
\providecommand{\taumm}{\tau_{\rm{mm}}}
\providecommand{\taumag}{\tau_{\rm{mag}}}

% ANGULAR MOMENTUM FLUX DEFINITIONS
\providecommand{\aflux}{\bm{\mathcal{F}}}
\providecommand{\afluxrs}{\aflux^{\rm{rs}}}
\providecommand{\afluxmc}{\aflux^{\rm{mc}}}
\providecommand{\afluxv}{\aflux^{\rm{v}}}
\providecommand{\afluxms}{\aflux^{\rm{ms}}}
\providecommand{\afluxmm}{\aflux^{\rm{mm}}}
\providecommand{\afluxmag}{\aflux^{\rm{mag}}}

% time-scales using tau
\providecommand{\taunu}{\tau_{\nu}}
\providecommand{\taukappa}{\tau_{\kappa}}
\providecommand{\taueta}{\tau_{\eta}}
\providecommand{\tauff}{\tau_{\rm ff}}
\providecommand{\tauomega}{\tau_\Omega}
\providecommand{\taun}{\tau_N}
\providecommand{\taues}{\tau_{\rm ES}}

% time-scales using P
\providecommand{\pes}{{P_{\rm{ES}}}}
\providecommand{\pburrow}{{P_{\rm{b}}}}
\providecommand{\pessun}{{P_{ {\rm ES}, \odot}}}
\providecommand{\pnu}{{P_{\nu}}}
\providecommand{\pkappa}{{P_{\kappa}}}
\providecommand{\peta}{{P_{\eta}}}
\providecommand{\prot}{{P_{\rm{rot}}}}
\providecommand{\pom}{{P_{\Omega}}}
\providecommand{\pff}{{P_{\rm ff}}}
\providecommand{\pequil}{{P_{\rm{eq}}}}
\providecommand{\pcyc}{{P_{\rm{cyc}}}}
\providecommand{\pcycm}{{P_{{\rm cyc}, m}}}

% ``non-dimensional" time-scales using P_hat
\providecommand{\pesnd}{{\hat{P}_{\rm{ES}}}}
\providecommand{\pburrownd}{{\hat{P}_{\rm{b}}}}
\providecommand{\pessunnd}{{\hat{P}_{ {\rm ES}, \odot}}}
\providecommand{\pnund}{{\hat{P}_{\nu}}}
\providecommand{\pkappand}{{\hat{P}_{\kappa}}}
\providecommand{\petand}{{\hat{P}_{\eta}}}
\providecommand{\protnd}{{\hat{P}_{\rm{rot}}}}
\providecommand{\pomnd}{{\hat{P}_{\Omega}}}
\providecommand{\pffnd}{{\hat{P}_{\rm ff}}}
\providecommand{\pequilnd}{{\hat{P}_{\rm{eq}}}}
\providecommand{\pcycnd}{{\hat{P}_{\rm{cyc}}}}
\providecommand{\pcycmnd}{{\hat{P}_{{\rm cyc}, m}}}

% time-scales using t
\providecommand{\tes}{{t_{\rm{es}}}}
\providecommand{\test}{{t_{\rm{es,t}}}}
\providecommand{\tcirc}{{t_{\rm{circ}}}}
\providecommand{\tburrow}{{t_{\rm{b}}}}
\providecommand{\tessun}{{t_{ {\rm es}, \odot}}}
\providecommand{\tnu}{{t_{\nu}}}
\providecommand{\tnut}{{t_{{\nu}\rm,t}}}
\providecommand{\tkappa}{{t_{\kappa}}}
\providecommand{\teta}{{t_{\eta}}}
\providecommand{\trot}{{t_{\rm{rot}}}}
\providecommand{\tomega}{{t_{\Omega}}}
\providecommand{\tbrunt}{t_N}
\providecommand{\tvs}{{t_{\rm vs}}}
\providecommand{\trs}{{t_{\rm rs}}}
\providecommand{\tnucz}{{t_{\nu,{\rm cz}}}}
\providecommand{\tkappacz}{{t_{\kappa,{\rm cz}}}}
\providecommand{\tetacz}{{t_{\eta,{\rm cz}}}}

\providecommand{\tff}{{t_{\rm ff}}}
\providecommand{\tequil}{{t_{\rm{eq}}}}
\providecommand{\tcyc}{{t_{\rm{cyc}}}}
\providecommand{\trun}{{t_{\rm{run}}}}
\providecommand{\tmax}{{t_{\rm{max}}}}
\providecommand{\tcycm}{{t_{{\rm cyc}, m}}}
\providecommand{\tsun}{t_\odot}

% dimensional time-scales using t^*
\providecommand{\tesdim}{{t_{\rm{es}}^*}}
\providecommand{\testdim}{{t_{\rm{es,t}}^*}}
\providecommand{\tcircdim}{{t_{\rm{circ}}^*}}
\providecommand{\tesshdim}{{t_{\rm{es,sh}}^*}}
\providecommand{\tburrowdim}{{t_{\rm{b}}^*}}
\providecommand{\tnudim}{{t_{\nu}^*}}
\providecommand{\tnutdim}{{t_{{\nu}\rm,t}^*}}
\providecommand{\tkappadim}{{t_{\kappa}^*}}
\providecommand{\tetadim}{{t_{\eta}^*}}
\providecommand{\trotdim}{{t_{\rm{rot}}^*}}
\providecommand{\tomegadim}{{t_{\Omega}^*}}
\providecommand{\tbruntdim}{{t_N^*}}

\providecommand{\tffdim}{{t_{\rm ff}^*}}
\providecommand{\tffdimsq}{{t_{\rm ff}^{*2}}}

\providecommand{\tequildim}{{t_{\rm{eq}}^*}}
\providecommand{\tcycdim}{{t_{\rm{cyc}}^*}}
\providecommand{\trundim}{{t_{\rm{run}}^*}}
\providecommand{\tmaxdim}{{t_{\rm{max}}^*}}
\providecommand{\tsundim}{t_{\odot}^*}
\providecommand{\tvsdim}{t_{\rm vs}^*}
\providecommand{\trsdim}{t_{\rm rs}^*}

\providecommand{\tnuczdim}{{t_{\nu,{\rm cz}}^*}}
\providecommand{\tkappaczdim}{{t_{\kappa,{\rm cz}}^*}}
\providecommand{\tetaczdim}{{t_{\eta,{\rm cz}}^*}}

\providecommand{\omcyc}{{\omega_{\rm{cyc}}}}
\providecommand{\omcycm}{{\omega_{{\rm{cyc}}, m}}}
% NON-DIMENSIONAL NUMBERS

% fluid non-d
\providecommand{\ra}{{\rm{Ra}}}
\providecommand{\ratwo}{{\rm{Ra_2}}}
\providecommand{\sigmaeff}{{\sigma_{\rm eff}}}
\providecommand{\sigmaest}{{\sigma_{\rm est}}}
\providecommand{\sigmaeffloc}{{\sigma_{\rm eff,loc}}}
\providecommand{\sigmazero}{{\sigma_0}}
\providecommand{\sigmazerosq}{{\sigma_0^2}}
\providecommand{\sigmashsq}{{\sigma^2_{\rm sh}}}
\providecommand{\sigmaloc}{{\sigma_{\rm loc}}}
\providecommand{\sigmadyn}{{\sigma_{\rm dyn}}}
\providecommand{\sigmadynest}{{\sigma_{\rm dyn,est}}}

\providecommand{\raf}{\ra_{\rm{f}}}
\providecommand{\rafloc}{\ra_{\rm{f,loc}}}
\providecommand{\ramod}{\ra^*}
\providecommand{\rafmod}{\raf^*}
\providecommand{\pr}{{\rm{Pr}}}
\providecommand{\prm}{{\rm{Pr_m}}}
\providecommand{\ek}{{\rm{Ek}}}
\providecommand{\ekt}{{\rm{Ek_T}}}
\providecommand{\ta}{{\rm{Ta}}}
\providecommand{\roc}{{\rm{Ro_c}}}
\providecommand{\rocsq}{{\rm{Ro_c^2}}}
\providecommand{\bu}{{\rm{Bu}}}
\providecommand{\bumod}{{\rm{Bu^*}}}
\providecommand{\buvisc}{{\rm{Bu_{visc}}}}
\providecommand{\burot}{{\rm{Bu_{rot}}}}
\providecommand{\fr}{{\rm Fr}}
\providecommand{\ma}{{\rm Ma}}

\providecommand{\di}{{\rm{Di}}}

%\providecommand{\he}{{\rm{He}}}

% output non-d
\providecommand{\ro}{{\rm{Ro}}}
\providecommand{\lo}{{\rm{Lo}}}

\providecommand{\re}{{\rm{Re}}}
\providecommand{\rem}{{\rm{Re_m}}}
\providecommand{\pe}{{\rm{Pe}}}

% ref non-d
\providecommand{\Nrho}{N_\rho}
\providecommand{\nrho}{n_\rho}
\providecommand{\Nrhocz}{N_\rho^{\rm cz}}
\providecommand{\Nrhorz}{N_\rho^{\rm rz}}

% UNITS
\providecommand{\gram}{{\rm{g}}}
\providecommand{\cm}{{\rm{cm}}}
\providecommand{\meter}{{\rm{m}}}
\providecommand{\km}{{\rm{km}}}
\providecommand{\erg}{{\rm{erg}}}
\providecommand{\kelvin}{{\rm{K}}}
\providecommand{\dyn}{{\rm{dyn}}}
\providecommand{\second}{{\rm{s}}}
\providecommand{\radsecond}{{\rm rad\ s^{-1}}}
\providecommand{\minute}{{\rm min}}
\providecommand{\hour}{{\rm hr}}
\providecommand{\ayear}{{\rm yr}}
\providecommand{\aday}{{\rm day}}
\providecommand{\days}{{\rm days}}

\providecommand{\yr}{{\rm{yr}}}
\providecommand{\gauss}{{\rm{G}}}
\providecommand{\kelv}{{\rm{K}}}
\providecommand{\unitentr}{{\rm{erg\ g^{-1}\ K^{-1}}}}
\providecommand{\unitdsdr}{{\rm{erg\ g^{-1}\ K^{-1}\ cm^{-1}}}}
\providecommand{\uniten}{\rm{erg}\ \cm^{-3}}
\providecommand{\unitprs}{\rm{dyn}\ \cm^{-2}}
\providecommand{\unitrho}{\gram\ \cm^{-3}}
\providecommand{\stoke}{\rm{cm^2\ s^{-1}}}

% MEAN FIELD THEORY
\providecommand{\meanb}{\overline{\bm{B}}}
\providecommand{\flucb}{\bm{B}^\prime}
\providecommand{\totb}{\bm{B}}

\providecommand{\meanv}{\overline{\bm{v}}}
\providecommand{\flucv}{\bm{v}^\prime}
\providecommand{\totv}{\bm{v}}

\providecommand{\emf}{\bm{\mathcal{E}}}
\providecommand{\meanemf}{\overline{\bm{\mathcal{E}}}}
\providecommand{\meanbpol}{\overline{\bm{B}_{\rm{pol}}}}

% key radii

% simulation boundaries
\providecommand{\rin}{{r_{\rm in}}}
\providecommand{\rout}{{r_{\rm out}}}
\providecommand{\rbot}{{r_{\rm b}}}
\providecommand{\rtop}{{r_{\rm t}}}
\providecommand{\rmin}{{r_{\rm min}}}
\providecommand{\rmax}{{r_{\rm max}}}

\providecommand{\rindim}{r_{\rm in}^*}
\providecommand{\routdim}{r_{\rm out}^*}
\providecommand{\rbotdim}{r_{\rm b}^*}
\providecommand{\rtopdim}{r_{\rm t}^*}

% physical boundaries
\providecommand{\rbcz}{r_{\rm bcz}}
\providecommand{\rtcz}{r_{\rm tcz}}
\providecommand{\rbrz}{r_{\rm brz}}
\providecommand{\rtrz}{r_{\rm trz}}
\providecommand{\rc}{r_{\rm c}}
\providecommand{\rnrhothree}{r_{3}}
\providecommand{\rtach}{r_{\rm t}}
\providecommand{\rov}{r_{\rm ov}}
\providecommand{\dtach}{\Delta_{\rm t}}
\providecommand{\dtachsun}{\Delta_{\rm t,\odot}}

\providecommand{\rbczdim}{{r_{\rm bcz}^*}}
\providecommand{\rtczdim}{{r_{\rm tcz}^*}}
\providecommand{\rbrzdim}{{r_{\rm brz}^*}}
\providecommand{\rtrzdim}{{r_{\rm trz}^*}}
\providecommand{\rcdim}{{r_{\rm c}^*}}
\providecommand{\rnrhothreedim}{{r_{3}^*}}
\providecommand{\rtachdim}{{r_{\rm t}^*}}
\providecommand{\dtachdim}{{\Delta_{\rm t}^*}}

% solar radii
\providecommand{\rsun}{R_\odot}
\providecommand{\rbczsun}{R_{\rm bcz}}
\providecommand{\rnrhothreesun}{R_{\rm 3}}
\providecommand{\rtachsun}{R_{\rm t}}

% SIMULATION CODES
\providecommand{\rayleigh}{\texttt{Rayleigh}}
\providecommand{\mesa}{\texttt{MESA}}
\providecommand{\dedalus}{\texttt{Dedalus}}
\providecommand{\rayleigha}{\texttt{Rayleigh 0.9.1}}
\providecommand{\rayleighb}{\texttt{Rayleigh 1.0.1}}

\providecommand{\eulag}{\texttt{EULAG}}
\providecommand{\eulagmhd}{\texttt{EULAG-MHD}}
\providecommand{\ash}{\texttt{ASH}}
\providecommand{\rsst}{\texttt{RSST}}
\providecommand{\rtdt}{\texttt{R2D2}}
\providecommand{\pencil}{\texttt{Pencil}}

\providecommand{\newtext}[1]{\textbf{#1}}

% and new macros
\renewcommand{\dimm}{^*}
\newcommand{\dimsq}{^{*2}}
%\renewcommand{\delta}{\delta}

% new macros
\newcommand{\omshell}{\Omega_{\rm shell}}
\newcommand{\omhor}{\Omega_{\rm hor}}
\newcommand{\lshell}{L_{\rm shell}}
\newcommand{\ltransr}{L_{{\rm trans},r}}
\newcommand{\ltranshor}{L_{{\rm trans},hor}}
\newcommand{\rcut}{r_{\rm cut}}
\newcommand{\rnssl}{r_{\rm NSSL}}
\newcommand{\betarz}{\beta\rz}
\newcommand{\drz}{_{\rm drz}}
\newcommand{\rdrz}{r\drz}
\newcommand{\rdrzdim}{r\drz\dimm}
\newcommand{\dOmdrz}{\Delta\Omega\drz}

\newcommand{\avdrz}[1]{\left\langle#1\right\rangle_{\rm drz}}
\newcommand{\avdrzt}[1]{\left\langle#1\right\rangle_{{\rm drz},t}}
\newcommand{\est}{_{\rm est}}

% additional specialized timescales
\newcommand{\tesfull}{t_{\rm es,full}}
\newcommand{\tnufull}{t_{\nu,{\rm full}}}
\newcommand{\tespartial}{t_{\rm es,partial}}
\newcommand{\tnupartial}{t_{\nu,{\rm partial}}}

% when adding text:
\newcommand{\lm}[1]{{\color{red}#1}}
\newcommand{\lk}[1]{{\color{blue}#1}}
\newcommand{\nb}[1]{{\color{orange}#1}}
\newcommand{\twopapers}{\citetalias{Matilsky2022} and \citetalias{Matilsky2024}}

% angular momentum notation 
\newcommand{\amomrz}{L\rz}
\newcommand{\amomcz}{L\cz}
\newcommand{\damom}{\Delta L}
\newcommand{\thalf}{t_{3/4}}
%\newcommand{\thalf}{t_{75}}

% when deleting text:
\newcommand\lmout{\bgroup\markoverwith{\textcolor{red}{\rule[0.5ex]{2pt}{1pt}}}\ULon}
\newcommand\lkout{\bgroup\markoverwith{\textcolor{blue}{\rule[0.5ex]{2pt}{1pt}}}\ULon}
\newcommand\nbout{\bgroup\markoverwith{\textcolor{orange}{\rule[0.5ex]{2pt}{1pt}}}\ULon}

% when making comments: format [INITIALS: A comment]
\newcommand{\com}[1]{{\color{purple}#1}}

\renewcommand{\heatradtilde}{\tilde{Q}_{\rm rad}}
\renewcommand{\avsph}[1]{\overline{#1}}
\renewcommand{\avspht}[1]{\left(\overline{#1}\right)_t}
\providecommand{\avvolt}[1]{\left\langle{#1}\right\rangle_{{\rm v}, t}}

\renewcommand{\newtext}{}

% abstract
\begin{abstract}
At the base of the Sun's convective zone, a narrow shear layer called the tachocline separates strong latitudinal differential rotation above from nearly rigid rotation in the radiative zone below. The observed thinness of the tachocline is a long-standing dynamical puzzle because the tachocline should have spread significantly due to inward-burrowing meridional circulation, also called ``radiative spreading." We recently presented the first pair of global simulations to reveal a statistically stationary tachocline confined against radiative spreading by the Maxwell stresses from the \newtext{large-scale} nonaxisymmetric modes of a dynamo, which penetrated into and below the tachocline through a novel magnetic skin effect. In the work presented here, we systematically examine how this ``dynamo confinement scenario" works against radiative spreading in a suite of simulations as the governing parameters trend in the direction of the true solar regime. We find that as the stable stratification of the radiative zone is made progressively stronger, the dynamo cycles get longer, the magnetic field consequently penetrates deeper due to the skin effect, and the tachocline becomes more confined. Furthermore, these results have interesting consequences for solar spin-down. In all of our radiatively spreading simulations, the tachocline region spins down due to the burrowing circulation. Below the tachocline, the Maxwell stresses transmit this spin-down further to rigidify the deeper radiative zone. We thus speculate that, in addition to confining the tachocline, the dynamo may provide a pathway to communicate spin-down from the near-surface layers to the deep interior. 
\end{abstract}

%% Keywords, take from 
%% https://astrothesaurus.org

% Frontmatter over
% Begin sections!

\section{Introduction} \label{sec:intro}
\setcounter{footnote}{0} % Resets the counter to 0

 The helioseismically observed solar tachocline is an internal shear layer at the base of the solar convective zone (CZ), in which strong latitudinal differential rotation transitions to almost rigid rotation in the radiative zone (RZ) below \citep{Brown1989,Howe2009,Basu2016}. Helioseismology gives the centroid of the tachocline as $\rtach\approx0.7\rsun$ (where $\rsun\define\sn{6.96}{10}$ cm is the radius of the Sun), which roughly coincides with the base of the CZ.\footnote{Throughout this work, we reserve the symbol ``$\coloneqq$" for definitions and use ``$\equiv$" to mean ``equal to [some constant] everywhere and for all time."} The helioseismic thickness $\Delta$ of the tachocline is quite small, with one source giving $\Delta\approx0.02\rsun$ \citep{Elliott1999}. It is likely that $\Delta$ is too small to be resolved, in which case the published values are really upper bounds, roughly equal to the helioseismic inversion kernel width \citep{Howe2009}.

The solar magnetic field exhibits a fairly regular 22-year cycle in the dominant polarity of the surface toroidal magnetic field \citep{Hathaway2015}. These polarity reversals must be caused by an interior large-scale dynamo, i.e., a process by which turbulent plasma motions spontaneously generate the large-scale magnetic field and maintain the polarity cycles against ohmic dissipation (for reviews, see \citealt{Brandenburg2005b} and \citealt{Charbonneau2014}). Exactly where inside the Sun the  field is produced remains an open question (e.g., \citealt{Vasil2024}), but, because it has the most intense shear of any region of the Sun, the tachocline is often thought to be the ``seat" of the solar dynamo, i.e., the the primary location where toroidal magnetic field can be produced from the shearing of poloidal field \citep{Parker1993, Charbonneau1997}.

The observed thinness of the tachocline poses a major dynamical problem which is still inadequately understood. The main issue is how the tachocline can remain thin, despite the tendency of thin internal shear layers to spread over time due to the action of diffusion. In the solar radiative interior there are two types of spreading processes that could be important, involving either viscous or thermal diffusion (\citealt{Spiegel1992}; hereafter \citetalias{Spiegel1992}). A turbulently enhanced ``eddy" viscosity, for instance, could spread the tachocline simply through viscous drag \newtext{(a process we call ``viscous spreading")}, although microscopic values of the viscosity would be far too weak to have an effect. The thermal diffusivity, however, can also spread the tachocline, even without being turbulently enhanced (by a process outlined shortly below). Because the thermal diffusivity is dominated by radiation in most of the solar interior, this spreading process is called ``radiative spreading" in the context of the tachocline. The radiative thermal diffusivity is quite large, and as a result, radiative spreading of the solar tachocline is thought to dominate over viscous spreading. 

Radiative spreading occurs because the tachocline's baroclinicity (i.e., the latitudinal temperature gradients which must balance the axial variations of the centrifugal force in the tachocline; see \citealt{Matilsky2023}) must spread downward through thermal diffusion. This burrowing baroclinicity drives a circulation, and the baroclinicity and circulation both spread downward appreciably on evolutionary time-scales, despite being significantly inhibited by the RZ's strong stable stratification (e.g., \citealt{Clark1973, Haynes1991}). Any rotational shear (which is present everywhere in the solar CZ) will cause the burrowing circulation to transport angular momentum. In the solar context, this angular momentum transport causes the tachocline to spread. Indeed, \citetalias{Spiegel1992} showed that, if the Sun were born on the main sequence with an infinitesimally thin tachocline ($\Delta=0$), radiative spreading should have increased $\Delta$ from zero to $\approx0.4\rsun$ after $\approx5$ Gyr. 

Radiative spreading, if unopposed, should therefore have effectively eliminated any vestige of rigid rotation in the observable RZ, in direct contradiction with the helioseismic deductions. The observation of a confined tachocline thus implies the presence of a large-scale axial torque in the radiative interior which stops the radiative spreading. We call any process by which this torque is generated a ``confinement scenario." Many such scenarios have been constructed over the last few decades (e.g., \citealt{Spiegel1992,Kumar1997,Gough1998,ForgcsDajka2001,ForgcsDajka2002,Garaud2002,ForgcsDajka2004,Strugarek2011a,Kim2007,Wood2012,AcevedoArreguin2013,Barnabe2017,Matilsky2024,Garaud2025}). The details of each scenario, as well as which one is ``correct" in the solar context remain subjects of considerable debate. However, broadly two main possibilities are now invoked by the community.

The first scenario invokes primarily hydrodynamic (HD) shear instabilities and posits that anisotropic stably stratified turbulence (i.e., turbulence with much stronger horizontal motion  than vertical motion) results in a correspondingly anisotropic turbulent momentum transport, or viscosity. \citetalias{Spiegel1992} showed that an idealized anisotropic viscosity could prevent the inward spread of latitudinal shear. Recent investigations of stratified turbulence have systematically quantified the scaling of the turbulent vertical and horizontal viscosity and thermal diffusivity \citep{Cope2020, Chini2022, Shah2024} and a version of \citetalias{Spiegel1992}'s original confinement scenario was recently shown to work under the modern framework \citep{Garaud2025}. Note that in the HD scenario, although the latitudinal shear is eliminated by the shear instabilities (and so the tachocline remains thin), the RZ may not become strictly rigid. Instead, a radially-dependent ``shellular" rotation profile can remain (e.g., \citealt{Zahn1992}). 
 % (herefater \citetalias{Gough1998}).
 
The second scenario is magnetohydrodynamic (MHD) and invokes a primordial magnetic field, which is itself confined to the RZ \citep{Gough1998}. This primordial magnetic field is thought to rigidify the whole RZ via magnetic tension (in other words, by Ferraro's law; \citealt{Ferraro1937}) and thus could confine the tachocline with no residual shellular rotation in the RZ.  This scenario has the distinct advantage of addressing two observations simultaneously: tachocline confinement and global rigidification of the RZ.
 
\citet{ForgcsDajka2001} (hereafter \citetalias{ForgcsDajka2001}) proposed a third confinement scenario, also MHD in nature, but in which the cyclic solar dynamo field penetrates to a skin depth and confines the tachocline via a time-varying mean magnetic torque. This third scenario was first explored in axisymmetric two-dimensional models with an imposed latitudinal shear profile and oscillating poloidal magnetic field at the top boundary. \citetalias{ForgcsDajka2001} found that, under certain conditions, the resulting mean magnetic torque was able to keep the tachocline confined against viscous spreading in a steady state (see also \citealt{ForgcsDajka2002,ForgcsDajka2004}). Later, a similar scenario was shown to work against a parameterized form of radiative spreading \citep{Barnabe2017}. These early studies were specifically designed to elucidate which parameters of the two-dimensional models could result in skin depths that confined a tachocline and so the global rigidification of the RZ was not systematically investigated.

In \citet{Matilsky2022} (hereafter \citetalias{Matilsky2022}) and \citet{Matilsky2024} (hereafter \citetalias{Matilsky2024}), we demonstrated the possibility of a significantly more general form of \citetalias{ForgcsDajka2001}'s oscillating magnetic field confinement scenario, which we now call the ``dynamo confinement scenario." This scenario was first revealed by simulations of a rotating three-dimensional spherical shell with convection (in an upper CZ), spontaneously excited dynamo action, and overshoot of the convection into an underlying RZ. For strong enough \newtext{large-scale} nonaxisymmetric dynamo modes \newtext{(dynamo modes with low values of the azimuthal wavenumber $m$, say $m=1,2,...$)}, a tachocline was confined against viscous spreading by the associated Maxwell stresses. Additionally, the whole simulated RZ was forced into rigid rotation.  These nonaxisymmetric modes were quasiperiodic, formed first at the interface between CZ and RZ, and could potentially penetrate deeply into the RZ via a novel type of magnetic skin effect. In this particular skin effect, each dynamo mode penetrates to its own skin depth, which is dependent on both the frequency $\omega$ and azimuthal wavenumber $m$ of the mode. The ultimate frequency determining the skin-depth is $\omega$ Doppler shifted by $m$ times the rigid rotation rate of the deep RZ. For nonaxisymmetric modes nearly corotating with the RZ, this Doppler-shifted frequency can be quite small and therefore the skin depth can become quite large (see Section \ref{sec:skin}, as well as Sections 6 and 7 of \citetalias{Matilsky2024}). 

The simulations described by \twopapers\ were the first fully global, three-dimensional, nonlinear simulations to achieve statistically stationary tachoclines with a self-consistent confinement mechanism, namely, the dynamo confinement scenario described above. However, for computational feasibility, the dominant spreading process in these earlier simulations was viscous instead of radiative. \newtext{In general, when radiative spreading dominates over viscous spreading for a given system (in a sense which will be made precise later on), we say that the system is in the ``radiative spreading regime"; conversely, when viscous spreading dominates over radiative spreading, we say that the system is in the ``viscous spreading regime." Dividing the parameter space into these two core regimes helps to identify which tachocline theories are most appropriate in the solar context.} The main goal of this paper is to \newtext{examine in detail which parts of the dynamo confinement scenario, which were first identified in the viscous spreading regime, carry over to the radiative spreading regime. The radiative spreading} regime has historically been very difficult to achieve because of the extremely high resolution required for low enough viscosity. \citet{Korre2024} performed the only previous global models that we are aware of with significant radiative spreading. They did so (in their purely HD simulations) by lowering both the viscosity and the level of stable stratification significantly compared to prior work. We use a similar approach here (in both HD and MHD models) to achieve even lower-viscosity simulations (and higher stable stratification) at the limit of what is computationally tractable. 

We have recently performed many such radiatively spreading dynamo simulations and have announced the first major results of this study for one HD/MHD pair of simulations in a brief Letter \citep{Matilsky2025a} (hereafter, \citetalias{Matilsky2025a}). We found that the dynamo confinement scenario of {\twopapers}, i.e., the rigidifying Maxwell stresses, nonaxisymmetric dynamo modes, and novel skin effect, all persisted in the new MHD simulation reported there. In the corresponding HD simulation (used as a control), radiative spreading propagated the shear throughout the entire RZ. 

In this longer article, we examine the full simulation suite, composed of many HD/MHD simulation pairs, and describe how both the dynamo confinement scenario and the tachocline itself behave as we vary the salient parameters. In particular, we explore the effect of increasing the level of stable stratification of the RZ (i.e., pushing the stratification towards the extremely high solar value) while staying within the radiative spreading regime. We find that the dynamo confinement scenario is largely robust and that, as the level of stable stratification is increased, the dynamo cycles become longer, the poloidal magnetic field penetrates deeper into the RZ, and the tachoclines become more confined. 

Finally, we find that our new radiatively spreading dynamo simulations are relevant to the phenomenon of stellar spin-down---the observation that stars slow their bulk rotation rate over time due to the action of magnetically driven torques at the surface (e.g., \citealt{Kraft1967, Skumanich1972}). Exactly how these spin-down torques are transmitted deeper into the star (requiring an associated outward transport of angular momentum) remains an open question, which is complicated further by poorly constrained knowledge of stars' surface magnetic field topologies and mass-loss rates  (e.g., \citealt{Matt2012,Finley2018,Wood2018b}). In the solar case, the simultaneous presence of an ostensibly steady-state tachocline and significant spin-down from when the Sun arrived on the main sequence (when it rotated, say, 50--100 times faster than it does now; e.g., \citealt{Spiegel1972}) poses an interesting paradox. On the one hand, the existence of the tachocline suggests that little angular momentum transport should occur between the CZ and RZ, because clearly the CZ's differential rotation (and associated angular momentum anomaly) has been unable to spread downward. On the other hand, the fact that the RZ rotates at rate also achieved by the CZ at mid-latitudes (and not at the extremely fast rotation rate from solar birth) indicates that substantial angular momentum \textit{has} been transferred from the RZ to the CZ throughout the Sun's lifetime. Our radiatively spreading dynamo simulations present a possible solution to this paradox, because they achieve both stationary confined tachocline and slowly rotating RZs through a combination of radiative spreading and Maxwell stresses. As discussed at length in Section \ref{sec:disc}, the simulations described here thus offer a possible mechanism by which stars can both maintain confined tachoclines while simultaneously transporting angular momentum outward (thus propagating spin-down torques inward).

In Section \ref{sec:num}, we describe our numerical experiment and chosen parameter space. In Section \ref{sec:basic}, we describe the typical flows, magnetic fields, and statistically stationary states of our simulations. In Section \ref{sec:meanflows}, we present our simulated mean flows and quantify the confinement of the tachocline. In Section \ref{sec:radspread}, we describe the form of radiative spreading achieved in our simulations and its relationship to RZ spin-down. In Section \ref{sec:skin}, we discuss the simulated dynamos' cycling behavior and confirm that the same nonaxisymmetric skin effect identified in \citetalias{Matilsky2024} occurs in the radiative spreading regime, especially for strong stable stratification. In Section \ref{sec:disc}, we discuss our results in terms of tachocline confinement and spin-down. 

\section{Numerical Setup}\label{sec:num}
We evolve the three-dimensional anelastic HD or MHD equations in spherical shells using the open-source {\rayleigh} code \citep{Featherstone2016a,Matsui2016,Featherstone2024}. We use spherical coordinates $r\dimm$ (dimensional radius\footnote{From this section onward, asterisks denote explicitly dimensional variables while the lack of asterisks denote their nondimensional counterparts. Our chosen nondimensionalization is given in Section \ref{sec:numeq}.}), $\theta$ (colatitude), and $\phi$ (azimuth angle), as well as cylindrical coordinates $\lambda\dimm\define r\dimm\cos\theta$ (dimensional cylindrical radius, or moment arm) and $z\dimm\define r\dimm\sin\theta$ (dimensional axial coordinate). We denote unit vectors by $\e$, with a subscript indicating the local coordinate direction of the unit vector (for example, $\erad$ is the radial unit vector). The shell extends from an inner radius $\rindim$ to an outer radius $\routdim$. The evolution equations are solved in a rotating frame with constant background angular velocity $\Omzerovec=\Omzero\ez$. See Figure \ref{fig:schematic} for a schematic representation of this simulation geometry.

\subsection{Simulation Geometry and Reference State}
The anelastic equations allow significant density contrast across the shell but disallow sound waves \citep{Batchelor1953,Ogura1962,Gilman1981,Clune1999}. Formally, the anelastic approximation consists of assuming a solenoidal mass flux and thermal perturbations that are small relative to a well-chosen ``background" or ``reference" state. In \rayleigh, this thermal background state is always spherically symmetric and time-independent \citep{Featherstone2016a}. The full background state consists of the density $\rhotilde\dimm$, temperature $\tmptilde\dimm$, pressure $\prstilde\dimm$, gravitational acceleration $\gravtilde\dimm$, buoyancy frequency $\brunttilde\dimm$, prescribed internal heating $\heatradtilde\dimm$ (meant to represent radiative heating of the bottom layers of the CZ), kinematic viscosity $\nutilde\dimm$, thermometric conductivity $\kappatilde\dimm$ (which acts on the entropy perturbations), and magnetic diffusivity $\etatilde\dimm$. We give the details of this largely solar-like reference state in Appendix \ref{ap:ref}.   

As shown schematically in Figure \ref{fig:schematic}, we divide the spherical shell into two layers, separated in radius at the point $\rbczdim$. We choose the background squared buoyancy frequency $\brunttilde\dimsq$ to be positive in the lower layer ($\rindim$ to $\rbczdim$, which we call the ``radiative zone"), making it stable to convection. We force convective instability in the upper layer ($\rbczdim$ to $\routdim$, which we call the ``convective zone") through the internal heating profile $\heatradtilde\dimm$, which heats roughly the bottom third of the CZ. Because there is no physical boundary between the CZ and RZ, strong downflows from the CZ can overshoot into the underlying RZ (shown schematically in Figure \ref{fig:schematic} as the thin layer in green).

\begin{figure*}
	\centering
	%\vspace*{10cm}
	%\hspace*{-1cm}
	\includegraphics[width=7.25in]{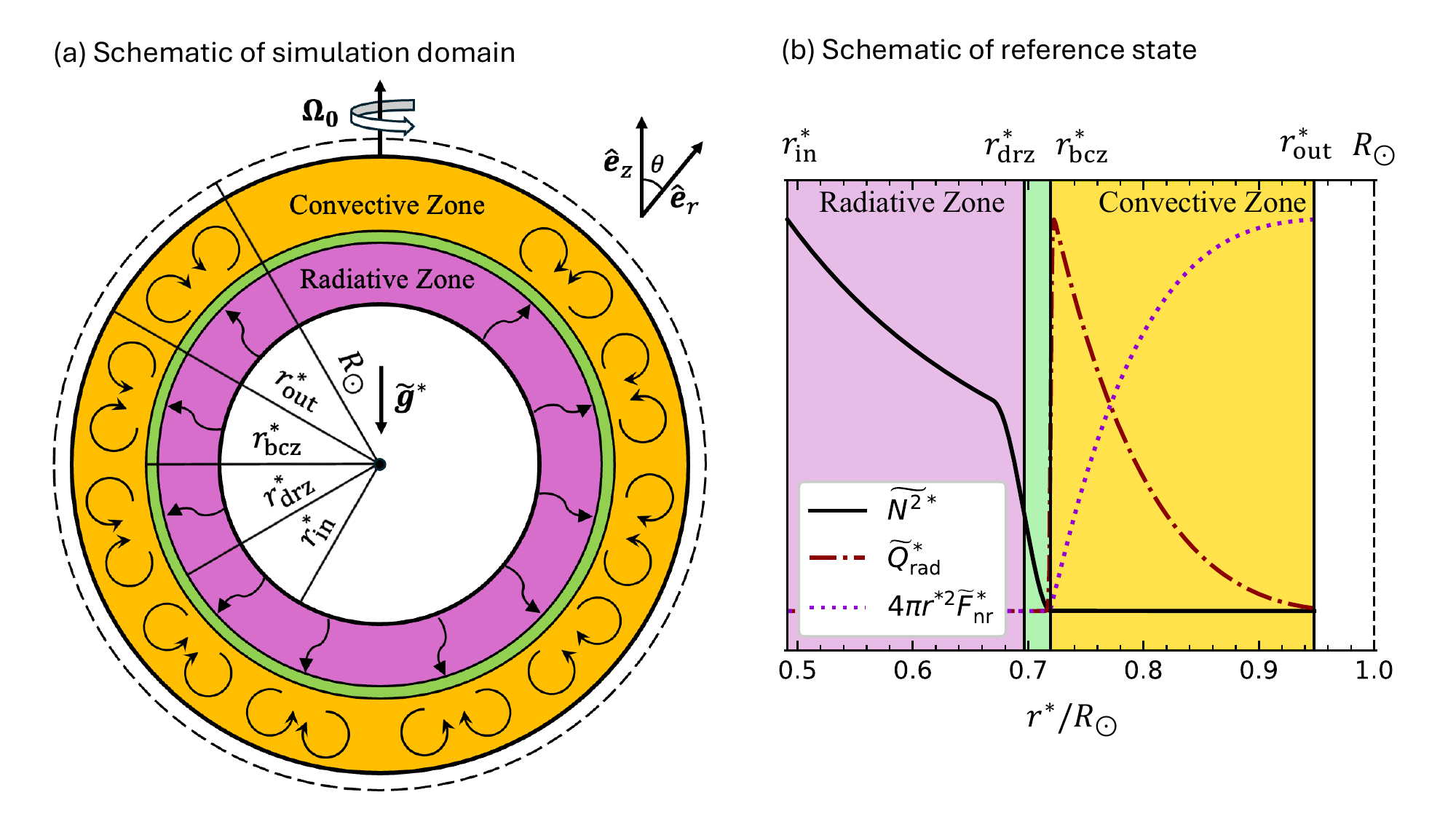}
	\caption{(a) Schematic of our three-dimensional spherical-shell simulation domain. The yellow region depicts the CZ with bulk overturning fluid motions represented by circular arrows. The CZ lies atop a thin stably stratified region (green) with significant nonlinear effects from overshoot and a deep RZ (purple) where the motion is essentially linear, represented by curvy ``radiation" arrows. (b) Schematic of the reference state which enforces the geometry of panel (a). Convection is forced in roughly the bottom third of the CZ by a prescribed internal heating $\heatradtilde\dimm$, which causes an accompanying ``nonradiative" energy flux $4\pi r\dimsq\fluxscalarnradtilde\dimm$ that convection and conduction must carry in a steady state to maintain thermal equilibrium (see \citealt{Featherstone2016a}). The stable stratification is enforced through the background positive squared buoyancy frequency $\brunttilde\dimsq$ in the RZ. These are the shapes of the profiles used for cases 0, 3, 4, and 6 and have been normalized to all lie on the same scale. The dashed circle (panel a) and line (panel b) mark the location of the true solar surface. }
	\label{fig:schematic}
\end{figure*}

We characterize the shell geometry by the aspect ratio of the CZ,
\begin{equation}\label{eq:beta}
	\beta\define\frac{\rbczdim}{\routdim}\equiv0.759
\end{equation}
and the aspect ratio of the RZ,
\begin{equation}\label{eq:betarz}
	\betarz \define \frac{\rindim}{\rbczdim} = \{0.683\ \text{or}\ 0.593\}.
\end{equation}
Note that $\beta$, $\rbczdim$, and $\routdim$ are fixed for all our simulations, but $\betarz$ and $\rindim$ take on one of two values, depending on the simulation (see Table \ref{tab:inputnondim}).

We characterize the degree of density stratification by the number of density scale-heights across the CZ,
\begin{equation}\label{eq:nrho}
\Nrho\define\ln\left[\frac{\rhotilde\dimm(\rbczdim)}{\rhotilde\dimm(\routdim)}\right]
\end{equation}
and the number of scale-heights across the RZ,
\begin{equation}\label{eq:nrhorz}
	\Nrhorz\define\ln\left[\frac{\rhotilde\dimm(\rindim)}{\rhotilde\dimm(\rbczdim)}\right].
\end{equation}

The CZ has the dimensional thickness 
\begin{align}
    H\define\routdim-\rbczdim,
\end{align}
which is a constant for all simulations.  

In the steady state, the convection and overshoot cause the thermal variables to develop spherically symmetric means over and above the background state, and the exact location of the base of the CZ (depending on how it is defined) deviates slightly from $\rbczdim$, as described in (e.g.) \citet{Korre2021}. However, these deviations are typically quite small and we neglect them for simplicity in our definitions of the CZ ($\rbczdim$ to $\routdim$) and RZ ($\rindim$ to $\rbczdim$). We also find that convective motions and nonlinear Reynolds stresses tend to die out around the fixed radius
\begin{align}
    {r\drz}\dimm\define\rbczdim-0.1H
\end{align}
for all simulations, despite variations in both the thickness of the overshoot layer (which can be substantial across our simulation suite because of large variations in the input parameters) and the true location of the base of the CZ. We thus call $(\rindim,\rdrzdim)$ the ``deep RZ." This is useful when we need to average over the RZ but wish to minimize the nonlinear effects of convective overshoot.

To explicitly identify our simulation domains with the solar interior, we choose $H=0.228\rsun$. This means that we simulate a sizable, but still incomplete, fraction of the true solar CZ, ignoring some of the tenuous near-surface layers (see Figure \ref{fig:schematic}). This helps make our radiative spreading simulations computationally tractable. We then calculate
\begin{subequations}\label{eq:radiidim}
\begin{align}
\rindim &= \frac{\beta\betarz}{1-\beta}H = \left\{0.491\ \text{or}\ 0.427\right\}\rsun,\\
\rdrzdim &= \left(\frac{\beta}{1-\beta}-0.1\right)H=0.697\rsun,\\
\rbczdim&=\frac{\beta}{1-\beta}H=0.720\rsun,\\ %\andd &\nonumber\\
\andd \routdim &= \frac{1}{1-\beta}H= 0.948\rsun.
\end{align}
\end{subequations}

We can then identify our simulation domains with the bottom $\sim$3 density scale-heights of the solar CZ  and the top $\sim$2 density scale-heights of the solar RZ according to the standard solar Model S (\citealt{ChristensenDalsgaard1996}; see Appendix \ref{ap:ref}). However, we note that this correspondence with Model S is rather rough, because our $\Nrho$ and $\Nrhorz$ deviate slightly from 3 and 2, respectively (see Table \ref{tab:inputnondim}).

We give details of the computational grid, including resolution, in Appendix \ref{ap:grid}.

\subsection{Types of Averages}
We use several types of averages in this work, denoted by angular brackets or overbars. Our averaging procedure is more precisely described in Appendix \ref{ap:averages}. If $\psi$ is a scalar quantity or a single component of a vector quantity, we denote an instantaneous azimuthal average by $\av{\psi}$. We also call $\av{\psi}$ the ``mean" $\psi$ and the deviation from this mean the ``fluctuating" $\psi$, which we denote with a prime, i.e., $\psi^\prime\define \psi - \av{\psi}$. We denote an instantaneous horizontal average over a spherical surface by $\avsph{\psi}$. We use $\avcz{\psi}$, $\avrz{\psi}$, and $\avdrz{\psi}$ to denote volume-averages over the CZ, RZ, and deep RZ, respectively. For a combined spatial and temporal average, we append a ``$t$" subscript to the averaging symbol, i.e., $\avt{\psi},\avspht{\psi},\avczt{\psi}$, and $\avdrzt{\psi}$. Unless otherwise specified, the temporal averaging interval is over the equilibrated state of the simulation.

\subsection{Equations of Motion}\label{sec:numeq}
We denote the vector velocity by $\vecu\dimm$ and the vector magnetic field by $\vecb\dimm$. We denote the small thermal perturbations about the background state by hats (here, we only need the pressure perturbation $\prshat\dimm$ and entropy perturbation $\entrhat\dimm$). To simplify the appearance of the MHD equations, we write the dimensional anelastic equations in Gaussian-CGS units:
\begin{align}
	\nabla\dimm\cdot(\rhotilde\vecu)\dimm &\equiv  0,
	\label{eq:contdim}
\end{align}
\begin{align}
	\nabla\dimm\cdot\vecb\dimm &\equiv  0,
	\label{eq:divb0dim}
\end{align}
\begin{subequations}
	\begin{align}
		\left(\rhotilde\frac{D\vecu}{Dt}\right)\dimm &= -2\rhotilde\dimm\Omzerovec\times\vecu\dimm\nonumber\\
        &\ \ \ -\left[\rhotilde\nabla \left(\frac{\prshat}{\rhotilde}\right)\right]\dimm +\frac{(\rhotilde\,\gravtilde \entrhat)\dimm}{\cp}\erad \nonumber\\
		&\ \ \ + \nabla\dimm\cdot (D_{ij}\dimm\e_i\e_j)	+ \frac{1}{4\pi}(\nabla\times\vecb)\dimm\times\vecb\dimm\nonumber,\\ \label{eq:momdim}\\
		\where D_{ij}\dimm &\define \left\{ 2\rhotilde\,\nutilde \left[e_{ij} - \frac{1}{3}(\nabla\cdot\vecu) \delta_{ij} \right]\right\}\dimm\label{eq:vstress}\\
		\andd e_{ij}\dimm &\define\frac{1}{2}\left(\pderiv{u_i}{x_j} + \pderiv{u_j}{x_i} \right)\dimm,\label{eq:ratestrain}
	\end{align}
\end{subequations}
\begin{align}
	\left(\rhotilde\tmptilde\frac{D\entrhat}{Dt}\right)\dimm =\ & -\cp\left[\rhotilde\tmptilde\frac{\brunttildesq}{\gravtilde} u_r\right]\dimm + \heatradtilde\dimm  + 
    \nabla\dimm\cdot\left(\kappatilde\,\rhotilde\tmptilde\nabla \entrhat\right)\dimm\nonumber \\
	&+ D_{ij}\dimm e_{ij}\dimm +  \frac{\etatilde\dimm}{4\pi}|\nabla\times\vecb|\dimsq,
	\label{eq:endim}
\end{align}
and
\begin{align}
	\pderiv{\vecb\dimm}{t\dimm}  =\ &\nabla\dimm\times (\vecu\times\vecb - \etatilde\nabla\times\vecb)\dimm. 
	\label{eq:inddim}
\end{align}
Here, $(D/Dt)\define(\pderivline{}{t})+\vecu\cdot\nabla$ is the material derivative, the indices $i$ and $j$ run over the three directions in any Cartesian coordinate system, repeated indices are summed, and $\cp$ is the specific heat at constant pressure (assumed constant for our reference state, as in a perfect gas).

We nondimensionalize the coordinates and fields using the following system of units:
\begin{subequations}\label{eq:nond}
\begin{align}
[\nabla\dimm] &= H^{-1},\\
[\partial/\partial t\dimm] &= \twoOmzero,\\
[\vecu\dimm] &= \twoOmzero H,\\
[\vecb\dimm] &= \sqrt{4\pi\rhocz}(\twoOmzero H),\\
[\prshat\dimm] &= \rhocz(\twoOmzero H)^2,\\
\andd [\entrhat\dimm] &=\Dentr \define \frac{\fluxnradcz H}{\rhocz\tmpcz\kappacz}.\label{eq:deltas}
\end{align}
\end{subequations}
Here, the ``cz" and ``rz" subscripts on a background-state quantity denote the volume-average of a dimensional background profile over the CZ or RZ, respectively (for example, $\rhocz$ is the volume-average of $\rhotilde\dimm$ over the CZ),
\begin{equation}\label{eq:fnrad}
\fluxscalarnradtilde\dimm(r\dimm) \define\frac{1}{r\dimsq}\int_{\rbcz\dimm}^{r\dimm}\heatradtilde\dimm\ofrprime \rprime^2d\rprime
\end{equation}
is the ``nonradiative" energy flux that convection and conduction must carry in the final statistically steady state (see \citealt{Featherstone2016a} and Figure \ref{fig:schematic}b). We thus compute $\Dentr$, the estimated entropy difference across the CZ, which is due to the imposed heating $\heatradtilde\dimm$ and its accompanying nonradiative energy flux $\fluxscalarnradtilde\dimm$.

We nondimensionalize the background state by scaling $\rhotilde\dimm$, $\tmptilde\dimm$, $\nutilde\dimm$, $\kappatilde\dimm$, $\etatilde\dimm$, and $\gravtilde\dimm$ by their volume-averages over the CZ, $\heatradtilde\dimm$ by $\fluxnradcz/H$, and $\brunttilde\dimm$ by $\bruntrz$. 

The nondimensional equations of motion are then

\begin{align}
	\Div(\rhotilde\vecu) &\equiv 0\label{eq:cont},\\
	\Div \vecb &\equiv 0\label{eq:divb0},
\end{align}
\begin{align}\label{eq:mom}
    \rhotilde\left(\matderiv{\vecu}\right) = &-\rhotilde\ez\times\vecu-\rhotilde\nabla\left(\frac{\prshat}{\rhotilde} \right) +\rocsq \rhotilde\, \gravtilde \entrhat\erad \nonumber\\
    & +\sqrt{\frac{\pr}{\raf}}\roc \Div (D_{ij}\e_i\e_j)\nonumber\\
    & +(\curl\vecb)\times\vecb,
\end{align}

\begin{align}\label{eq:heat}
	\rhotilde\tmptilde \matderiv{\entrhat} = &- \frac{\sigma^2}{\pr\rocsq} \rhotilde\tmptilde \frac{\brunttildesq}{\gravtilde} u_r+ \frac{\roc}{\sqrt{\pr\raf}} \heatradtilde\nonumber\\
    &+\frac{\roc}{\sqrt{\pr\raf}}\Div(\rhotilde \tmptilde \kappatilde \nabla \entrhat)  \nonumber\\
	&+ \sqrt{\frac{\pr}{\raf}}\frac{\di}{\roc}\left( D_{ij}e_{ij} + \frac{\etatilde}{\prm} |\curl\vecb|^2\right),
\end{align}
and
\begin{align}\label{eq:ind}
	\pderiv{\vecb}{t} = \curl(\vecu\times\vecb) - \sqrt{\frac{\pr}{\raf}}\frac{\roc}{\prm} \curl(\etatilde\curl\vecb).
\end{align}	

\begin{table*}
    \caption{Simulation input parameters, which are described more fully in the main text. The column ``TBC" (top boundary condition) refers to the two types of outer top thermal boundary condition described via Equation \eqref{eq:upper_thermal_BC}. The types of HD and MHD initial conditions (under column headers ``HIC" and ``MIC," respectively) are described in the text. The ``diff. type" column header refers to whether the background diffusivities $\nutilde$, $\kappatilde$, and $\etatilde$ all increase with height like $\rhotilde^{-1/2}$ or are all constant (note that in either case, $\pr$ and $\prm$ are spatially constant). The solar parameters are calculated according to the scheme in Appendix \ref{ap:dim}. The solar $\ek$, $\raf$, and $\roc$ are calculated from quantities averaged over the solar CZ, whereas $\bu$, $\sigma$, $\pr$, and $\prm$ are calculated from quantities averaged over the solar tachocline. For completeness, we note that in the solar CZ, $\pr = \sn{1.3}{-6}$ and $\prm=\sn{2}{-4}$ and in the solar tachocline, $\ek=\sn{2.6}{-14}$. } 
\label{tab:inputnondim}
\centering
\begin{tabular}{*{16}{l}}
\multicolumn{16}{c}{\textbf{Input parameters for the simulations}} \\ % This acts as a title
\hline
Name & $\bu$ & $\sigma$ & $\pr$ & $\prm$ & $\roc$ & $\raf$ & $\ek$ & $\beta$ & $\betarz$ & $\Nrho$ & $\Nrhorz$ & TBC & HIC & MIC  & diff. type\\
\hline
Case 0 & 5830 & 76.4 & 1 & 4  & 0.4 & 5.62e5 & 5.34e-04 & 0.759 & 0.683 & 3 & 2.1 & FF & noise & noise & $\propto\rhotilde^{-1/2}$\\
\hline
Case 1 & 0.222 & 0.471 & 1 & 4 & 0.5 & 5e6 & 2.24e-04 & 0.759 & 0.593 & 2 & 1.69 & FE & noise & H1  & constant\\
Case 2 & 0.222 & 0.471 & 1 & 4 & 0.5 & 1e6 & 5e-04 & 0.759 & 0.593 & 2 & 1.69 & FE & noise & H2  & constant \\
Case 3 & 0.166 & 0.288 & 0.5 & 4 & 0.4 & 5.62e5 & 3.77e-04 & 0.759 & 0.683 & 3 & 2.10 & FF & M0 & M0   & $\propto\rhotilde^{-1/2}$\\
Case 4 & 0.828 & 0.288 & 0.1 & 4 & 0.4 & 5.62e5 & 1.69e-04 & 0.759 & 0.683 & 3 & 2.10 & FF & M0 & M0  & $\propto\rhotilde^{-1/2}$\\
Case 5 & 2.22 & 1.49 & 1 & 4 & 0.5 & 1e6 & 5e-04 & 0.759 & 0.593 & 3 & 1.86 & FE & noise & H5 & constant\\
Case 6 & 33.1 & 2.88 & 0.25 & 4 & 0.4 & 5.62e5 & 2.67e-04 & 0.759 & 0.683 & 3 & 2.10 & FF & M0 & M0  & $\propto\rhotilde^{-1/2}$\\
\hline
Sun & 2.2e4 & 0.13 & 8e-7 & 7e-3 & 0.42 & 2e20 & 3.4e-14 & 0.713 & 0 & 11.4 & 6.7 & n/a & n/a & n/a & n/a \\
\hline
\end{tabular}
\end{table*}

\subsection{Control Parameters}

The control parameters appearing in Equations \eqref{eq:cont}--\eqref{eq:ind} are:
\begin{subequations}\label{eq:control}
\begin{align}
     \pr &\define \frac{\nucz}{\kappacz} \five \text{(thermal Prandtl number)},\label{eq:controlpr}\\
     \prm &\define \frac{\nucz}{\etacz} \five \text{(magnetic Prandtl number)},\\
     \raf &\define \frac{\gravcz H^3}{\nucz\kappacz}\left(\frac{\Dentr}{\cp}\right)= \frac{\fluxnradcz \gravcz H^4}{\rhocz\tmpcz\cp\nucz\kappacz^2}\nonumber\\
     &\five \text{(flux-based Rayleigh number)},\\
      \roc &\define \frac{\sqrt{\gravcz(\Dentr/\cp)}}{\twoOmzero} =\ek\sqrt{\frac{\ra}{\pr}}\nonumber\\
      &\five \text{(convective Rossby number)}\label{eq:controlroc},\\
      &\nonumber\\
      &\text{and}\nonumber\\
       \sigma &\define \left(\frac{\bruntrz}{\twoOmzero}\right)\sqrt{\frac{\nucz}{\kappacz}}=\sqrt{\bu\pr}\nonumber\\
       &\five\text{(sigma  parameter).}\label{eq:controlsigma}
\end{align}
\end{subequations}

In Equations \eqref{eq:controlroc} and \eqref{eq:controlsigma} we have implicitly defined the Ekman number
\begin{align}\label{eq:controlek}
    \ek&\define \frac{\nucz}{\twoOmzero H^2}
\end{align}
and the buoyancy number
\begin{align}\label{eq:controlbu}
    \bu\define\frac{\bruntrz^2}{4\Omzero^2}.
\end{align}

The five independent parameters in Equations \eqref{eq:control} fully characterize the system and we thus treat $\ek$ and $\bu$ as dependent parameters. The ``dissipation number" 
\begin{equation}\label{def:di}
\di\define\frac{\gravcz H}{\cp\tmpcz}
\end{equation} 
is $O(1)$ due to the virial theorem and is also not an independent control parameter, but can be calculated from $\Nrho$, $\beta$, and the ratio of specific heats $\gamma$ (see Appendix \ref{ap:ref}). 

The physical meanings of the various control parameters in Equations \eqref{eq:control} become clearer when we write each control parameter as the ratio of dimensional time-scales. This is done explicitly for all our control parameters in Appendix \ref{ap:timescales}.

Although the equations are solved nondimensionally and we report all results in nondimensional form, we can always consider an equivalent dimensional system. This ``redimensionalization" process is outlined in Appendix \ref{ap:dim}, where we explicitly show how to translate each simulation's nondimensional quantities and time-scales into units appropriate for the solar context.

The input parameters for our simulations and the Sun are given in Table \ref{tab:inputnondim}. For the Sun, we calculate the nondimensional numbers using the dimensional parameters given in Appendix \ref{ap:dim}. 

\subsection{Parameter Space}\label{sec:par}
 
The relative importance of the two spreading processes identified by \citetalias{Spiegel1992} (viscous and radiative) can be characterized  by the following time-dependent nondimensional diagnostic parameter. We call this the dynamical sigma parameter:  
\begin{align}\label{eq:sigmadyn}
	\sigmadyn(t)\define \left|\frac{\int_0^t \avdrz
    {\tauv}(t^\prime)dt^\prime}{\int_0^t \avdrz{\taumc}(t^\prime)dt^\prime}\right|^{1/2},
\end{align}
where $\tauv$ is the viscous torque density and $\taumc$ is the torque density due to the meridional circulation. The dynamical sigma parameter measures the ratio of the cumulative angular momentum transport (from $t=0$ to the current time $t$) into or out of the deep RZ by the viscous spreading to the cumulative transport by the radiative spreading. We consider angular momentum transport to or from the deep RZ to minimize the nonlinear influence of overshoot dynamics like large Reynolds stresses. When $\sigmadyn(t)<1$, we say that the shear is ``radiatively spreading" and when $\sigmadyn(t)>1$, we say that the shear is ``viscously spreading." Note that often a given system exhibits different types of spreading at different times. For example, the final steady-state torque balance in simulated RZs with no magnetic field is often a balance between $\taumc$ and $\tauv$, a situation also referred to as the ``final diffusion layer" by \citet{Clark1973}. For such HD cases, the system might start out radiatively spreading ($\sigmadyn(t)\ll1$ at early times), but will always end up balanced ($\sigmadyn(t)\rightarrow 1$ at late times). Other factors, like magnetic fields and residual Reynolds stresses even for $r<r\drz$, complicate the spreading processes further. 

%Note that in our simulations, angular momentum transport always occurs from the deep RZ into the upper layers, thus spinning the bulk RZ down. In this work, we thus take ``radiative spreading" to mean that most of the RZ spin-down occurs through burrowing meridional circulation. In other words, if $\sigmadyn(t)<1$ once most of the angular momentum transport out of the deep RZ has occurred (a time we refer to as $\thalf$; see Section \ref{sec:radspread}), we \newtext{classify a given simulation as being in the radiative spreading regime. If $\sigmadyn(t)>1$ at $t=\thalf$, we classify the simulation as being in the viscous spreading regime.} 

Note that in our simulations, angular momentum transport always occurs from the deep RZ into the upper layers, thus spinning the bulk RZ down. In this work, we thus take ``radiative spreading" to mean that most of the RZ spin-down occurs through burrowing meridional circulation.  \newtext{
Since we define the time where most of the angular momentum transport out of the deep RZ has already occurred  as $t=\thalf$ (see Section \ref{sec:radspread}), if 
$\sigmadyn(t=\thalf)<1$ then we classify a given simulation as being in the radiative spreading regime.  If $\sigmadyn(t=\thalf)>1$, we classify the simulation as being in the viscous spreading regime.} 

Because $\sigmadyn(t)$ is a time-dependent emergent property of the system, it is useful to have an input parameter that suggests a priori which spreading regime the system is likely to be in (i.e., whether $\sigmadyn(\thalf)<1$ or $\sigmadyn(\thalf)>1$ is most likely to be achieved). As shown by \citetalias{Spiegel1992} (and earlier by \citealt{Clark1973}), radiative spreading occurs on the Eddington-Sweet time-scale $\tesdim$, whereas viscous spreading occurs on the viscous diffusion time-scale $\tnudim$ (see Appendix \ref{ap:timescales}). It is therefore typical to attempt to control the spreading regime (e.g., \citealt{Garaud2008a, Garaud2009, Wood2012, AcevedoArreguin2013, Korre2024}) by choosing the input sigma parameter, because
\begin{equation}\label{eq:sigma}
\sigma=\sqrt{\frac{\tesdim}{\tnudim}}=\sqrt{\dfrac{\text{radiative spreading time-scale}}{\text{viscous spreading time-scale}}}. 
\end{equation}
(See Equations \ref{eq:timescalesdim} and \ref{eq:controltimeratio}.) For $\sigma$ low enough, we expect the system to be in the radiative spreading regime, whereas for $\sigma$ too high, we expect the simulation to be in the viscous spreading regime. Although there clearly must be a critical order-unity value $\sigma_c$ that separates the two regimes a priori, it is not obvious that $\sigma_c=1$ is the exact boundary between regimes (a situation which is common for other dimensionless numbers as well). In fact, it is well-known that radiative spreading occurs much more rapidly than viscous spreading at early times due to the higher order (i.e., more acute response to changes in the boundary-layer width) of a radiatively spreading tachocline compared to a viscously spreading one (see \citealt{Clark1973} and \citetalias{Spiegel1992}). We thus at least expect $\sigma_c>1$, which we confirm in Appendix \ref{ap:sigmac} by estimating $\sigma_c\approx10$ for our specific suite of simulations.

Because $\sigma=\sqrt{\pr\bu}$, both the relative strengths of the diffusivities (measured by $\pr$) and the strength of stable stratification of the RZ (measured by $\bu$) are important for radiative spreading. Low $\sigma$ can be achieved by low $\pr$ or low $\bu$. For the solar tachocline, $\bu=\sn{2.2}{4}$, $\pr\approx\sn{8}{-7}$, and $\sigma\approx0.13$ (see Table \ref{tab:inputnondim}). In other words, the Sun achieves low $\sigma$, despite being very strongly stably stratified, by having extremely low $\pr$. We would ideally explore the same parameter space in global simulations, but this is currently impractical because of the extreme resolution required for such a low $\pr$ (wherein the low viscosity leads to very small-scale turbulent structures that require a very fine grid to resolve). Instead, we take advantage of the fact that $\sigma_c>1$ to access the radiative spreading regime even for some cases where $\pr$ is not $\ll1$, but is still $<1$.

We consider seven pairs of HD/MHD simulations in total, the parameters of which are listed in Table \ref{tab:inputnondim}. These are named Cases H0--H6/M0--M6 and when we refer to an HD/MHD pair, we omit the letter (e.g., ``Case 1" refers to the pair of Cases H1 and M1). Each HD case has the same HD parameters as the corresponding MHD case, except, in some cases, for the type of initial condition. Case 0 is basically a nondimensional version of the viscously spreading pair of simulations discussed in \citetalias{Matilsky2022}.  Cases 1--6 are all in the radiative spreading regime, as we verify in Section \ref{sec:radspread}. Case 6 is the same pair of simulations considered in \citetalias{Matilsky2025a}.

Figure \ref{fig:input} shows our chosen path through parameter space for our simulation suite \newtext{and the relative location of the Sun. The horizontal line $\sigma=\sigma_c\approx10$ roughly separates this parameter space into the viscous spreading and radiative spreading regimes (at a value of $\sigma$ greater than unity, as discussed in Appendix \ref{ap:sigmac}). Quantitatively, the Sun is further into the radiative spreading regime than any of the simulations (it has the lowest $\sigma$), having both the highest stratification (highest $\bu$) and the lowest $\pr$. All the simulations are quite far from solar (2--3 orders of magnitude in one or more of our key parameters $\sigma$ and $\bu$, and, of course, very far off in others, such as $\raf$). However, Cases 1--6 (which all lie firmly in the radiative spreading regime) at least have the proper \textit{ordering} of key time-scales, e.g., $\tomegadim<\tesdim<\sigma_c\tnudim$ (where $\tomegadim\define(\twoOmzero)^{-1}$ is the rotational time-scale). Due to numerical limitations, the \textit{separations} of these time-scales are necessarily much different than the separations expected in the Sun. For instance, for Case 6, $\tomegadim\approx0.8$ days, $\tesdim\approx2000$ yr, and $\sigma_c\tnudim\approx2500$ yr, whereas for the Sun, $\tomegadim\approx2$ days, $\tesdim\approx140$ Gyr, and $\sigma_c\tnudim\approx30$ Tyr (see Table \ref{tab:time-scales_dim}).  }

\begin{figure}
	\centering
	%\vspace*{10cm}
	%\hspace*{-1cm}
	\includegraphics[width=3.4375in]{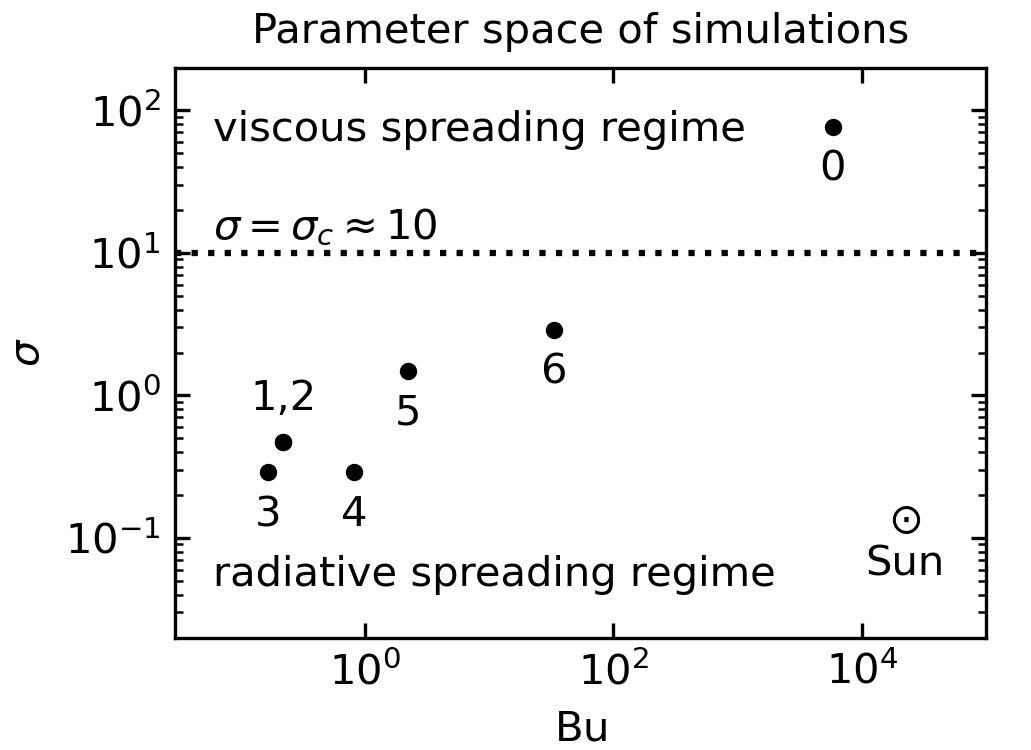}
	\caption{Location in parameter space ($\bu$ and $\sigma$) of the different simulations and the Sun. Cases 1 and 2 have the same $\bu$ and $\sigma$, but different $\raf$ (see Table \ref{tab:inputnondim}). The viscous and radiative spreading regimes are separated at $\sigma=\sigma_c\approx10$ (see Appendix \ref{ap:sigmac}). }
	\label{fig:input}
\end{figure}

\subsection{Initial and Boundary Conditions}
We use several types of initial conditions for our simulations, which are given under the ``HIC" headings (for the HD cases) and the ``MIC" headings (for the MHD cases) in Table \ref{tab:inputnondim}. ``Noise" means that $\entrhat$ (and $\vecb$ for the MHD cases) is initialized from weak $O(10^{-4}\text{--}10^{-2})$, pseudorandom, small-scale noise everywhere in the shell, while all other fields are initialized from zero. We sometimes initialize simulations from the equilibrated checkpoints of a different simulation, and this type of initial condition is denoted in Table \ref{tab:inputnondim} by the initializing case's name. When initializing an HD case from an MHD case (Cases H3, H4, and H6), the fields $\vecu$, $\entrhat$, and $\prshat$ are initialized from the MHD case, while the MHD case's $\vecb$ is simply unused. When initializing an MHD case from an HD case (Cases M1, M2, and M5), $\vecu$, $\entrhat$, and $\prshat$ are initialized from the corresponding HD case, while $\vecb$ is initialized from a combination of noise in the CZ and zero in the RZ.  

Cases H0 and M0 are, respectively, the nondimensional versions of ``the HD case" and ``the MHD case" from \citetalias{Matilsky2022} (or equivalently, ``Case H" and ``Case 4.00" from \citetalias{Matilsky2024}) and behave identically to their previously published dimensional counterparts. In particular, Case M0 has a very well-confined tachocline (but recall that it is in the viscous spreading regime). Cases 3, 4, and 6 are intended to explore the effect of a dynamo on a previously confined tachocline that begins to radiatively spread, which is a (possibly magnetized) version of the situation envisioned by \citetalias{Spiegel1992}. Cases 3, 4, and 6 (both HD and MHD) are therefore all initialized from the confined-tachocline state of Case M0.

Cases 1, 2, and 5 are initialized more traditionally (e.g., \citealt{Brown2010,Augustson2015,Guerrero2016a,Matilsky2020a}). The HD cases are initialized from random noise, whereas the MHD cases are initialized from the late stages of their corresponding equilibrated HD counterparts with the addition of CZ noise for the magnetic field. Therefore, in Cases 1, 2, and 5, the dynamo develops upon a differential rotation profile that has already radiatively spread into the RZ.

All our cases share the following boundary conditions at both the top and bottom: impenetrable ($u_r\equiv0$), stress-free ($\pderivline{}{r}(u_\theta/r) \equiv0$ and $\pderivline{}{r}(u_\phi/r)\equiv0$), and potential magnetic field ($\vecb=\nabla\Phi$, where $\nabla^2\Phi\equiv0$ outside the simulation domain). At the bottom, all the cases have the thermal boundary condition
\begin{equation}\label{eq:lower_thermal_BC}
    \pderiv{\entrhat}{r}\bigg{|}_{\rin}\equiv0,
\end{equation}
whereas at the top, the cases are either ``fixed-flux" (FF; Cases 3, 4, and 6):
\begin{equation}\label{eq:upper_thermal_BC}
    \pderiv{\entrhat}{r}\bigg{|}_{\rout} = -\frac{\int_{\rin}^{\rout}\heatradtilde(r)r^2dr}{(\rhotilde\tmptilde\kappatilde r^2)|_{r=\rout}},
\end{equation}
or ``fixed-entropy" (FE; Cases 1, 2, and 5): 
\begin{align}
    \entrhat|_\rout\equiv0.
\end{align}
(See the column with heading ``TBC" in Table \ref{tab:inputnondim}.)

\section{Basic Simulation Properties}\label{sec:basic}
In Figure \ref{fig:cutout3d}, we show (at a typical late time in Case M6) spherical and meridional surfaces cut out of the simulation domain exhibiting (a) the fluctuating axial vorticity $\omzprime$ (we define the vector vorticity $\vecom\define\curl\vecu$) and (b) the colatitudinal component of the poloidal magnetic field, $\btheta$. The nature of the flows and fields shown here is representative of all our cases. There is a complex pattern of relatively slowly-moving axially-aligned Taylor columns around the equator in the CZ, which penetrate to varying degrees into the RZ and excite strong horizontal flows there in the form of relatively fast-moving Rossby waves (see \citealt{Blume2024} for a clear identification of Rossby waves in a similar simulation). The poloidal magnetic field is strongly nonaxisymmetric, but still contains a large-scale envelope (dominated by $m=(0,1,2)$, where $m$ is the azimuthal wavenumber). The field has much smaller spatial and temporal scales in the CZ than it does in the RZ, where only the slowly-evolving $m=(0,1,2)$ envelope survives. 

%\com{NB: add appropriate symbol e.g. $Ro_{\rm fluc}$ or say all in words?}
In Appendix \ref{ap:outputdim}, we describe the general dynamical regime of our simulation suite in terms of the rotational constraint, turbulence level, and partition of the various forms of energy (see Table \ref{tab:outputnd}). In general, all our simulations have moderately rotationally constrained CZs (i.e., the Rossby numbers for the fluctuating flow $\ro_{\rm fluc}$ are $\lesssim1$) and the CZs are moderately turbulent (i.e., the Reynolds numbers for the fluctuating flow $\re_{\rm fluc}$ are $O(100)$). In the RZs, the flows are very rotationally constrained (the Rossby numbers for the mean flows $\ro_{\rm mean}$, along with $\ro_{\rm fluc}$, are both $O(0.1)$) but are still reasonably turbulent ($\re_{\rm fluc}=O(10$--$100)$). 

\begin{figure*}
	\centering
	%\vspace*{10cm}
	%\hspace*{-1cm}
	\includegraphics[width=7.25in]{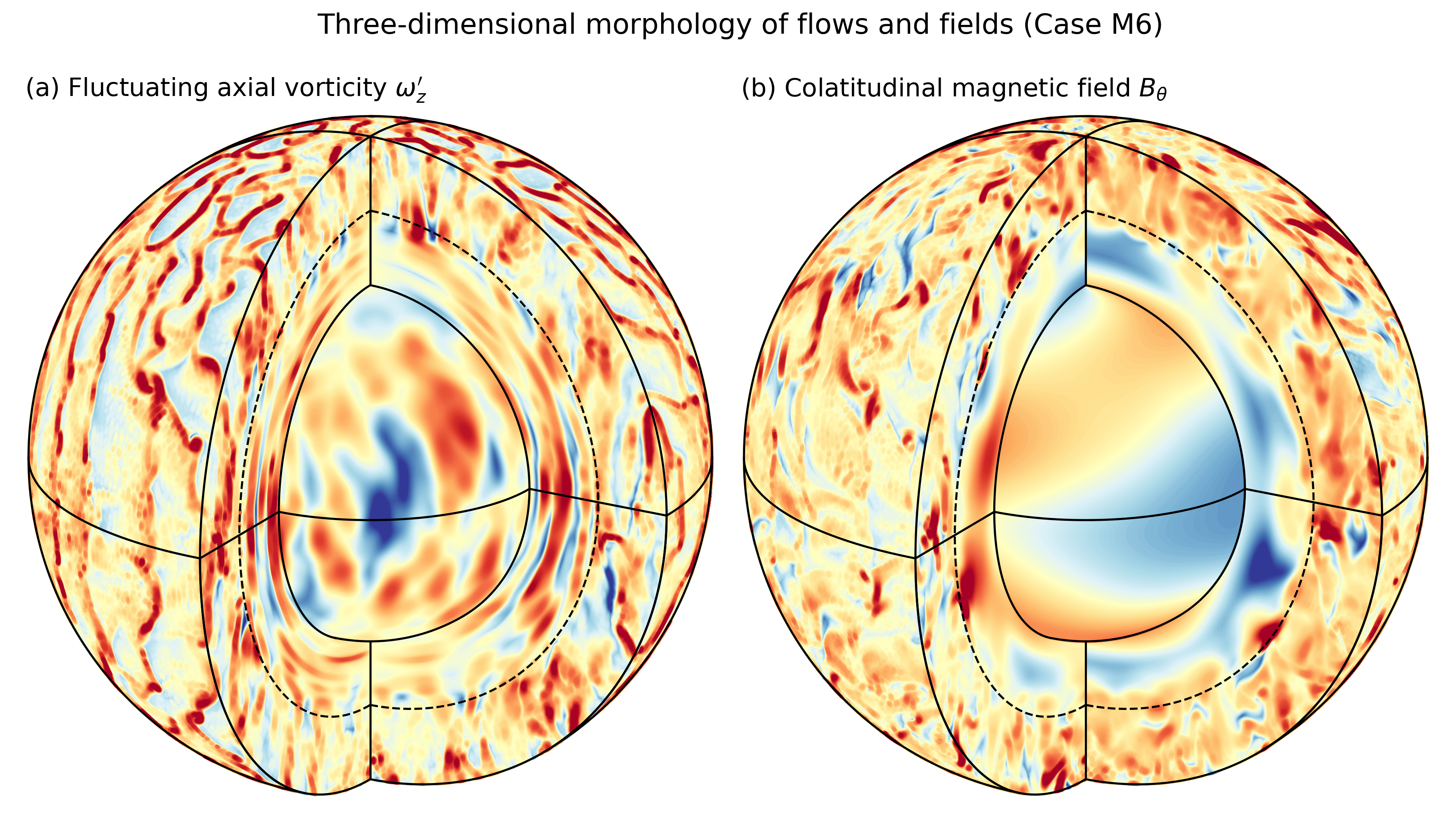}
	\caption{Spherical and meridional surfaces cut out of the simulation domain showing snapshots of the simulated flows in Case M6 for (a) the fluctuating axial vorticity $\omzprime$ and (b) the colatitudinal component of the poloidal magnetic field $\btheta$.  Red/yellow/blue tones indicate positive/zero/negative values. Each snapshot is taken toward the end of the simulation ($t=\trun$). The inner and outer spherical surfaces bounding the shell are taken just above and below the inner and outer domain boundaries $\rin$ and $\rout$, respectively. The meridional surfaces are located $90^\circ$ apart, at $\phi=-30^\circ$ and $\phi=60^\circ$ (with $\phi=0^\circ$ being the central viewing longitude; \newtext{the central viewing latitude is $20^\circ$ north}). Each radial level is normalized separately by the rms of the field at that level, leading to enhanced signal from the deeper regions, whose flows and fields are in fact much weaker-amplitude than they are higher up. The interface $r=\rc$ is marked by dashed black curves. \newtext{An animated version of this figure is available in the online journal. In the 26-second animation, the short-term evolution of $\omzprime$ ($t=46542$ to $t=46702$) and long-term evolution of $\btheta$ ($t=0$ to $t=17023$) are shown side by side. Animated versions of the separate panels are also available for download at \url{https://doi.org/10.5281/zenodo.18265773} and for viewing at \url{https://www.youtube.com/shorts/z425Iy9Loc4} (short-term evolution of $\omzprime$), \url{https://www.youtube.com/shorts/BQfHSm4_I0o} (long-term evolution of $\btheta$), and \url{https://www.youtube.com/shorts/WjBdOqFB_Yo} (short-term evolution of $\btheta$).}}
	\label{fig:cutout3d}
\end{figure*}

\section{Mean Flow Profiles}\label{sec:meanflows}
\subsection{Differential Rotation}
We define the nondimensional time-dependent mean rotation rate $\Omega$ in the meridional plane (in the nonrotating ``lab" frame) via
\begin{equation}\label{eq:omega}
\Omega(r,\theta,t) \define \frac{1}{2} + \frac{\av{\uphi}}{\lambda}.
\end{equation}
Note that the dimensional counterpart is just
\begin{equation}\label{eq:omegadim}
\Omega^* = (\twoOmzero)\Omega =  \Omzero + \frac{\av{\uphi^*}}{\lambda^*}
\end{equation}
and that the background frame rate is simply $1/2$ in nondimensional units. 

In this work, we discuss the ``differential rotation," defined by
\begin{align}\label{eq:diffrot}
2\Omega-1 = \frac{2\av{\uphi}}{\lambda} = \frac{\Omega\dimm}{\Omzero}-1.
\end{align}

\begin{figure*}
	\centering
	%\vspace*{10cm}
	%\hspace*{-1cm}
	\includegraphics[width=7.25in]{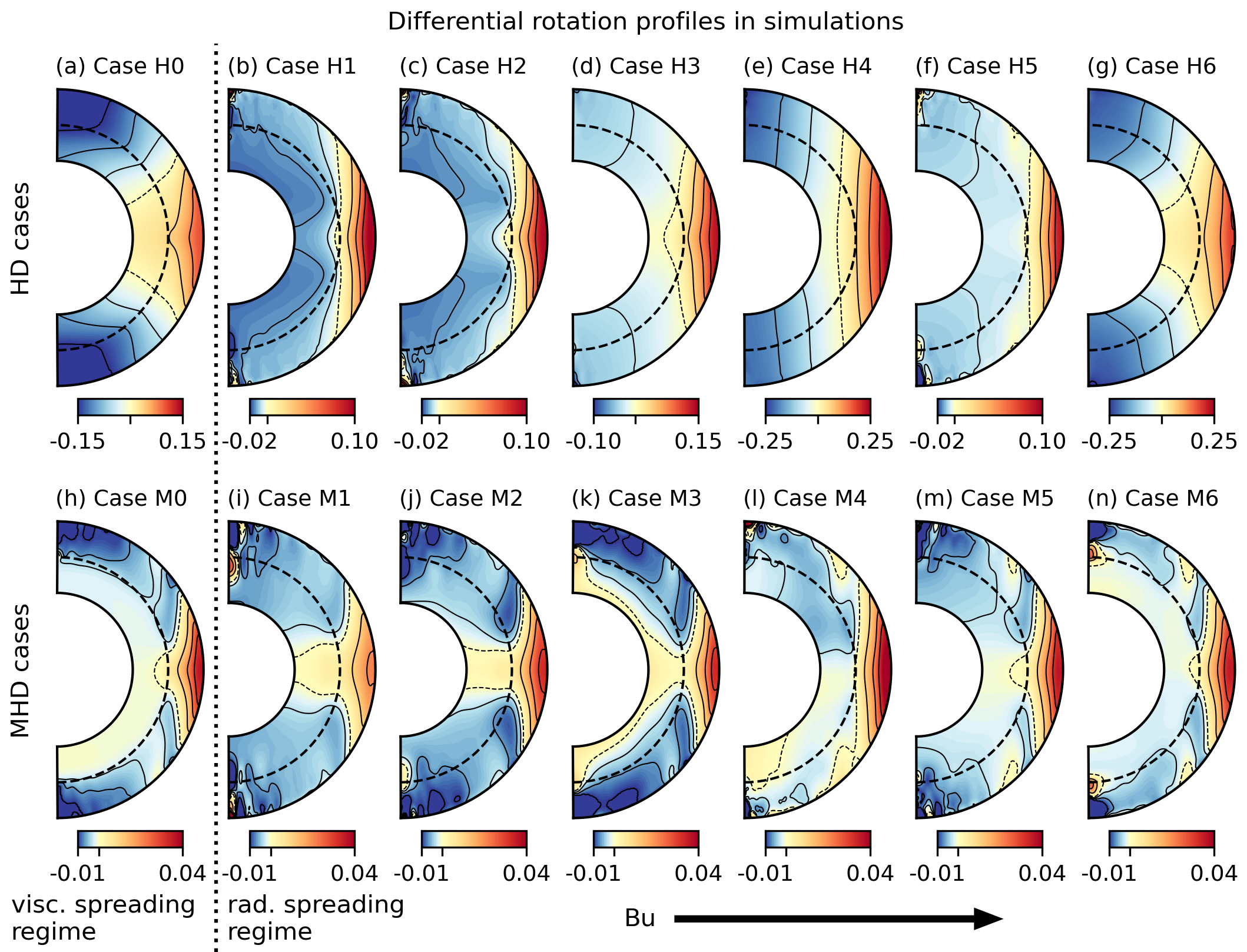}
	\caption{Steady-state differential rotation $\avt{2\Omega-1}$ for the full simulation suite, plotted in the meridional plane. Red/yellow/blue tones indicate positive/zero/negative values. The outer ticks in each color bar mark the the saturation values (which are labeled) and the inner tick marks zero. Contours are equally spaced (for positive and negative values separately) and the zero contour is dashed. The dashed interior semicircle marks $r=\rbcz$. A dotted vertical line separates the viscous spreading regime (Case 0) from the radiative spreading regime (Cases 1--6). In the radiative spreading regime, the stratification ($\bu$) mostly increases to the right, as shown schematically by the arrow.  }
	\label{fig:diffrot}
\end{figure*}

%\com{LK: this figure looks much better and clearer now!}
Figure \ref{fig:diffrot} shows the equilibrated differential rotation profiles $\avt{2\Omega-1}$ in the meridional plane for all our cases.  Overall, the profiles are solar-like, with fast equatorial regions and slower polar regions. It is also clear that the differential rotation in the HD cases has largely spread from the CZ into the RZ. We verify in Section \ref{sec:radspread} for several cases that the spreading is radiative, whereas Case 0 appears to be the only viscously spreading case. 

We characterize the steady-state differential rotation contrasts in the RZ and CZ by the standard deviation of the the temporally averaged rotation rate in the appropriate region:
\begin{subequations}\label{eq:dr_contrasts}
\begin{align}
\dOmrz &\define \sqrt{\avrz{\avt{\Omega}^2} -\ \Omrz^2}\label{eq:drrz}\\
\andd \dOmcz &\define \sqrt{\avcz{\avt{\Omega}^2} -\ \Omcz^2}\label{eq:drcz},
\end{align}
\end{subequations}
where
\begin{align}\label{eq:bulkrates}
    \Omrz \define \avrzt{\Omega}\five\andd  \Omcz \define \avczt{\Omega}
\end{align}
are the bulk rotation rates of the RZ and CZ, respectively.

We then define the ``tachocline confinement ratio" $f$ as
\begin{align}\label{eq:f}
    f\define \frac{\dOmrz}{\dOmcz}. 
\end{align}
Lower $f$ indicates a more strongly confined tachocline. Cases 1--6 are ordered (in case number) by decreasing $f$.
 
Finally, we define the thickness of the RZ's shear layer (the tachocline thickness, if there is a tachocline) by
\begin{align}\label{eq:dtach}
    \Delta\define\frac{\dOmrz}{\avrz{\abs{\avt{\nabla\Omega}}}}.
\end{align}

\begin{figure*}
	\centering
	%\vspace*{10cm}
	%\hspace*{-1cm}
	\includegraphics[width=7.25in]{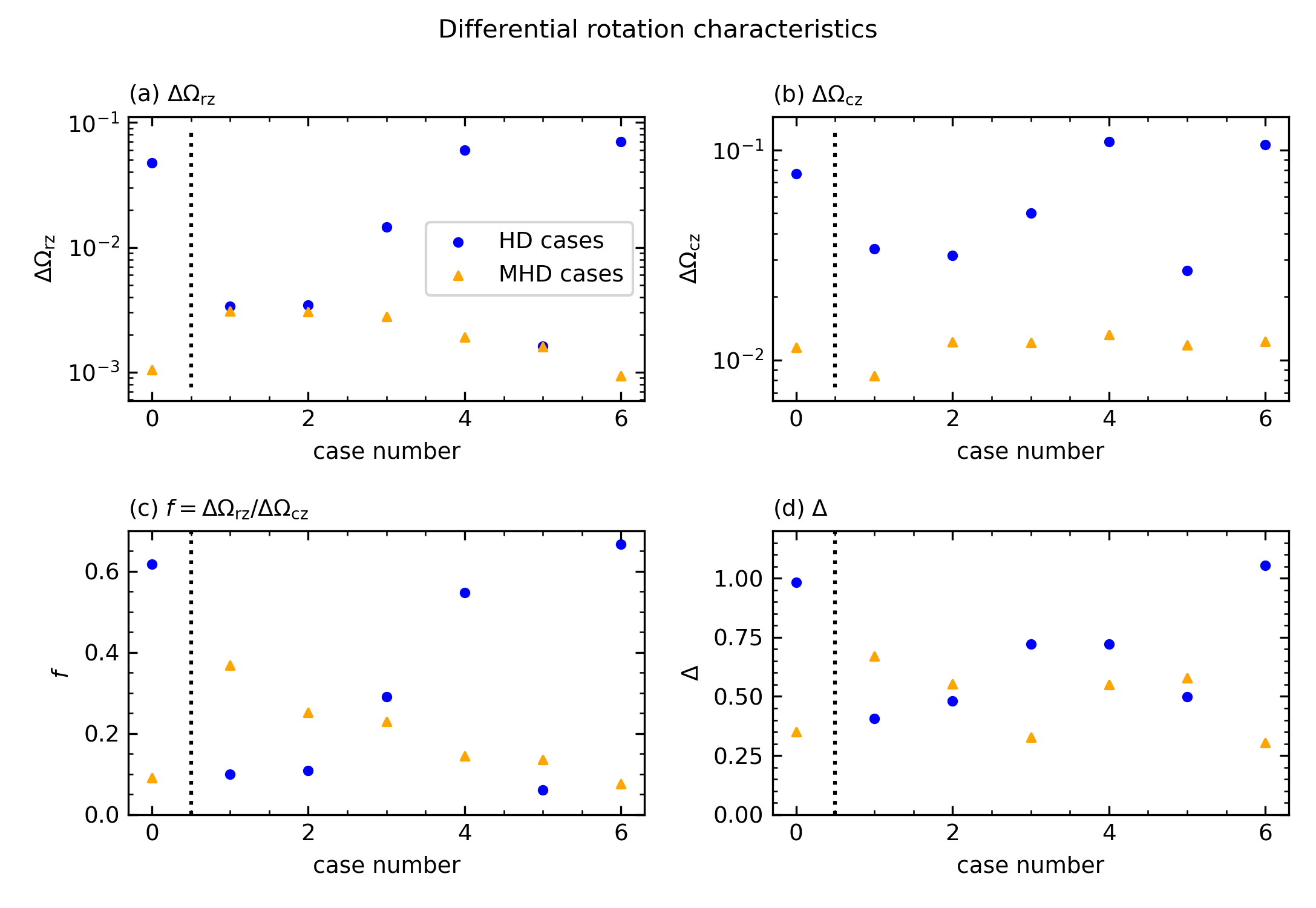}
	\caption{(a) Differential rotation contrast in the RZ defined in Equation \eqref{eq:drrz}, plotted with respect to case number. (b) Differential rotation contrast in the CZ defined in Equation \eqref{eq:drcz}. (c) The confinement ratio $f$ defined in Equation \eqref{eq:f}. Note that because of the chosen ordering, $f$ monotonically decreases for the MHD cases within the radiative spreading regime. (d) The RZ's shear-layer thickness defined in Equation \eqref{eq:dtach}. }
	\label{fig:confinement_ratio} 
\end{figure*}

Figure \ref{fig:confinement_ratio} shows $\dOmrz$, $\dOmcz$, $f$, and $\Delta$ plotted for all our cases. As is also visible in Figure \ref{fig:diffrot}, the dynamo tends to eliminate differential rotation everywhere, causing the rotation contrasts in both the CZ and RZ to be significantly smaller in each MHD case compared to the corresponding HD case. This is consistent with the idea that Maxwell stresses, wherever they have significant amplitude, tend to eliminate shear through magnetic tension. 

Within the radiative spreading regime (i.e., excluding Case 0), for high case numbers with strong stable stratification in the RZ, the dynamo eliminates the differential rotation much more effectively in the RZ than in the CZ. Figure \ref{fig:confinement_ratio}a shows that $\dOmrz$ declines sharply from $\approx\sn{3}{-3}$ for the less stratified MHD cases to $\approx$$10^{-3}$ for the most stratified cases. 
However, $\dOmcz\approx\sn{6}{-2}$ is almost a constant for all the radiatively spreading MHD cases (Figure \ref{fig:confinement_ratio}b). This results in significantly more well-confined tachoclines (i.e., lower values of $f$) for the highly-stratified cases (Figure \ref{fig:confinement_ratio}c).

The shear-layer thickness $\Delta$ (Figure \ref{fig:confinement_ratio}d) exhibits no clear trend with case number or type of simulation (HD or MHD). This is because of the wide variation of topology in the equilibrated differential rotation profiles (see Figure \ref{fig:diffrot}), which furthermore shows the difficulty of choosing a quantitative definition of a ``solar-like confined tachocline" that works for all different types of shear layers. For instance, if the definition is simply ``low values of $f$ and $\Delta$," then Cases H1, H2, and H5 have better-confined tachoclines than do Cases M1, M2, and M5, respectively. We thus emphasize the importance, at least in simulations, of choosing a tachocline definition that is based on dynamics as well as kinematics. For our radiative spreading cases, we incorporate the dynamics into our definition of tachocline confinement by focusing on the $f$-values for the MHD cases only, which all have Maxwell stresses that attempt to rigidify the RZ and stop radiative spreading (see Section \ref{sec:radspread}).

\subsection{Meridional Circulation}
The other component of the mean flow is the meridional circulation, which is fully characterized by the vector mean mass flux $\rhoumer$, where
\begin{equation}\label{eq:umer}
\umer\define\urad\erad + \utheta\etheta
\end{equation}
is the vector meridional flow. 

Because $\Div \rhoumer \equiv0$ (taking the zonal average of Equation \ref{eq:cont}), we define a streamfunction $\Psi(r,\theta,t)$ such that $\rhoumer=\nabla\times[\Psi(\ephi/\lambda)]$, or
\begin{equation}\label{eq:psi}
\rhourad=\frac{1}{r^2\sin\theta}\pderiv{\Psi}{\theta}\ \ \text{and}\ \ \rhoutheta=- \frac{1}{r\sin\theta}\pderiv{\Psi}{r}.
\end{equation}
The contours of $\Psi$ give the streamlines of $\av{\umer}$ and the density of the contours gives the magnitude of the mass flux $\lvert\rhoumer\rvert$. The integration constant is chosen such that $\Psi$ vanishes on the boundaries of the meridional plane. The contours $\Psi\equiv0$ thus give the boundaries of the circulation cells and the sign of $\Psi$ gives the sense of circulation within a cell. Looking into the page (i.e., along positive $\phi$), $\Psi>0$ means clockwise circulation and $\Psi<0$ means counterclockwise circulation.

\begin{figure*}
	\centering
	%\vspace*{10cm}
	%\hspace*{-1cm}
	\includegraphics[width=7.25in]{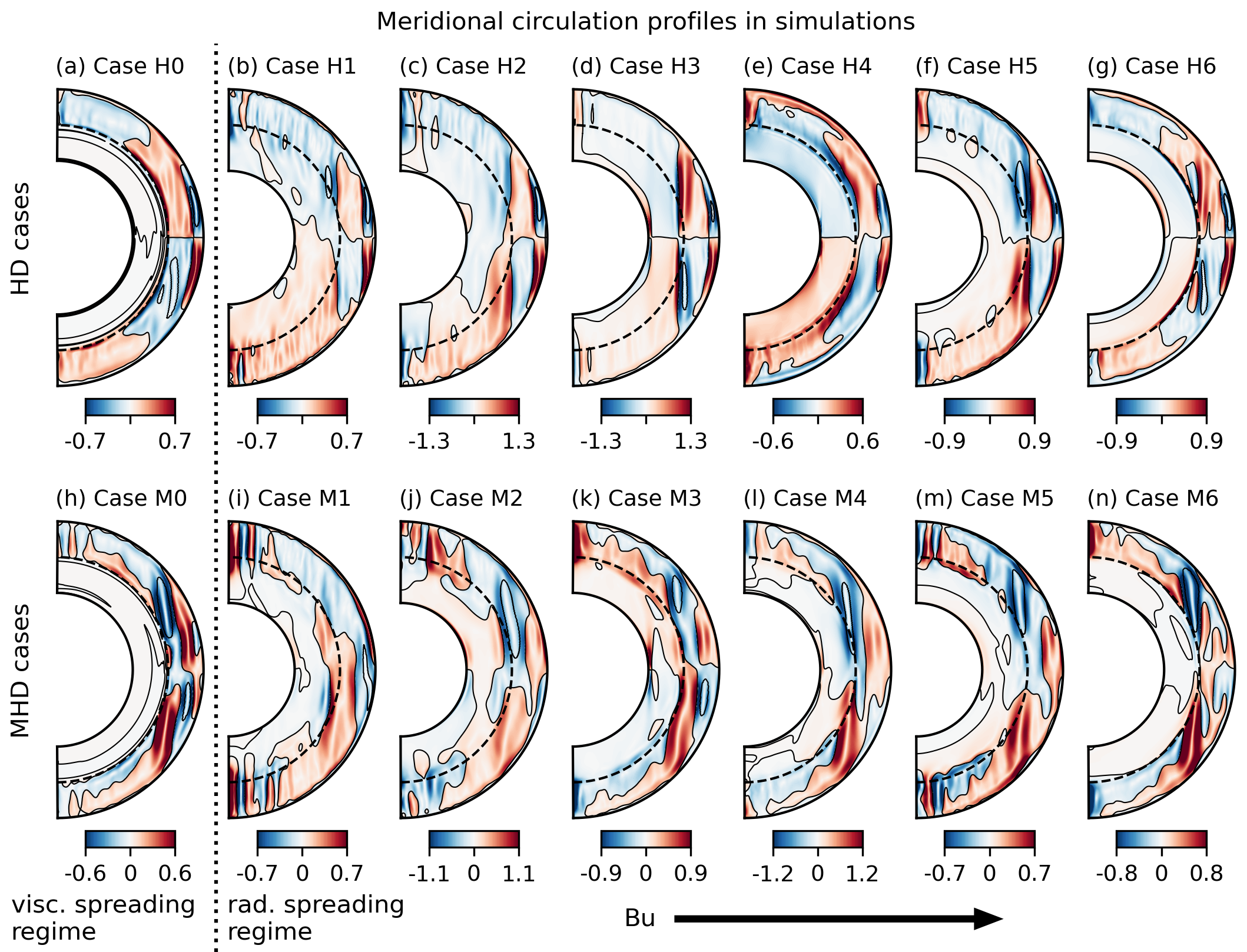}
	\caption{Time-averaged meridional circulation profiles $\lvert\rhoumer_t\rvert\sgn{\avt{\Psi}}$ (see Equations \ref{eq:umer} and \ref{eq:psi}) for the full simulation suite, plotted in the meridional plane. Red tones indicate clockwise circulation and blue tones indicate counterclockwise circulation. The solid contours mark ${\avt{\Psi}}\equiv0$, which are the circulation cell boundaries and also streamlines of $\umer$. The dashed semicircles mark $r=\rbcz$. }
	\label{fig:rhoutheta}
\end{figure*}

Figure \ref{fig:rhoutheta} shows the equilibrated contour plots of $\lvert\rhoumer_t\rvert\sgn{\avt{\Psi}}$ (which gives both the magnitude of the mass flux and the sense of the circulation) for all our cases. Each CZ has a complicated multicellular circulation pattern that often has cell boundaries oriented parallel to the rotation axis. This pattern penetrates by varying degrees into the RZ below. In general, the penetration is weaker for the high case numbers, which have stronger stable stratification in the RZ. However, clear trends in the equilibrated circulation profiles with respect to the input parameters are not apparent, and it is not obvious how to tell whether a given simulation is in the radiative spreading regime from analyzing the circulation profiles alone. We thus turn to an investigation of the time-dependent dynamics in order to more precisely characterize radiative spreading in our simulations. 

\section{Radiative Spreading in Terms of Spin-Down}\label{sec:radspread}
In our simulations, a primary effect of the shear profile spreading inward is that the bulk of the RZ slows its bulk rotation rate to be less than the frame rate; in other words, the RZ spins down over time. This is predominantly due to either viscous or radiative spreading, which manifest in the torque balance as negative values for the torques (volume-averaged over the deep RZ) due to the viscosity or meridional circulation, respectively. In this section, we focus on this spin-down behavior to confirm that all our cases except Case 0 lie in the radiative spreading regime. 

\subsection{Torque Balance during Spin-Down}
The specific angular momentum density of the fluid is 
\begin{equation}\label{eq:amom}
	\amom\define \lambda\left(\frac{\lambda}{2} + \av{\uphi}\right)=\lambda^2\Omega.
\end{equation}

Multiplying the $\phi$-component of the zonal mean of Equation \eqref{eq:mom} by $\lambda$ yields the evolution equation for $\amom$:
\begin{subequations}\label{eq:torque}
	\begin{align}
		\rhotilde\pderiv{\amom}{t}=&
		-\Div[\rhotilde (\lambda \av{u^\prime_\phi \umerprime} 
		+\amom\av{\umer} -\ek\nutilde\lambda^2\nabla\Omega)\nonumber\\
		&- \lambda (\av{ B_\phi^\prime\bpol^\prime} +  \av{\bphi}\av{\bpol} )]\\
		=\ & \taurs + \taumc + \tauv + \taums + \taumm,
	\end{align}
\end{subequations}
where we have defined the appropriate time-dependent torque densities:
\begin{subequations}\label{eq:torques}
	\begin{align}
		\taurs \define &\ -\Div (\rhotilde (\lambda \av{u^\prime_\phi \umerprime})\nonumber\\
		&\ \text{(Reynolds-stress torque)}, \label{eq:taurs}\\
		\taumc \define  &\ -\rhoumer\cdot\nabla\amom \nonumber\\
		&\ \text{(meridional-circulation torque)},\label{eq:taumc}\\
		\tauv \define  &\ \ek \Div(\rhotilde\nutilde\lambda^2\nabla\Omega)\nonumber\\ 
		&\ \text{(viscous torque)},\label{eq:tauv}\\
		\taums \define  &\ \Div( \lambda \av{ B_\phi^\prime\bpol^\prime})\nonumber\\ 
		&\ \text{(Maxwell-stress torque)}, \label{eq:taums}\\
		\andd \taumm \define &\ \Div( \lambda \av{\bphi}\av{\bpol})\nonumber\\
		&\ \text{(mean magnetic torque)}.\label{eq:taumm}
	\end{align}
\end{subequations}
Here, $\bpol\define \brad\erad+\btheta\etheta$ is the vector poloidal magnetic field.

To study the two spreading processes in terms of spin-down, we volume-average Equation \eqref{eq:torque} over the deep RZ ($r<{r\drz}$) and then perform a cumulative integral in time from $t=0$ to the current time $t$. Note that this same cumulative averaging process was used to define $\sigmadyn(t)$ in Equation \eqref{eq:sigmadyn}. Figure \ref{fig:torque_tint} shows the evolution of each term in the cumulatively-time-integrated and deep-RZ-averaged Equation \eqref{eq:torque} for Case 0, which is in the viscous spreading regime and Case 6, which is in the radiative spreading regime. 

For Case 0 (Figure \ref{fig:torque_tint}(a,b)), it is clear that it is almost exclusively the viscosity that is responsible for spinning the RZ down. In the HD case, this spreading mechanism continues unopposed until the differential rotation reaches a state with $\avt{\taunu}\equiv0$ (seen in Figure \ref{fig:torque_tint}a as an asymptote in the evolution of $\int_0^t\avdrz{\tauv}(t^\prime)dt^\prime$).  In the MHD case, by contrast, the torque from the Maxwell stresses (with some small assistance from mean magnetic tension) becomes positive and counters most of the viscous torque, allowing only a small amount of angular momentum to spread.

For Case H6 (Figure \ref{fig:torque_tint}c), the RZ spins down almost exclusively due to radiative spreading, as evidenced by the negative integrated torque from the inward-burrowing meridional circulation, $\int_0^t\avdrz{\taumc}(t^\prime)dt^\prime$. However, near the very beginning of the simulation (for $t\lesssim2500$), viscosity also contributes to the spin-down ($\int_0^t\avdrz{\tauv}(t^\prime)dt^\prime<0$), before ultimately working against the radiative spreading ($\int_0^t\avdrz{\tauv}(t^\prime)dt^\prime>0$) for $t\gtrsim2500$. At late times, a steady-state balance is reached between $\taumc$ and $\tauv$, which represents the final diffusion layer of \citealt{Clark1973} and occurs after an interval on the order of the Eddington-Sweet time-scale.

In Case M6 (Figure \ref{fig:torque_tint}d), the final diffusion layer never occurs because the dynamo changes the ultimate torque balance. Instead, the RZ spins down only slightly, at very early times, due to a combination of strong radiative spreading and weaker viscous spreading. This combined radiative and viscous spreading is stopped very quickly by the magnetic torque, which is mostly from the Maxwell stresses, but also contains a finite contribution from the mean magnetic tension. It is this ultimate balance between $\taumc$ and $\taums$ (or more accurately, a four-way balance between $\taumc$, $\taums$, $\tauv$, and $\taumm$) that leads to the well-confined tachocline in this case. 

\subsection{Defining the Radiative Spreading Regime}
It is important to note that we do not achieve ``pure" radiative spreading (the $\sigmadyn(t)\equiv0$ case considered by \citetalias{Spiegel1992} in their Equations 4.12--4.16), because viscous effects are always (unfortunately) still quite substantial. To quantify these issues, we use Equation \eqref{eq:sigmadyn} to compute $\sigmadyn(t)$ for several of our cases and plot the results in Figure \ref{fig:sigma_dyn_tint}(a,b). In the HD cases (Figure \ref{fig:sigma_dyn_tint}a), except for an initial adjustment period wherein $\int_0^t\av{\taumc}(t^\prime)dt^\prime$ and/or $\int_0^t\av{\tauv}(t^\prime)dt^\prime$ approach zero at different times, $\sigmadyn(t)$ increases monotonically from a very low value to a near-unity value. This reflects the inherently greater speed of radiative spreading compared to viscous spreading. For example, under the idealized circumstances of \citetalias{Spiegel1992} (see their Equation 4.9), the radiative-spreading torque scales like $\Delta(t)^{-4}$, whereas the viscous torque scales like $\Delta(t)^{-2}$ (where $\Delta(t)$ is a time-varying shear-layer thickness; see Equation \ref{eq:delta}). Radiative spreading thus always dominates over viscous spreading at early times, when the tachocline is still thin. Our HD models are, of course, more complex than the scenario considered by \citetalias{Spiegel1992} because we have Reynolds stresses, nonlinear circulations, and nonzero initial boundary-layer thicknesses. Nonetheless, we still see an enhanced speed of radiative spreading compared to viscous spreading, as evidenced by Figure \ref{fig:sigma_dyn_tint}a at early times. At late times, however, the situation changes. After roughly an Eddington-Sweet time-scale, the two transport processes become comparable and in the HD cases, they balance to achieve a final diffusion layer for which $\sigmadyn(t)\sim1$.

\begin{figure*}
	\centering
	%\vspace*{10cm}
	%\hspace*{-1cm}
	\includegraphics[width=7.25in]{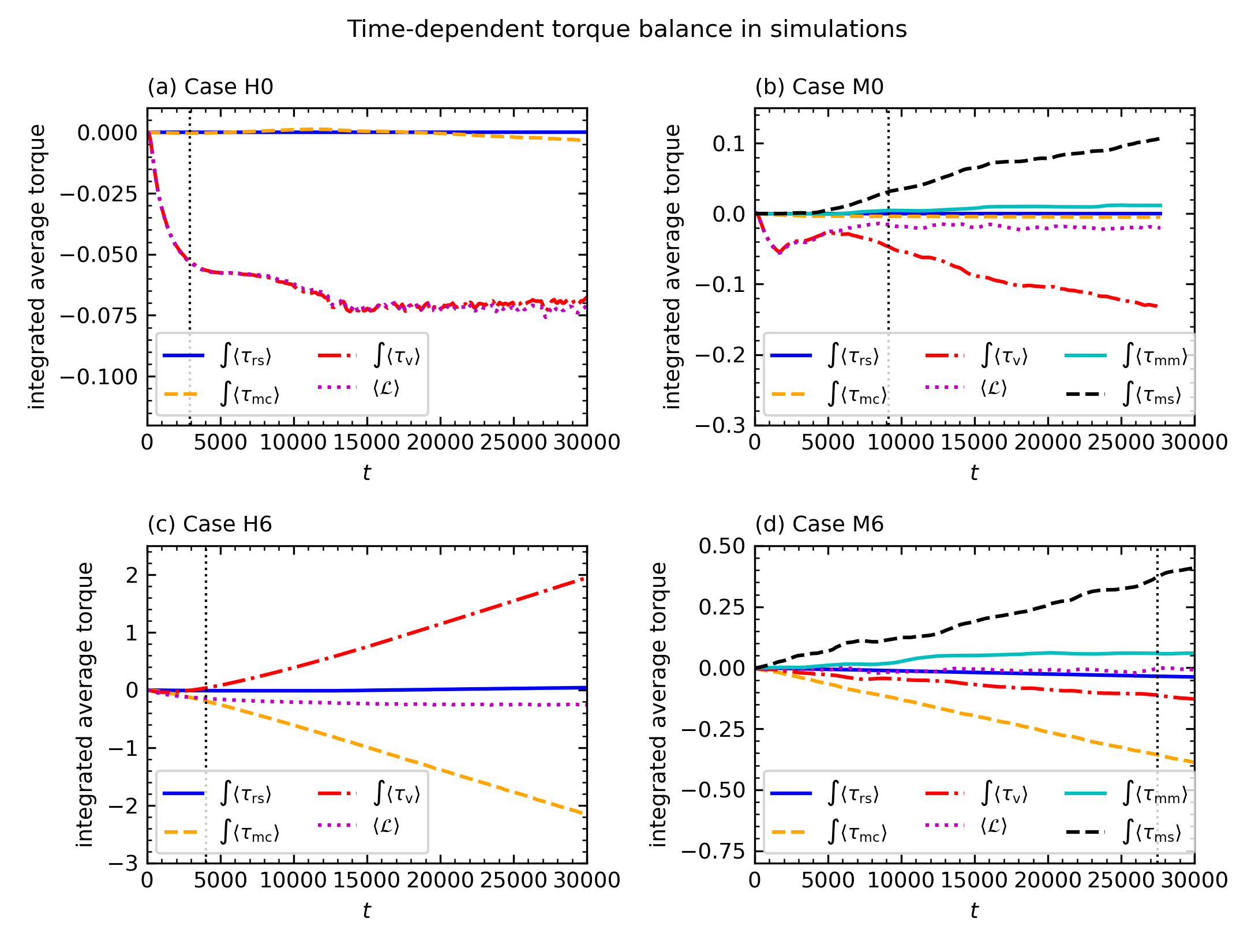}
	\caption{Cumulatively-time-integrated torque densities from Equation \eqref{eq:torques} as functions of time, after volume-averaging over the deep RZ, for (a) Case H0, (b) Case M0, (c) Case H6, and (d) Case M6. In the legend, we label the curves by the torque densities, which are understood to be shorthand for the time-integrated and volume-averaged quantities. For instance, $\int\av{\taurs}$ is shorthand for $\int_0^t\avdrz{\taurs}(t^\prime)dt^\prime$. The vertical dotted lines correspond to the values of $\thalf$ for each case.}
	\label{fig:torque_tint}
\end{figure*}

% \com{NB: we need to be clear then that the regime is defined by the ** HD case **}

For all of our HD simulations, the bulk of the RZ always spins down over time due to a transport of angular momentum out of the deep RZ. Because angular momentum is conserved in our simulations, the cumulative angular momentum that has been transported out of the deep RZ at a given time $t$ during RZ spin-down is simply
\begin{align}
	\damom(t)&\define \avdrz{\rhotilde\amom}(t) - \avdrz{\rhotilde\amom}(0),\label{eq:damom}
\end{align}
which is a deficit for the deep RZ and a surplus for the upper layers. (Note that we enforce machine-precision conservation of total angular momentum in our simulations.) Figure \ref{fig:sigma_dyn_tint}(c,d) shows the evolution of the angular momentum deficit for the same cases and time intervals as those shown in the $\sigmadyn(t)$ plots of Figure \ref{fig:sigma_dyn_tint}(a,b). For the HD cases (Figure \ref{fig:sigma_dyn_tint}c), the $\damom(t)$ curves decrease relatively quickly at early times (again reflective of the inherent enhanced speed of radiative spreading), but then asymptote to some finite negative value at late times.

It therefore seems reasonable to define the viscous and radiative spreading regimes for the HD cases based on which spreading process causes more of the total end-of-simulation angular momentum transport. More precisely, to determine in which spreading regime our HD cases lie, we consider $\sigmadyn(t)$ at the (somewhat arbitrary) time $\thalf$ at which $75\%$ of the end-of-run angular momentum transport out of the deep RZ has occurred, i.e.,
\begin{align}\label{eq:thalf}
\damom(\thalf)=\frac{3}{4}\, \Delta L (\trun).
\end{align}
The value of $\sigmadyn(\thalf)$ then indicates which spreading process has caused most of the total bulk spin-down of the RZ, as measured by the net angular momentum deficit of the deep RZ at the end of the run. We thus define our HD simulations to be in the radiative spreading regime when $\sigmadyn(\thalf)<1$ and to be in the viscous spreading regime when $\sigmadyn(\thalf)>1$.

The concepts of the radiative and viscous spreading regimes can only be defined unambiguously when (1) there is a spreading process actually occurring and (2) viscosity and burrowing circulation are the dominant terms in the torque balance. Both of these conditions are violated by the MHD cases, for which Maxwell stresses shut down both spreading processes very quickly and do so by playing a big role in the torque balance. For the MHD cases, it therefore makes more sense to define the viscous and radiative spreading regimes by which torque (viscosity or meridional circulation, respectively) is larger in the steady state, which is  measured by $\sigmadyn(\infty)$ as opposed to $\sigmadyn(\thalf)$. Fortunately, however, there is no ambiguity in these two definitions for our simulation suite. In the MHD cases (Figure \ref{fig:sigma_dyn_tint}b), $\sigmadyn(t)$ remains $<1$ for all the radiatively spreading cases and quickly asymptotes to its late-time value. For compactness, we therefore use $\sigmadyn(\thalf)$ to define the regime boundaries for both the HD and MHD cases alike. Note that the time $\thalf$ is less meaningful for the MHD cases than for the HD cases. As shown in Figure \ref{fig:sigma_dyn_tint}d,  $|\Delta L(t)|$ always remains quite small for the MHD cases and does not monotonically increase at late times. This is because little angular momentum is transported out of the deep RZ; the Maxwell stresses simply preserve the initial confined-tachocline state.

Table \ref{tab:thalf} shows the values of $\thalf$ and $\sigmadyn(\thalf)$ for the cases considered in Figure \ref{fig:sigma_dyn_tint}. The values of $\sigmadyn(\thalf)$ clearly confirm that the cases that we were previously demarking as radiatively spreading cases from our input parameters (i.e., all cases apart from Case 0; see Figure \ref{fig:input}) are indeed such cases by this measure. As might be expected, the spin-down for the radiative spreading cases always occurs after some fraction of an Eddington-Sweet time-scale. Although viscosity plays a role at times, ultimately it is the burrowing meridional circulation that transports most of the angular momentum out of the RZ for Cases 1--6 (both HD and MHD).

The time-dependent thickness of the tachocline (or more generally the ``RZ shear length-scale" $\Delta(t)$) is also an important quantity because of its role in setting the critical value $\sigma_c$, which divides the radiative spreading regime from the viscous spreading regime (see Appendix \ref{ap:sigmac}). Explicitly, we define this length-scale by
\begin{align}\label{eq:delta}
    \Delta(t) \define \frac{\dOmdrz(t)}{\avdrz{|\nabla\Omega|}(t)},
\end{align}
where
\begin{align}
    \dOmdrz(t) &\define \sqrt{\avdrz{\Omega^2} - \avdrz{\Omega}^2}
\end{align}
is the time-dependent differential rotation contrast in the deep RZ.
%Compare this to Equation \eqref{eq:drrz}.
%\com{NB. I hate $\Delta$s and $\nabla$'s in the same equation!}

Figure \ref{fig:sigma_dyn_tint}(e,f) shows $\Delta(t)$ for the same cases and time intervals as those shown in the $\sigmadyn(t)$ and $\Delta L(t)$ plots of Figure \ref{fig:sigma_dyn_tint}(a--d).  In the HD cases (Figure \ref{fig:sigma_dyn_tint}e), $\Delta(t)$ starts out small, but asymptotes to a significantly larger order-unity value, meaning that the initially-confined tachocline ultimately spreads to a width comparable to that of the CZ and thus effectively disappears. By contrast, in the MHD cases (Figure \ref{fig:sigma_dyn_tint}f), the tachocline mostly stays confined and so $\Delta$ stays roughly constant at its initial value of $\approx0.3$ inherited from the end of Case M0.

Table \ref{tab:thalf} also shows the values of $\Delta(\thalf)$ for all cases. Interestingly, despite the disappearance of the tachocline in the HD cases, the values of $\Delta(\thalf)$ are comparable for all cases, both HD and MHD, all being $\lesssim$0.5 (with an average of $\approx$$0.3$).  For the HD cases, these relatively low values of $\Delta(\thalf)$ come from the fact that $\Delta(t)$ appears to increase nearly linearly with time initially, while $|\Delta L(t)|$ increases much faster. Note that in the pure radiative spreading regime (i.e., the situation where $\sigmadyn(t)\equiv0$ for which \citetalias{Spiegel1992} calculated explicit similarity solutions for radiative spreading), $\Delta(t)\propto t^{1/4}$, whereas in the pure viscous spreading regime, $\Delta(t)\propto t^{1/2}$. Evidently, the mixture of radiative spreading, viscous spreading, and nonlinear effects in our simulations causes an emergent behavior in which $\Delta(t)\propto t$ at early times in the HD simulations. 

In Appendix \ref{ap:sigmac}, we show how the relatively low values of $\Delta(\thalf)$ can be used to estimate the critical value $\sigma_c\approx10$ in our simulations. The fact that $\sigma_c$ is relatively high is the fundamental reason why we can still achieve radiative spreading for $\sigma$ as high as $\approx3$. Physically, we can see from the results of this section that $\sigma_c$ is large because most of the angular momentum transport out of the deep RZ  occurs while the shear layer is still thin.

\begin{table}
    \caption{Values of $\thalf$, $\sigmadyn(\thalf)$, and $\Delta(\thalf)$ for Cases 0, 3, 4, and 6.}
\label{tab:thalf}
\centering
\begin{tabular}{*{5}{l}}
%\multicolumn{5}{c}{\textbf{Characteristics of RZ spin-down in simulations}} \\ % This acts as a title
\hline
Name & $\thalf$ & $\thalf/\tes$ & $\sigmadyn(\thalf)$ & $\Delta(\thalf)$ \\
\hline
Case H0 & 2910 & 8.70e-06 & 143 & 0.391 \\
Case H3 & 4020 & 0.599 & 0.274 & 0.321 \\
Case H4 & 5670 & 0.377 & 0.488 & 0.275 \\
Case H6 & 7410 & 0.00779 & 0.518 & 0.507 \\
\hline
Case M0 & 9160 & 2.74e-05 & 12.2 & 0.282 \\
Case M3 & 27500 & 4.09 & 0.269 & 0.312 \\
Case M4 & 8250 & 0.548 & 0.0547 & 0.305 \\
Case M6 & 26500 & 0.0278 & 0.311 & 0.264 \\
\hline
\end{tabular}
\end{table}

\begin{figure*}
	\centering
	%\vspace*{10cm}
	%\hspace*{-1cm}
	\includegraphics[width=7.25in]{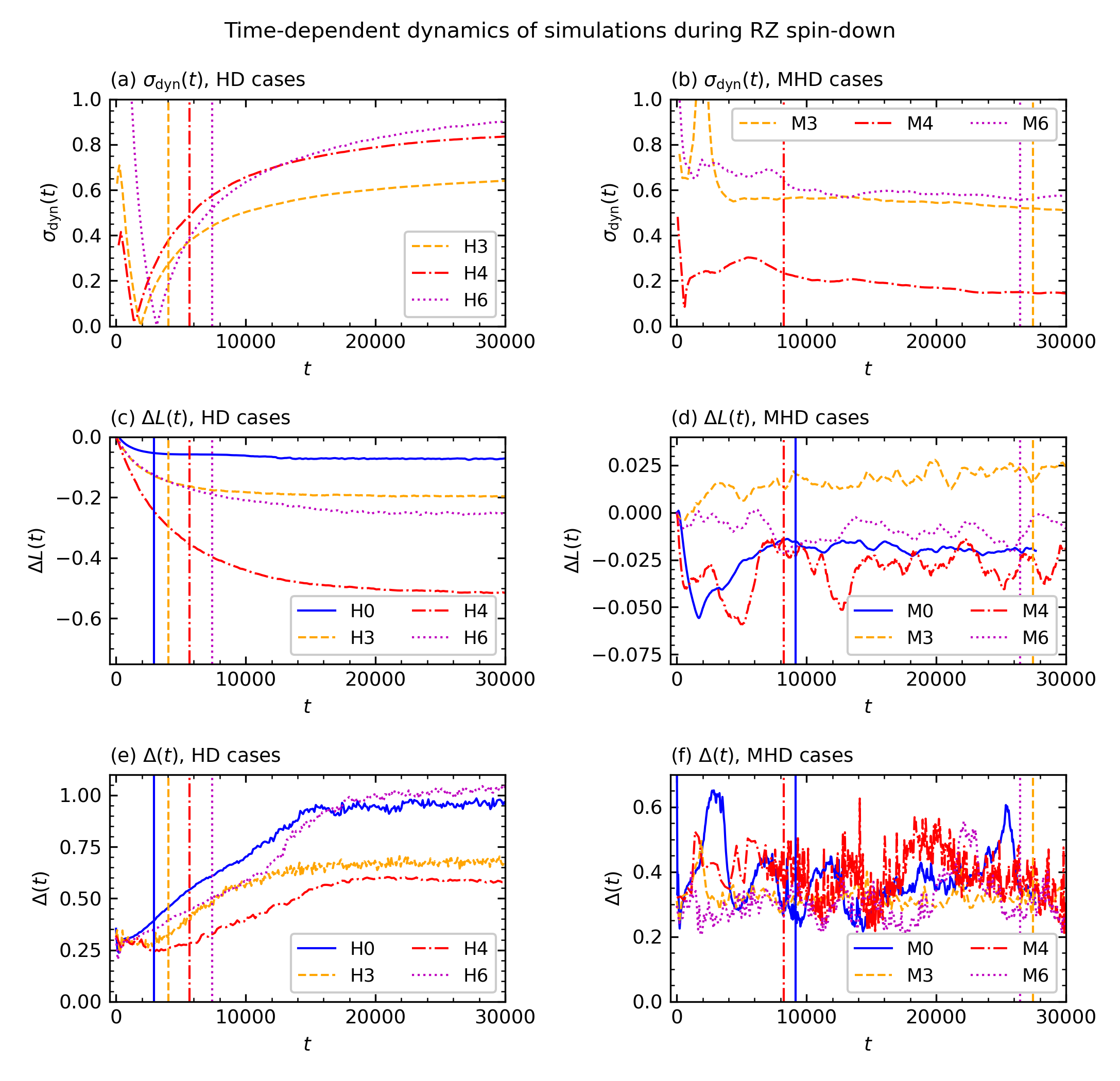}
	\caption{(a) $\sigmadyn(t)$ for Cases H3, H4, and H6 and (b) Cases M3, M4, and M6. (c) Angular momentum deficit of the RZ, $\Delta L(t)$, for Cases H0, H3, H4, and H6 and (d) Cases M0, M3, M4, and M6. (e) and (f): Shear-layer thickness $\Delta(t)$ for the same cases as panels (c) and (d), respectively. See Equations \eqref{eq:sigmadyn}, \eqref{eq:damom}, and \eqref{eq:delta}. The vertical colored lines correspond to the values of $\thalf$ for each case. }
	\label{fig:sigma_dyn_tint}
\end{figure*}

\begin{figure}
	\centering
	%\vspace*{10cm}
	%\hspace*{-1cm}
	\includegraphics[width=3.4375in]{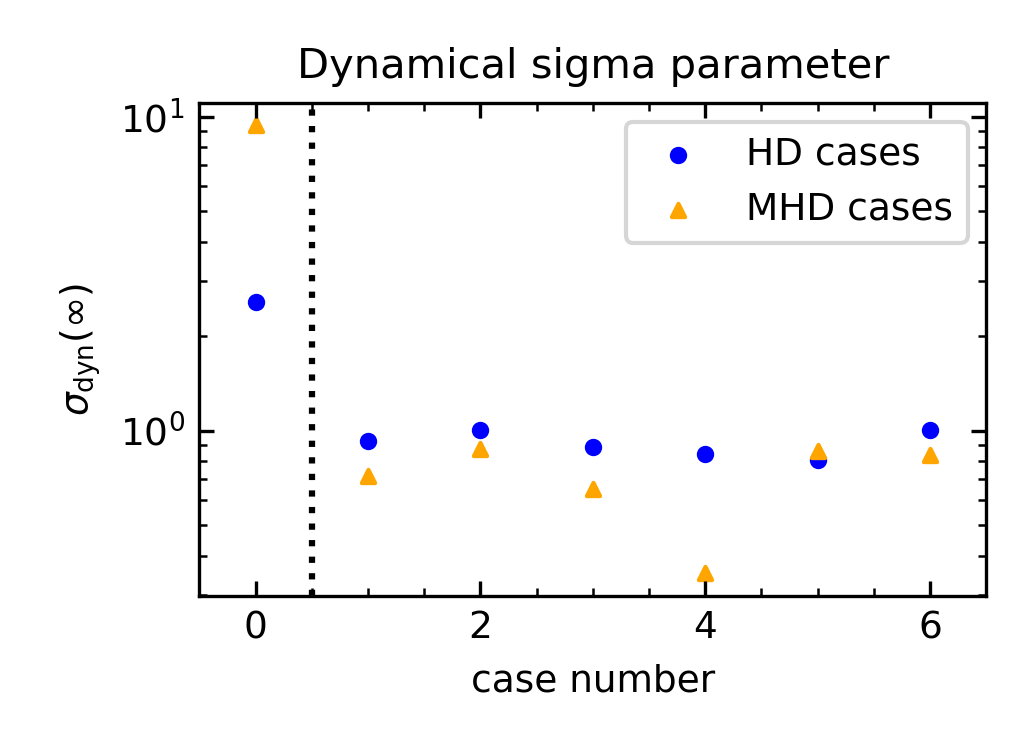}
	\caption{Equilibrated-state values of $\sigmadyn$, denoted by $(\sigmadyn(\infty)$, for all of our cases, plotted with respect to case number. We approximate $\sigmadyn(\infty)\approx [\avdrzt{\tauv}/\avdrzt{\taumc}]^{1/2}$, where the temporal average is over the equilibrated state.}
	\label{fig:sigma_dyn}
\end{figure}

\subsection{The Difference between Time-Dependent and Steady-State Dynamics}

The preceding discussion makes the important point that the dynamics of the late-time steady state do not necessarily match the dynamics at early times. It is thus not always possible to determine the spreading regime of a given system by analyzing its steady state. For instance, when there is no magnetism, Figure \ref{fig:sigma_dyn_tint}a shows that the burrowing circulation ultimately comes into balance with the viscosity in a final diffusion layer, i.e., the equilibrated-state value of $\sigmadyn$, which we denote by $\sigmadyn(\infty)$, is $\approx1$. This is despite all these cases clearly being in the radiative spreading regime as defined by early-time measures.

In fact, the steady-state torque balances for all the radiative spreading cases of a given type (either HD or MHD) appear to be similar, despite their markedly different positions in parameter space as characterized by $\sigmadyn(\thalf)$. Figure \ref{fig:sigma_dyn} shows the values of $\sigmadyn(\infty)$ for most of our cases. It is clear that the HD cases all have $\sigmadyn(\infty)\approx1$, consistent with the final diffusion layer interpretation. The MHD cases have systematically lower $\sigmadyn(\infty)$,\footnote{This is because the Maxwell stresses do not allow the final diffusion layer to form. Instead, as exemplified by Figure \ref{fig:torque_tint}d, there is a primary balance between $\taums$ and $\taumc$, with $\tauv$ allowed to be significantly smaller. } but again this value is approximately a constant. Finally, Case H0, which is very far into the viscous spreading regime, nonetheless has $\sigmadyn(\infty)\lesssim3$. Thus, although $\sigmadyn(\infty)\lesssim1$ \textit{may} indicate the past action of radiative spreading, it is far from a sure indicator. 

\section{Skin Effect}\label{sec:skin}
The viability of the dynamo confinement scenario for the tachocline (and the depth to which the RZ can be rigidified by Maxwell stresses) depends on how far the convective dynamo's magnetic field can penetrate into the deep RZ. For primarily oscillatory fields like the ones in the simulations discussed here (see Figure \ref{fig:timelat}), the distance by which the field can spread from the base of the CZ is determined by a skin effect. Because the large-scale magnetic fields in our simulations are predominantly nonaxisymmetric, here we derive the skin depths appropriate for nonaxisymmetric fields. Note that the toroidal magnetic field $\bphi$ is mostly generated directly from the poloidal magnetic field $\bpol$ through mean shear (see \citetalias{Matilsky2022}). Thus, only the amplitude of $\bpol$ is determined through the skin effect, whereas the amplitude of $\bphi$ depends on both $\bpol$ and the rotational shear profile $\avt{2\Omega-1}$.

We partition $\bpol$ into its contributions from each azimuthal wavenumber $m$,
\begin{align}\label{eq:mspec}
    \bpol &= \sum_m\bpol^m e^{im\phi},
\end{align}
where $m$ ranges across its discrete values $-\lmax + 1$ to $\lmax$ (see Appendix \ref{ap:grid}). Note that we have chosen the decomposition in Equation \eqref{eq:mspec} such that Parseval's theorem takes the form $\av{|\bpol|^2} = \sum_m |\bpol^m|^2$. 

Figure \ref{fig:moll} shows the vector components of the poloidal magnetic field and its power spectrum in azimuthal wavenumber in the upper RZ for several cases in the radiative spreading regime. As was also seen in Figure \ref{fig:cutout3d}, it is clear that the field is primarily nonaxisymmetric with complicated small-scale patterns amidst a strong large-scale $m=(0,1,2)$ envelope. The $m=(0,1,2)$ components are clearly visible in the power spectra. Overall, these field structures, which are the root cause of the Maxwell stresses confining the tachoclines, are similar to the dynamo modes that confined tachoclines in the viscous spreading regime as well (see \twopapers.)

\begin{figure*}
	\centering
	%\vspace*{10cm}
	%\hspace*{-1cm}
	\includegraphics[width=7.25in]{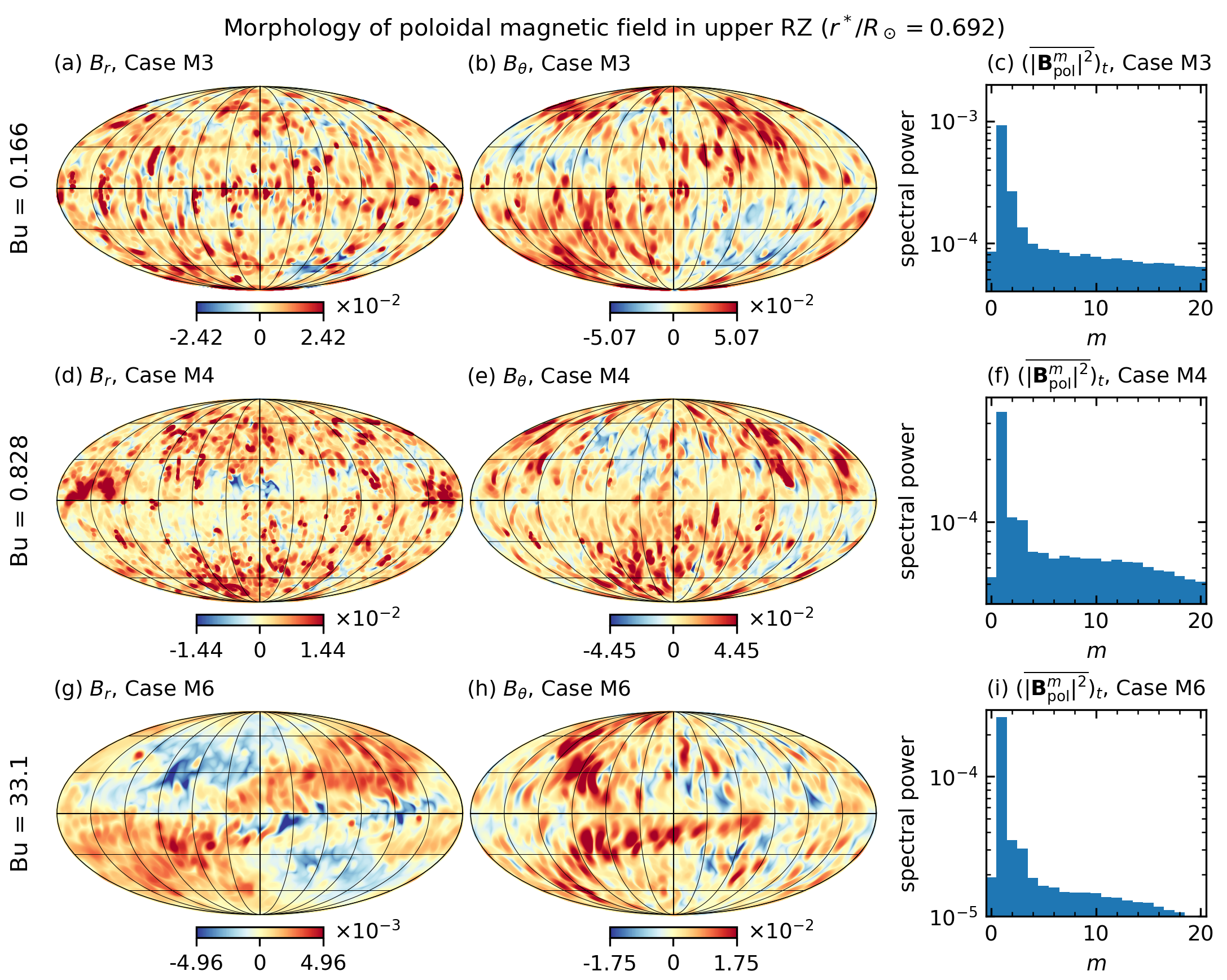}
	\caption{Mollweide projections and power spectra in azimuthal wavenumber for the poloidal magnetic field $\bpol$ in the upper RZ ($r\dimm/\rsun=0.692$) for (a--c) Case M3, (d--f) Case M4, and (g--i) Case M6. The power spectra $|\bpol^m|^2$ (Equation \ref{eq:mspec}) have been horizontally and temporally averaged over the equilibrated state.  }
	\label{fig:moll}
\end{figure*}

The nonaxisymmetric fields also exhibit complicated quasi-periodic dynamo cycles. Generalizing Equation \eqref{eq:mspec}, we further partition $\bpol$ into its components from each azimuthal wavenumber $m$ and angular frequency $\omega$:
\begin{align}\label{eq:tspec}
    \bpol &= \sum_m\sum_\omega \bpol^{m\omega}e^{i(m\phi - \omega t)},
\end{align}
where the frequency components are obtained from a discrete Fourier transform over the full simulation (from $t=0$ to $t=\trun$) and range from the negative Nyquist frequency to the positive Nyquist frequency. Our sign convention is chosen such that $\bpol^{m\omega}$ represents the amplitude of the field at azimuthal wavenumber $m$ moving in longitude at phase speed $\omega/m$. For positive $m$, positive $\omega$ represents prograde movement relative to the background rotating frame and negative $\omega$ represents retrograde movement. For the longitudinally-moving wave components $\bpol^{m\omega}$ defined in Equation \eqref{eq:tspec}, we note that Parseval's theorem takes the form $\avt{|\bpol|^2} = \sum_m \sum_\omega |\bpol^{m\omega}|^2$, where here the temporal average is over the whole simulation. 

Figure \ref{fig:timelat} shows the time-latitude diagram of $\mathrm{Re}(B_\theta^1)$ for Cases M3, M4, and M6 in the upper RZ and the corresponding power spectra in frequency. There are quasi-cyclic dynamo cycles in each case, with the dominant period increasing with increasing stable stratification. This trend is visible in the frequency spectra, which show predominantly retrograde movement of the nonaxisymmetric magnetic field (negative frequency), with a central frequency that decreases in magnitude (i.e., become less negative) as the stable stratification increases.

\begin{figure*}
	\centering
	%\vspace*{10cm}
	%\hspace*{-1cm}
	\includegraphics[width=7.25in]{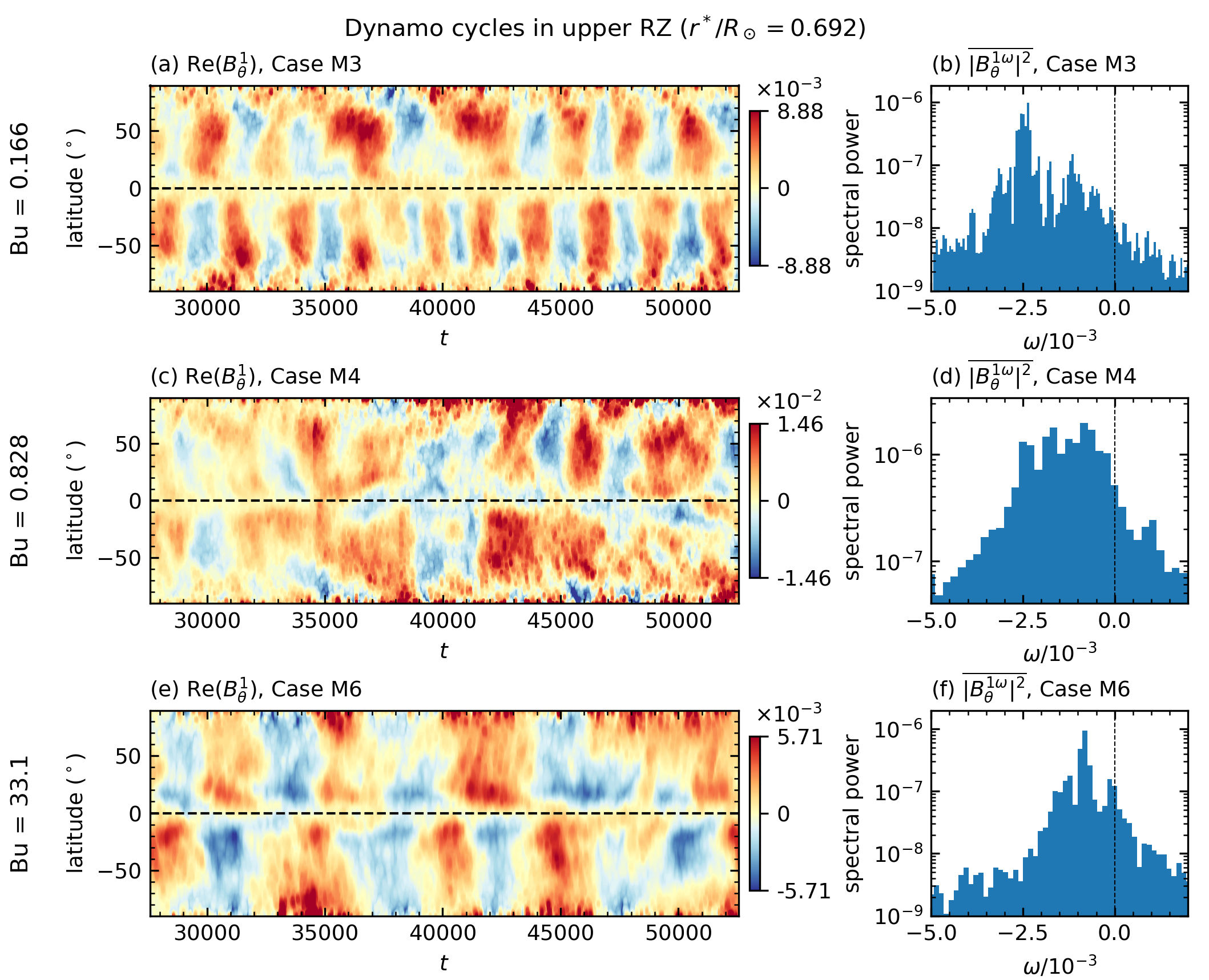}
	\caption{Time-latitude diagrams over the interval $t=(27500,52500)$ of $\mathrm{Re}(\btheta^1)$ in the upper RZ ($r\dimm/\rsun=0.692$) and the corresponding truncated horizontally averaged power spectra in frequency $\avsph{\abs{\btheta^{1\omega}}^2}$ for (a--b) Case M3, (c--d) Case M4, and (e--f) Case M6. The full time series $t=(0,\trun)$ was used to obtain the frequency spectra and a Hanning window function was employed to reduce high-frequency noise. For Cases M3, M4, and M6, respectively, the frequency resolution (the width of the bars) is $1/\trun =\sn{6.06}{-5}$, $\sn{2.08}{-4}$, and $\sn{1.35}{-4}$ and the Nyquist frequencies are $1/(2\Delta t) = 0.151$, $0.247$, and $0.147$, where $\Delta t$ is the average sampling cadence. }
	\label{fig:timelat}
\end{figure*}

\begin{figure}
	\centering
	%\vspace*{10cm}
	%\hspace*{-1cm}
	\includegraphics[width=3.4375in]{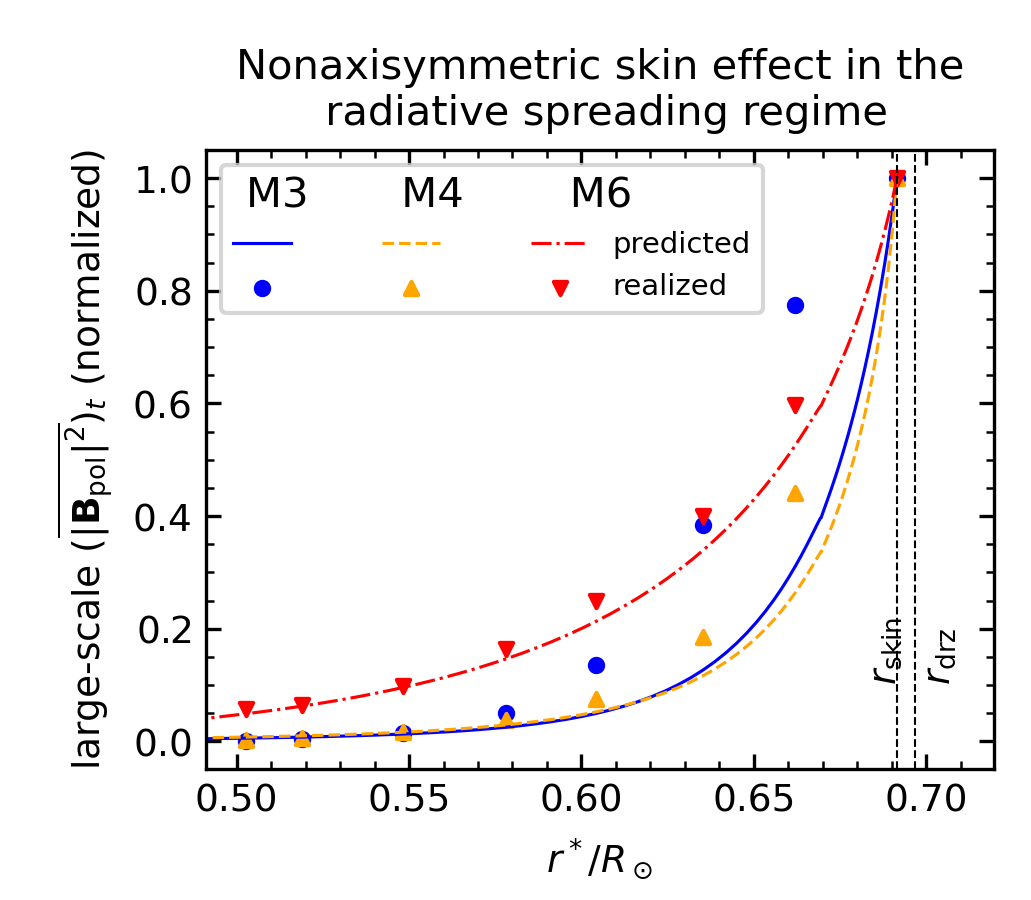}
	\caption{Scatter symbols: Realized amplitude of $\avt{|\bpol|^2}$ in the RZ summed over the largest scales $m=(0,1,2)$ for Cases M3, M4, and M6. Solid curves: Profile for $\avspht{|\bpol|^2}$ predicted by the nonaxisymmetric quasicyclic skin effect (Equation \ref{eq:ampskin}). Each profile is normalized by its value at $r\dimm=r_{\rm skin}\dimm = 0.692\rsun$. }
	\label{fig:skin}
\end{figure}

In \citetalias{Matilsky2024} and \citetalias{Matilsky2025a}, we showed that nonaxisymmetric, cycling dynamo modes like the ones depicted in Figures \ref{fig:moll} and \ref{fig:timelat} were able to penetrate into the RZ by way of a unique type of skin effect. This skin effect is novel because each frequency component penetrates to its own skin depth (and separately produces a tachocline-confining Maxwell stress; see Figure 3 from \citetalias{Matilsky2022}). Furthermore, the frequency determining the skin depth is Doppler-shifted by the bulk rotation rate of the RZ because it is the local cycle frequency in the frame of the RZ that ultimately sets the skin depth.

Following \citetalias{Matilsky2024}, we assume that, in a frame rotating at rate $\Omrz$, there is a some radius $r_{\rm skin}$ in the upper RZ below which in the induction Equation \eqref{eq:ind}, the nonlinear emf term $\curl(\vecu\times\vecb)$ is negligible. In that case, only diffusion plays a role in spreading $\bpol$ inward. Assuming that the spatial scale of the magnetic field $\bpol$ is much smaller in the vertical direction than in the horizontal directions and that the fluid below $r=r_{\rm skin}$ rotates rigidly at the rate $\Omrz$, Equation \eqref{eq:ind} becomes
\begin{align}\label{eq:ind_diffonly}
    \pderiv{\bpol}{t} +\left(\Omrz-\frac{1}{2}\right)\pderiv{\bpol}{\phi} \approx\ &\frac{\ek}{\prm} \etatilde \pderiv{^2\bpol}{r^2}\nonumber\\
    &r\leq r_{\rm skin},
\end{align}
where we recall that all quantities, including $\bpol$, are measured in rotating frame (which rotates with the background frame rate $1/2$). Since this equation is linear in $\bpol$, we use Equation \eqref{eq:tspec} to derive the differential equation for each mode $\bpol^{m\omega}$ separately:  
\begin{align}\label{eq:indskin}
	-i\left[\omega-m\left(\Omega\rz-\frac{1}{2}\right)\right]\bpol^{m\omega} \approx\ &\frac{\ek}{\prm}\etatilde\pderiv{^2\bpol^{m\omega}}{r^2}\nonumber\\
    &r\leq r_{\rm skin}.
\end{align}

Because $\etatilde$ varies with radius in some of our simulations, in order to make a more mathematically tractable problem, we follow \citet{Garaud1999} and define
\begin{align}\label{eq:reta}
	r_\eta \define r\inn + \frac{\int_{r\inn}^{r}\etatilde(r^\prime)^{-1/2}dr^\prime}{\int_{r\inn}^{r_{\rm skin}}\etatilde(r^\prime)^{-1/2}dr^\prime}.
\end{align}
Note that $r_\eta$ is a monotonically increasing function of $r$ and is equal to $r$ at $r=r\inn$ and $r=r_{\rm skin}$.  Again assuming rapid radial variation of the magnetic field components $\bpol^{m\omega}$, Equation \eqref{eq:indskin} becomes
\begin{subequations}\label{eq:indskin2}
	\begin{align}
	-i\left[\omega-m\left(\Omega\rz-\frac{1}{2}\right)\right]\bpol^{m\omega} \approx\ &\frac{\ek}{\prm}\etatilde\const \pderiv{^2\bpol^{m\omega}}{r_\eta^2}\nonumber\\
    &r\leq r_{\rm skin},\\
		\where \etatilde\const \define &\left[ \frac{r_{\rm skin}-r\inn}{\int_{r\inn}^{r_{\rm skin}}\etatilde(r^\prime)^{-1/2}dr^\prime} \right]^2
	\end{align}
\end{subequations}
is an intermediate value of $\etatilde(r)$ in the RZ.

Equation \eqref{eq:indskin2} has the exact solution
\begin{subequations}\label{eq:ampskin}
\begin{align}
	\overline{|\bpol^{m\omega}|^2}(r) &\approx \overline{|\bpol^{m\omega}|^2}(r_{\rm skin})\exp{\left[-2\left(\frac{r_{\rm skin}-r_\eta}{\delta^{m\omega}} \right)\right]},\\
	\where \delta^{m\omega} &\define\ \sqrt{\frac{2\ek\etatilde\const}{\prm|\omega-m(\Omega\rz-1/2)|} }\label{eq:skindepthmw}
\end{align}
\end{subequations}
is the skin depth unique to each $m$ and $\omega$. 

The ``skin-predicted" amplitude of large-scale $\avspht{|\bpol|^2}$ is then found by summing the amplitudes found in Equation \eqref{eq:ampskin} over all $\omega$ and low $m$. Here we confine ourselves to $m=(0,1,2)$. To minimize interference from nonlinear overshoot effects (recalling that $r\drz\dimm/\rsun=0.696$), we choose $r_{\rm skin}\dimm/\rsun=0.692$. Figure \ref{fig:skin} shows the large-scale $\avspht{|\bpol|^2}$ for both the skin-predicted and actually-realized values for Cases M3, M4, and M6. Equation \eqref{eq:ampskin} does a poor job of predicting the field strength for Case M3, a reasonable job for Case M4, and a very good job for Case M6. Overall, it thus seems that the skin effect is most relevant for stronger stable stratification. This might not be surprising, since for weak stratification, overshooting flows penetrate much deeper and can create a strong emf even in the deep RZ, which then makes Equation \eqref{eq:ind_diffonly} a poor assumption.

Furthermore, it is important to notice that the large-scale $\bpol$ in Case M6, although it is the weakest of all the cases toward the top of the RZ (see Figure \ref{fig:moll}), is perhaps surprisingly the strongest deep in the RZ. This is because the longer dominant cycle period of the dynamo allows deeper penetration of the poloidal field via the nonaxisymemtric skin effect.

\section{Discussion}\label{sec:disc}
We have explored global simulations of a CZ--RZ system with self-excited mean flows and magnetic fields in the solar-like radiative spreading dynamical regime where burrowing circulation causes the CZ's differential rotation to spread inward and spin the bulk of the RZ down. We have shown that the dynamo tachocline confinement scenario first explored in \twopapers\ continues to hold in this more solar-appropriate but computationally challenging regime. 

Furthermore, by analyzing the simulations at progressively higher levels of stable stratification while staying within the radiative spreading regime, we have found that large values of the stable stratification (large $\bu$), as are achieved in the Sun and other low-mass stars, are key to obtaining more solar-like thin tachoclines. As the stable stratification increases, the tachocline becomes more confined, the dynamo cycle becomes longer and more closely approximated by the nonaxisymmetric skin effect, and the skin depth becomes larger, allowing the dynamo's poloidal magnetic field to penetrate deeper into the RZ. 

These results indicate that the Maxwell stresses from the nonaxisymmetric skin effect may be a robust mechanism to confine solar and stellar tachoclines. Additionally, the fact that the dynamo cycle becomes longer as the stable stratification tends toward more realistic stellar values means that these same Maxwell stresses may penetrate deep enough to rigidify larger portions of the RZ than just the region immediately below the tachocline. This model may therefore have the benefit, as did the model of \citet{Gough1998}, of serendipitously providing a mechanism for rigidifying the deep RZ as well as confining the tachocline, with the additional bonus of self-consistent verification by simulations. %\newtext{Indeed, one could speculate on a possible synergy between the dynamo confinement scenario presented here and the simplified magnetic tachocline models which invoked a primordial magnetic field (e.g., \citealt{Gough1998} and \citealt{Rudiger2007}; the former was formulated for the radiative spreading regime, the latter for the viscous spreading regime). For example, it would be interesting to add a primordial magnetic field to one of our simulations (say Case M6) and see how the magnetic field strength (which would now be due to both primordial magnetism and a dynamo) compared to the estimates necessary for tachocline confinement provided by \citealt{Gough1998} and \citealt{Rudiger2007}.}

\newtext{Looking forward, one could speculate on a possible synergy between the dynamo confinement scenario presented here and the simplified magnetic tachocline models which invoked a primordial magnetic field (e.g., \citealt{Gough1998} and \citealt{Rudiger2007}; the former was formulated for the radiative spreading regime, the latter for the viscous spreading regime). For example, it would be interesting to add a primordial magnetic field to one of our simulations (say Case M6) and see how the magnetic field strength (which would then be due to both primordial magnetism and a dynamo) compared to the estimates necessary for tachocline confinement provided by \citealt{Gough1998} and \citealt{Rudiger2007}.}

\newtext{Another natural question is whether the field strength and nonaxisymmetric topology required for dynamo confinement of the solar tachocline is stable or unstable. There is a plethora of instabilities that may arise in the tachocline, from hydrodynamic and hydromagnetic shear instabilities (e.g., \citealt{Watson1981,Charbonneau1999b,Cally2003}) to baroclinic instabilities (e.g., \citealt{Gilman2014,Gilman2017}), and instabilities of the magnetic field in the presence of rotation (e.g., \citealt{Tayler1973, Kitchatinov2007,Skoutnev2025b}). Given that our simulations are highly nonlinear, the presence of relatively stationary magnetic cycles means that the fastest-growing instabilities have already saturated. It is thus challenging to assess which instabilities created the magnetic field configurations that we observe during the nonlinear evolution of our simulations. We also reemphasize that the simulations presented here are rather far from the appropriate asymptotic solar regime. Thus, if the dynamo confinement scenario were to operate in the Sun, the requisite magnetic structures might differ significantly from those presented here. }%To fully assess whether a dynamo confinement scenario is realistic for the Sun (which would of course require the necessary nonaxisymmetric cycling modes to be stable), one would likely need to extend the simulations presented here much closer toward the solar regime and then perform linear stability analyses about the resulting field configurations.} 
%\textcolor{orange}{Does even doing that make sense?}

%Nevertheless, our simulations provide confirmation that largely horizontal magnetic fields with $m=1$ or $m=2$ are at least in principle capable of confining tachoclines.
%  \textcolor{orange}{Alternative to your trailing phrase: to differentially rotate in a shellular manner} \com{[LM: I don't think that's quite right...shellular rotation is granted by Spiegel and Zahn ($\Omega=\Omega(r)$), but Ferraro's law is even less restrictive ($\Omega=\Omega(r,\theta)$).]}
% textcolor{orange}{I wonder if this isn't more important?  Maybe we should put this up front in this paragraph?} \com{[LM: Agreed. Moving it]}

The solar dynamo is typically characterized by its \textit{axisymmetric} ($m=0$) interior dipolar toroidal field. The presence of such a mean field is inferred observationally from the ``butterfly diagram" (the observed time-latitude behavior of emergent mean field; e.g., \citealt{Hathaway2015}) and is the basic assumption of ``mean field theory," from whence most of our theoretical intuition about large-scale dynamo processes is derived (e.g., \citealt{Radler1986}). By contrast, the dynamo confinement scenario invokes the less well-studied \textit{nonaxisymmetric} modes as the key mechanism for tachocline confinement. Indeed, \citetalias{Matilsky2024} found that (in the viscous spreading regime) primarily axisymmetric dynamo fields were not capable of tachocline confinement, but large-scale nonaxisymmetric fields were able to. Furthermore, nonaxisymmetric magnetic fields have long been recognized as a mechanism to induce rigid rotation (e.g., \citealt{Mestel1987}) because they restrict the freedom granted by Ferraro's law \citep{Ferraro1937} for differential rotation contours to align with axisymmetric poloidal field lines. Finally (and perhaps most importantly), there is significant observational evidence for large-scale nonaxisymmetric interior fields. One of the most prominent examples comes in the form of ``active longitudes" or ``activity nests" (e.g., \citealt{Maunder1905,Svalgaard1975,Henney2002,Norton2025}), which are preferred solar (Carrington) longitudes at which active regions emerge and flaring activity occurs. These preferred longitudes are often observed to lie on nearly opposite sides of the Sun (e.g., \citealt{Bai2003}), possibly with opposite polarity for the bipolar magnetic regions on either side as well (e.g., \citealt{Berdyugina2003,Mordvinov2004}). Such structures are consistent with an underlying large-scale, nonaxisymmetric $(m=1)$ toroidal dynamo field. We thus do not believe it unreasonable to speculate that the Sun's large-scale nonaxisymmetric magnetic fields (which are mostly hidden from direct observation), are responsible for both tachocline confinement and the rigid rotation of the RZ. To verify such a hypothesis, future work might assess possible correlations between the long-term variability of active longitudes, the nonaxisymmetric dynamo cycles achieved in global simulations with confined tachoclines, and the recent helioseismic observations of quasi-cyclic variability in the solar tachocline itself (e.g., \citealt{Basu2024,Korzennik2026,Basu2026}). 

While \newtext{we believe that our results are intriguing}, they rely on a small number of simulations that were somewhat arbitrarily scattered in parameter space. This was mainly due to the fact that our work is the first to explore overshooting convective dynamos in the radiative spreading regime. A wide variety of parameters (nondimensional numbers, initial conditions, shell geometries, and background profiles) were thus sampled in order to determine how best to access the correct dynamics. Now knowing that radiative spreading would likely still be achieved in slightly less turbulent simulations with higher $\sigma$ (because $\sigma_c\approx10$), a more systematic walk through parameter space would be to keep $\sigma$ fixed at a value slightly larger than unity and then to progressively increase $\bu$ while decreasing $\pr$. This approach could more clearly elucidate the trends in tachocline behavior and RZ rigidification as stellar values of the stable stratification and diffusivities are approached. 

As we noted in \citetalias{Matilsky2025a}, one drawback of the dynamo confinement scenario is that it relies on an essentially diffusive mechanism: the nonaxisymmetric skin effect. Although we have shown that the field's nonaxisymmetry can lead to significantly enhanced skin depths for field nearly corotating with the RZ (due to the frequency relevant for the skin-depth being Doppler-shifted from the dynamo cycle frequency), the molecular values of the diffusivities in most stars are still too small to yield large enough skin depths for the rigidification of significant portions of the RZ. Thus, for tachocline confinement and/or rigidification of the RZ to be possible, a turbulently enhanced diffusivity probably needs to be invoked. There is significant precedent for assuming turbulent enhancement of the diffusivities in both stellar astrophysics (e.g., \citealt{Zahn1992}) and atmospheric physics (e.g., \citealt{Pedlosky1987,Vallis2017}) and the processes leading to the enhancement of the viscosity and the thermal diffusivity are now becoming better understood (e.g., \citealt{Prat2014b, Lignieres2019, Chini2022, Skoutnev2023, Shah2024, Garaud2025}). However, we must acknowledge that much is still unknown about such turbulent enhancement, especially for the magnetic diffusivity in the presence of strong Maxwell stresses.

%(for a recent review of the ``solar lithium problem" see, e.g., \citealt{Buldgen2019})

\newtext{One prominent issue with assuming turbulent enhancement of the magnetic diffusivity $\eta_T\dimm$ (or of any diffusivity for that matter) is a strong tension with the problem of solar lithium abundance. Lithium is burned if it reaches temperatures of $\approx$ $\sn{2.6}{6}$ K, which are achieved only $d\sim0.06\rsun$ below the base of the CZ (e.g., \citealt{Rudiger2001}). Thus, in order to slow the transport of lithium across the tachocline, which would otherwise lead to a decay of lithium faster than the observed surface time-scale of $t_{\rm decay}\dimm\sim1$ Gyr, there must be an upper bound on lithium's turbulent diffusivity (which we denote by $D_T\dimm$). In other words, we must have $D_T\dimm < d^2/t_{\rm decay}\dimm = 500\ \stoke$. Meanwhile, for the dynamo confinement scenario to work in the Sun, the turbulently enhanced skin depths must be as large as the tachocline depth $\Delta\dimm\approx0.02\rsun$, leading to the requirement that $\eta_T\dimm>\Delta\dimsq\abs{\omega_{\rm cyc}\dimm - m(\Omega_{\rm rz}\dimm-\Omega_0)}$, where $\omega_{\rm cyc}\dimm$ is the (dimensional) dynamo cycle period (or a range of cycle periods for irregular cycles). Using Figures \ref{fig:timelat}(f) and \ref{fig:omrz}, we estimate (for the nondimensional variables) the typical cycle period (for $m=1$) and the rotation rate of the RZ in Case M6: $-\sn{2}{-3}\lesssim\omega_{\rm cyc}\lesssim10^{-3}$ and $\Omrz-1/2\approx-10^{-3}$. Then, using Table \ref{tab:time-scales_dim}, we obtain $\eta_T\dimm > (\sn{1.4}{9}\ \cm)^2 (10^{-3})(\sn{1.5}{-5}\ \second^{-1})= \sn{3}{10}\ \stoke$. Thus, for our dynamo confinement scenario to be consistent with a solution to the lithium problem, the magnetic turbulent diffusivity must nominally be eight orders of magnitude larger than the turbulent diffusivity of lithium ($\eta_T\dimm\gg D_T\dimm$). Of course, our estimate of $\eta_T^*$ is derived from our simulations and is likely not representative of the solar value. Nevertheless, this simple analysis clearly reveals the tension between the dynamo confinement scenario (or really any theory which assumes turbulent enhancement of diffusivity of any kind) and the solar lithium problem.}
% This large degree of separation is dependent on the actual tachocline width, as well as the locations of the tachocline and magnetic pumping regions relative to the $T=\sn{2.6}{6}$ K surface, all of which are uncertain. Nevertheless the tension between the dynamo confinement scenario (or really any theory which assumes turbulent diffusivity enhancement of any kind) and the lithium problem is clear.

\begin{figure}
	\centering
	%\vspace*{10cm}
	%\hspace*{-1cm}
	\includegraphics[width=3.4375in]{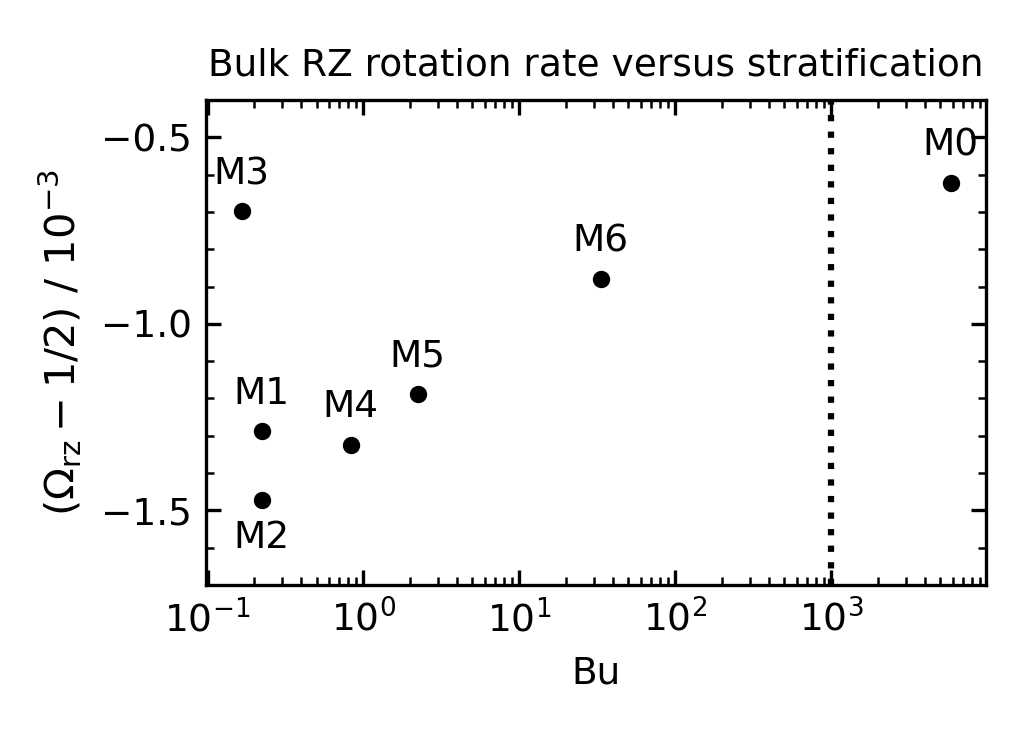}
    \vspace{-0.7cm}
	\caption{Difference of the bulk rotation rate of the RZ from the background frame rate, $\Omrz-1/2$, as a function of $\bu$ for the MHD cases (see Equation \ref{eq:bulkrates}). The dotted line separates the radiative spreading cases on the left from the lone viscous spreading case on the right. }
	\label{fig:omrz}
\end{figure}

\newtext{We assume for now that the dynamo} skin depths can be significantly enhanced, whether through turbulence, the Doppler effect, or both. It is then intriguing to imagine the wider implications for RZ spin-down from a dynamo confinement scenario. The Sun used to spin significantly faster, but was slowed down over time due to the action of mass loss and magnetic torques at the surface. However, as we noted in Section \ref{sec:intro}, it is not obvious exactly how this spin-down might be communicated down below the CZ, especially because the existence of the tachocline suggests little angular momentum transport between the CZ and RZ. A long-standing problem in solar physics is thus why the solar RZ is not helioseismically observed to rotate much faster. Given the results of this paper, dynamo processes within the radiative spreading regime offer a potential pathway to spin the RZ down, even in the presence of a tacholine. Figure \ref{fig:omrz} shows the average bulk RZ rotation rate in the equilibrated state for our simulations. All of the RZs are appreciably spun down, although the cases with the best-confined tachoclines tend to be spun down the least (save for Case M3).

However, one might expect the spin-down process to be effective only within the tachocline region (if a tachocline were to be present). Surprisingly, these simulations show that, even in the tachocline cases, the Maxwell stresses provide a mechanism to spin the RZ down even below the tachocline. Figure \ref{fig:torque_vs_radius} shows the equilibrated horizontally averaged differential rotation and torque balance for Case M6 in both the RZ and CZ. In the CZ, the Reynolds stresses work in concert with the meridional circulation to spin up the upper CZ and slow down the lower CZ, but they are strongly opposed by the Maxwell stresses. In the tachocline region, the negative meridional-circulation torque radiatively spreads downward and causes the spin-down of the tachocline region. However, because there is a tachocline, the Maxwell stresses prevent the radiative spreading from penetrating deeper. Below the confined tachocline, it is the Maxwell-stress torque itself that becomes negative (below $r/\rsun\approx0.64$) and communicates the spin-down even deeper into the RZ. 

\begin{figure}
	\centering
	%\vspace*{10cm}
	%\hspace*{-1cm}
	\includegraphics[width=3.4375in]{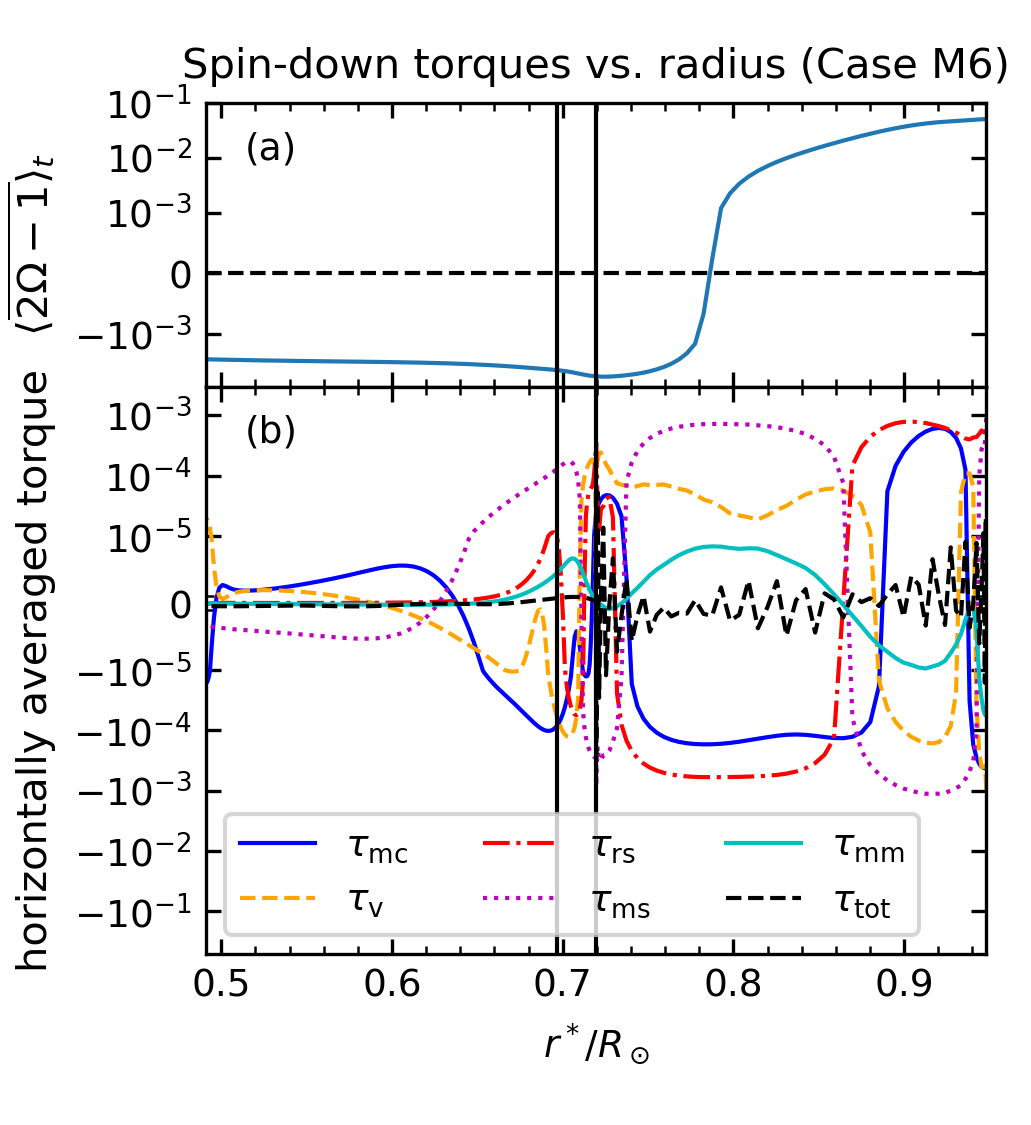}
    \vspace{-0.7cm}
	\caption{(a) Horizontally averaged differential rotation, $\avt{\overline{2\Omega-1}}$, as a function of radius for Case M6. The scale is linear in the range $(-10^{-3},10^{-3})$ and logarithmic elsewhere. (b) Horizontally and temporally averaged torques from Equation \eqref{eq:torques} as functions of radius. The scale is linear in the range $(-10^{-5},10^{-5})$ and logarithmic elsewhere. For both panels, the temporal average is over the equilibrated state. Also, $\tau_{\rm tot}\define \taumc+\tauv+\taurs+\taums+\taumm$. The vertical lines mark ${r\drz}$ and $\rbcz$.  }
	\label{fig:torque_vs_radius}
\end{figure}

These simulations are currently the only nonlinear solutions of the MHD equations that self-consistently establish a tachocline in the solar-like radiative spreading regime. If they are indeed representative of the Sun's interior dynamics, then, in summary, the picture we have is the following. Reynolds stresses and meridional circulation both work to create a solar-like differential rotation with a relatively fast equator and slow high-latitude regions. This requires both spin-up of the upper CZ (most of which is at relatively low latitudes by volume) and compensatory spin-down of the lower CZ. Due to radiative spreading, the meridional circulation burrows inward and communicates the spin-down of the CZ's base to the upper RZ. Then Maxwell stresses from the dynamo, transported inward via the nonaxisymmetric skin effect, halt the radiative spreading but in doing so, end up rigidifying the deep RZ by propagating the spin-down even further below the tachocline. If this picture is accurate, then the very presence of the Sun's tachocline may have significant implications for which physical processes are expected to cause RZ spin-down.  

Finally, it is intriguing to speculate on a possible synergistic relationship between the solar dynamo and tachocline. It has long been thought that the tachocline is an essential component of the dynamo. Our new simulations suggest that the converse is also true: namely, that the dynamo is an essential ingredient for not only a confined tachocline, but possibly also the rigidification of the deeper RZ.

% We thank an anonymous reviewer for highly insightful commentary.
\begin{acknowledgments}
We thank P. Garaud, J. Toomre, N. Featherstone, B. Hindman, C. Blume, J. Pedlosky, V. Skoutnev, and A. Strugarek for helpful discussions. This work was primarily supported by the COFFIES DRIVE Science Center (NASA grant 80NSSC22M0162), with additional support from NSF award AST-2202253 and NASA grants 80NSSC18K1125, 80NSSC18K1127, 80NSSC19K0267, 80NSSC20K0193, 80NSSC21K0455, 80NSSC24K0270 and 80NSSC24K0125. Computational resources were provided by the NASA Advanced Supercomputing (NAS) Division at Ames Research Center. {\rayleigh} is supported by the Computational Infrastructure for Geodynamics (CIG) through NSF awards NSF-0949446 and NSF-1550901. For reproducibility, the {\rayleigh} input files and final checkpoints for each simulation \newtext{are publicly accessible at \url{https://doi.org/10.5281/zenodo.18265773} (Cases 0--3) and \url{https://doi.org/10.5281/zenodo.18275692} (Cases 4--6). }
\end{acknowledgments}

%\clearpage
%\newpage
\appendix
\twocolumngrid
\restartappendixnumbering

\section{Reference State for Cases 0, 3, 4, and 6}\label{ap:ref}

This section describes the background state used for Cases 0, 3, 4, and 6. For Cases 1, 2, and 5, see \citet{Korre2024}. 

According the standard solar structure Model S \citep{ChristensenDalsgaard1996}, the base of the solar CZ lies at $\rbczsun\define\sn{4.96}{10}$ cm and the third density scale-height is crossed at $\rnrhothreesun\define\sn{6.58}{10}\ \cm$. Since we choose $\beta=0.759\approx\rbczsun/\rnrhothreesun$ (Equation \ref{eq:beta}), the $\Nrho=3$ simulations' CZs can be thought of as coinciding with the lowermost three density scale-heights of the solar CZ according to Model S. We thus identify the dimensional values $\rbczdim$ and $\routdim$ approximately with $\rbczsun$ and $\rnrhothreesun$, respectively, as we did in Equation \eqref{eq:radiidim}. We then choose our unit of length to be
\begin{align}\label{eq:czdepth}
    H&\define \routdim - \rbczdim = 0.228\rsun = \sn{1.59}{10}\ \cm \nonumber\\
    &\approx \rnrhothreesun - \rbczsun.
\end{align}
Note that $\rnrhothreesun - \rbczsun=0.232\rsun = \sn{1.61}{10}\ \cm$, which is slightly greater than $H$ and therefore the location of our CZ ($\rbczdim$, $\routdim$) does not correspond exactly with the location of the solar CZ's lowermost three density scale-heights ($\rbczsun$, $\rnrhothreesun$). This appears to be due to historical rounding error in prior calculations $\rbczsun$ and/or $\rnrhothreesun$ (in particular, $\rbczdim$ is usually taken to be $\sn{5.00}{10}\ \cm= 0.718\rsun$ instead of the more precise $\sn{4.96}{10}\ \cm = 0.713\rsun$; e.g., \citealt{Brun2002}). 
	
In terms of the dimensional background state, the perfect gas law is
\begin{align}\label{eq:idgasdim}
    \prstilde\dimm = \left[\frac{(\gamma-1)\cp}{\gamma}\right]\rhotilde\dimm\tmptilde\dimm,
\end{align}
hydrostatic balance is
\begin{align}\label{eq:hydrdim}
    \frac{d\prstilde\dimm}{dr\dimm}=-\rhotilde\dimm\gravtilde\dimm,
\end{align}
and the first law of thermodynamics is
\begin{align}\label{eq:firstlawdim}
    \frac{1}{\cp}\left(\frac{d\entrtilde}{dr}\right)\dimm=\frac{d\ln\tmptilde\dimm}{dr\dimm}-\left(\frac{\gamma-1}{\gamma}\right)\frac{d\ln\rhotilde\dimm}{dr\dimm}.
\end{align}
Here
\begin{equation}\label{eq:gamma}
	\gamma\define\frac{\cp}{\cv}\equiv\frac{5}{3}
\end{equation}
is the ratio of specific heats, where $\cp$ and $\cv$ are the specific heats at constant pressure and volume, respectively. 

After the nondimensionalization described in Section \ref{sec:numeq}, Equations \eqref{eq:idgasdim}--\eqref{eq:firstlawdim} take the form
\begin{align}\label{eq:idgas}
    \prstilde&= \rhotilde\tmptilde,
\end{align}
\begin{align}\label{eq:hydr}
    \frac{d\prstilde}{dr}=-\di\left(\frac{\gamma}{\gamma-1}\right)\rhotilde\,\gravtilde,
\end{align}
and 
\begin{align}\label{eq:firstlaw}
    \dsdrtilde=\frac{1}{\gamma}\frac{d\ln\tmptilde}{dr}-\left(\frac{\gamma-1}{\gamma}\right)\frac{d\ln\rhotilde}{dr},
\end{align}
where $\entrtilde\define\entrtilde\dimm/\cp$ (unique only up to an arbitrary constant) and $\prstilde\define\prstilde^*/p\cz$. Recall that $\di\define\grav\cz H/\cp T\cz$. We combine Equations \eqref{eq:idgas}--\eqref{eq:firstlaw} to find
\begin{align}
    \frac{d\tmptilde}{dr}-\left(\dsdrtilde\right)\tmptilde &= -\di\,\gravtilde,
\end{align}
which has the exact solution 
\begin{align}\label{eq:tmptilde}
    \tmptilde &= e^{\entrtilde}\left[\tmptilde(\rbcz) - \di\int_{\rbcz}^r \gravtilde(x)  e^{-\entrtilde(x)}dx\right].
\end{align}
Note that the actual background entropy $\entrtilde$ is only relevant up to an integration constant. In Equation \eqref{eq:tmptilde}, we have therefore chosen, without loss of generality, 
\begin{equation}\label{eq:entrtilde}
    \entrtilde(r)\define \int_{\rbcz}^r\frac{d\entrtilde}{dr^\prime}dr^\prime
\end{equation}
such that $\entrtilde(\rbcz)\equiv0$.% (strictly, in this Appendix, the ``$\equiv$" refers only to Cases 3, 4, and 6, not all cases.) 

We then eliminate $\prstilde$ from Equations \eqref{eq:idgas}  and \eqref{eq:hydr} to yield
\begin{align}\label{eq:rhotilde}
    \rhotilde &= \rhotilde(\rbcz) \exp{\left[-\left(\frac{\gamma}{\gamma-1}\right)\entrtilde\right]}\tmptilde^{1/(\gamma-1)}.
\end{align}

We choose the gravity $\gravtilde\dimm$ to be due to a centrally concentrated mass:  
\begin{align}\label{eq:gravdim}
	\gravtilde\dimm\propto \frac{1}{r\dimsq}. 
\end{align}
For $\gravtilde\ofr\propto1/r^2$ and the normalization condition $(4\pi/V\cz)\int_{\rbcz}^{\rout}\gravtilde(r)r^2dr=1$ (where $V\cz\define (4\pi/3)(\rout^3-\rbcz^3)=(4\pi/3)[(1-\beta^3)/(1-\beta)^3]$ is the nondimensional volume of the CZ), we require
\begin{align}
    \gravtilde(r) = \left[ \frac{1-\beta^3}{3(1-\beta)^3}    \right]\frac{1}{r^2}.
\end{align}

The background squared buoyancy frequency $\brunttilde\dimsq$ is given by
\begin{align}\label{eq:nsqdim}
	\brunttilde\dimsq\define\frac{\gravtilde\dimm}{\cp}\left(\dsdrtilde\right)\dimm
\end{align} 
and its nondimensional counterpart (to satisfy the normalization condition $\avrz{\brunttilde}=1$) is
\begin{align}\label{eq:nsqnd}
	\brunttilde^2 \define \frac{\gravtilde\,\dsdrtildeline}{\avrz{\sqrt{\gravtilde\,\dsdrtildeline}}^2}.
\end{align} 
%where $\entrtilde\dimm$ is the background entropy and $\cp$ is the specific heat at constant pressure.

To model the transition from convective stability to instability at $r=\rbcz$, we choose $\dsdrtildeline$ to be 0 in the CZ, a constant positive (near-unity) value in the RZ, and continuously matched in between. We accomplish this via quartic matching: 
\begin{equation}\label{eq:dsdrquart}
    \dsdrtilde  = \begin{cases}
        0.453\ten  r\leq \rbcz - \delta& \\
        0.453\left\{1 - \left[1 - \left(\dfrac{r-\rbcz}{\delta}\right)^2\right]^2\right\} & \nonumber\\
         \ten \rbcz - \delta < r < \rbcz & \nonumber\\
        0  \ten r\geq \rbcz. &
    \end{cases}
\end{equation}

The value 0.453 ensures that our nondimensional background entropy stratification $\dsdrtildeline$ matches the dimensional $(d\entrtilde/dr)\dimm=(\cp/H)\dsdrtildeline$ specified in \citetalias{Matilsky2022}. It comes from estimating $(\dsdrtildeline)\dimm\approx0.01\ \unitdsdr$ near the top of the solar RZ (e.g., \citealt{Brun2011}, their Fig. 1), $H=\sn{1.59}{10}\ \cm$ as in Equation \eqref{eq:czdepth}, and $\cp\approx\sn{3.50}{8}\ \unitentr$. More fundamentally, the chosen nondimensional amplitude of $\dsdrtildeline$ in the RZ sets the number of density scale-heights across the RZ, $\Nrhorz$. For Cases 0, 3, 4, and 6, $\Nrhorz=2.10$.

The background $\rhotilde(r)$ and $\tmptilde(r)$ are now completely determined via Equations \eqref{eq:tmptilde}--\eqref{eq:dsdrquart}. 

There are three equations relating $\rhotilde(\rbcz)$, $\tmptilde(\rbcz)$, $\di$, $\gamma$, $\beta$, and $\Nrho$: two from our  choice of nondimensionalization---$(4\pi/V\cz)\int_{\rbcz}^{\rout}\rhotilde(r)r^2dr=1$ and $(4\pi/V\cz)\int_{\rbcz}^{\rout}\tmptilde(r)r^2dr=1$---and one from the definition \eqref{eq:nrho} $\Nrho\define\ln{[\rhotilde(\rbcz)/\rhotilde(\rout)]}$. Thus, $\rhotilde(\rbcz)$, $\tmptilde(\rbcz)$, and $\di$ may all be regarded as functions of $\gamma$, $\beta$, and $\Nrho$. 

Because $\dsdrtildeline\equiv0$ in the CZ, explicit formulae can be derived for $\di$ and $\tmptilde(\rbcz)$:
\begin{align}
    & \di \define \frac{\gravcz H}{\cp\tmpcz} = \nonumber\\
    &\frac{3 \beta (1 - \beta)^2 (1 - e^{-\Nrho/n})} 
    { \frac{3\beta}{2} (1 - \beta^2) (1 - e^{-\Nrho/n}) - (1-\beta^3)(\beta-e^{-\Nrho/n})}\label{eq:di}    
\end{align}
and
\begin{align}
    &\tmptilde(\rbcz)=\nonumber\\
    & \frac{(1-\beta^3)(1-\beta)}{ \frac{3\beta}{2} (1 - \beta^2) (1 - e^{-\Nrho/n}) - (1-\beta^3)(\beta-e^{-\Nrho/n})},\label{eq:tmp0}
\end{align}
where 
\begin{equation}\label{eq:polyn}
    n\define \frac{1}{1-\gamma}=\frac{3}{2}
\end{equation}
is the ``polytropic index" (but note that only the CZ has a polytropic stratification). The equation for $\rhotilde(\rbcz)$ is transcendental and requires a numerical solution. For completeness:  $\rhotilde(\rbcz)=2.68$, $\tmptilde(\rbcz)=2.04$, and $\di=1.72$.  Note that $\di=O(1)$ is a consequence of the virial theorem.

We note that the shape of the background diffusivity profiles can be arbitrarily chosen without affecting thermal or hydrostatic equilibrium. For Cases 0, 3, 4, and 6, we choose all diffusivities to increase with height according to $\rhotilde^{-1/2}$ (see Table \ref{tab:inputnondim}). Note that in Cases 1, 2, and 5 all the difffusivities are constant.

Finally, we choose the background radiative heating $\heatradtilde$ to be proportional to the background pressure in the CZ (meaning that most of the luminosity---which is of course what $\heatradtilde$ represents---is injected in the bottom $\sim$$1/3$ fo the CZ) and to taper to zero in the RZ:
\begin{equation}\label{eq:heatrad}
    \heatradtilde \propto \left[1 + \tanh{\left(\frac{r-\rbcz}{\drad}\right)}\right]\prstilde,
\end{equation}
where we choose $\drad\equiv0.005$ to ensure a sharp cutoff at the CZ--RZ interface.  

Because $\heatradtilde\dimm=(\fluxnradcz/H)\heatradtilde$, $\fluxscalarnrad\dimm=\fluxnradcz\fluxscalarnrad$, and $\fluxscalarnradtilde\dimm\define(H/r^2)\int_r^{\rout}\heattilde\dimm\ofrprime\rprime^2d\rprime$ (see Equation \ref{eq:fnrad}), normalization of the heat flux---$(4\pi/V\cz)\int_{\rbcz}^{\rout}\fluxscalarnrad r^2dr=1$---requires
\begin{equation}\label{eq:heatnorm}
    \frac{4\pi}{V\cz}\int_{\rbcz}^{\rout}\int_r^{\rout}\heatradtilde\ofrprime \rprime^2d\rprime dr=1,
\end{equation}
which sets the proportionality constant in Equation \eqref{eq:heatrad}.

\section{Computational Grid}\label{ap:grid}
{\rayleigh} is pseudospectral: for the linear terms, it evolves the equations using spherical harmonics to represent the horizontal variation of the fields and Chebyshev polynomials to represent the vertical variation. $\lmax$ is the maximum degree of the spherical harmonics and $\nmax$ is the maximum degree of the Chebyshev polynomials. Each grid is dealiased by a factor of 2/3, that is, if $N_r$ and $N_\theta$ represent the number of collocation points in $r$ and $\theta$, respectively, then $\nmax=\lfloor(2/3)N_r\rfloor$ and $\lmax=\lfloor(2/3)N_\theta\rfloor$. In all cases, the number of collocation points in longitude is $N_\phi=2N_\theta$ and the spherical-harmonic truncation is triangular. Note that this means we use $\nmax+1$ Chebyshev polynomials, $\lmax+1$ spherical-harmonic degrees, and $2\lmax$ spherical-harmonic orders, with the integer azimuthal wavenumber $m$ (for a given $\ell$) ranging from $-\ell+1$ to $+\ell$. Table \ref{tab:grid} contains the resolutions and domain boundaries for all cases considered here. 

\begin{table*}
    \caption{Grid resolution for the cases considered here. For some cases, we use three stacked Chebyshev collocation grids in radius. For those cases, we give number of collocation points in each domain as a list, ordered by increasing radii of the subdomains. The dimensional radial domain boundaries (in units of $\rsun$) is always one greater than the number of collocation grids and is given in the final column of the table. }
\label{tab:grid}
\centering
\begin{tabular}{*{7}{l}}
\multicolumn{7}{c}{\textbf{Properties of the simulation grid}} \\ % This acts as a title
\hline
Name & $N_r$ & $N_\theta$ & $N_\phi$ & $\nmax$ & $\lmax$ & (domain boundaries)$/R_\odot$ \\
\hline
Case H0 & (64, 64, 64) & 384 & 768 & (42, 42, 42) & 256 & (0.491, 0.670, 0.720, 0.948) \\
Case M0 & (64, 64, 64) & 384 & 768 & (42, 42, 42) & 256 & (0.491, 0.670, 0.720, 0.948) \\
\hline
Case H1 & 192 & 1344 & 2688 & 128 & 896 & (0.427, 0.948) \\
Case M1 & 192 & 1152 & 2304 & 128 & 768 & (0.427, 0.948) \\
Case H2 & 192 & 1152 & 2304 & 128 & 768 & (0.427, 0.948) \\
Case M2 & 192 & 1152 & 2304 & 128 & 768 & (0.427, 0.948) \\
Case H3 & (64, 64, 64) & 384 & 768 & (42, 42, 42) & 256 & (0.491, 0.670, 0.720, 0.948) \\
Case M3 & (64, 64, 64) & 384 & 768 & (42, 42, 42) & 256 & (0.491, 0.670, 0.720, 0.948) \\
Case H4 & (64, 64, 64) & 1152 & 2304 & (42, 42, 42) & 768 & (0.491, 0.670, 0.720, 0.948) \\
Case M4 & (64, 64, 64) & 1152 & 2304 & (42, 42, 42) & 768 & (0.491, 0.670, 0.720, 0.948) \\
Case H5 & 192 & 1104 & 2208 & 128 & 736 & (0.427, 0.948) \\
Case M5 & 192 & 1152 & 2304 & 128 & 768 & (0.427, 0.948) \\
Case H6 & (64, 64, 64) & 768 & 1536 & (42, 42, 42) & 512 & (0.491, 0.670, 0.720, 0.948) \\
Case M6 & (64, 64, 64) & 512 & 1024 & (42, 42, 42) & 341 & (0.491, 0.670, 0.720, 0.948) \\
\hline
\end{tabular}
\end{table*}

\section{Detailed Averaging Procedure}\label{ap:averages}
Let $\psi=\psi(r,\theta,\phi,t)$ denote a scalar quantity (or a single component of a vector quantity) dependent on position and time. Then 
\begin{align}\label{eq:av}
\av{\psi}(r,\theta,t)\define \frac{1}{2\pi}\int_0^{2\pi}\psi(r,\theta,\phi,t)d\phi
\end{align}
is the instantaneous longitudinal average of $\psi$.

We denote averages over spherical surfaces with an overbar: 
\begin{align}\label{eq:avsph}
\overline{\psi}(r,t)\define \frac{1}{4\pi}\int_0^{2\pi}\int_0^{\pi}\psi(r,\theta,\phi,t)\sin\theta d\theta d\phi.
\end{align}

We define a volume-average over a spherical shell bounded by radii $r_1$ and $r_2$  by 
\begin{align}\label{eq:avcz}
\av{\psi}_{\rm v}(t; r_1, r_2)\define &\frac{3}{4\pi(r_2^3-r_1^3)}\times\nonumber\\
\int_0^{2\pi}\int_0^{\pi}\int_{r_1}^{r_2}&\psi(r,\theta,\phi,t)r^2\sin\theta dr d\theta d\phi.
\end{align}

We then define instantaneous volume-averages over the CZ, RZ, and deep RZ by 
\begin{subequations}
\begin{align}
\avcz{\psi}(t) &\define \av{\psi}_{\rm v}(t; \rbcz, \rout),\\
\avrz{\psi}(t) &\define \av{\psi}_{\rm v}(t; \rin, \rbcz),\\
\andd \avdrz{\psi}(t) &\define \av{\psi}_{\rm v}(t; \rin, {r\drz}),
\end{align}
\end{subequations}
respectively. 

We never consider temporal averages of quantities that were not also spatially averaged in some way. We therefore denote an additional temporal average over the interval $t_1$ to $t_2$ by appending a ``$t$" subscript to the appropriate averaging symbol, i.e., by using the notation $\avt{\psi}$, $\avaltspht{\psi}$, $\avczt{\psi}$, $\avrzt{\psi}$, and ${\av{\psi}}_{{\rm drz},t}$. For example, 
\begin{align}\label{eq:avt}
\avt{\psi}(r,\theta)\define \frac{1}{2\pi(t_2-t_1)}\int_0^{2\pi}\int_{t_1}^{t_2}\psi(r,\theta,\phi,t)d\phi dt.
\end{align}
The averaging interval in Equation \eqref{eq:avt} is over a long portion of the equilibrated state of the simulation unless otherwise specified.

In reality, the integrals in Equations \eqref{eq:av}--\eqref{eq:avt} are replaced with sums over the appropriate Gaussian weights when computing spatial averages from the simulation data. For the temporal averages, we replace the operator $1/(t_2-t_1)\int_{t_1}^{t_2}dt$ with a simple (unweighted) average over the relevant time series, on the grounds that our sample rate is usually slower than the frequencies of various waves and convective realizations. 

\section{Time-Scales}\label{ap:timescales}
 We define the dimensional time-scales appearing in our CZ--RZ system (most of which are associated with fairly obvious physical processes) as follows: 
%\FloatBarrier

\begin{subequations}\label{eq:timescalesdim}
\begin{align}
    \tbruntdim &\define \bruntrz^{-1}\five\text{(gravity-wave time-scale)},\\
    \tomegadim &\define (\twoOmzero)^{-1}\five\text{(rotational time-scale)},\\
\tffdim&\define \sqrt{\frac{H\cp}{\gravcz\Dentr}}\five\text{(convective free-fall time-scale)},\\
    \tkappaczdim &\define \frac{H^2}{\kappacz}\five\text{(CZ thermal diffusion time-scale)},\\
    \tetaczdim &\define  \frac{H^2}{\etacz}\five\text{(CZ magnetic diffusion time-scale)},\\
    \tnuczdim &\define  \frac{H^2}{\nucz}\five\text{(CZ viscous diffusion time-scale)},\\
    \tesdim&\define \frac{\rbcz\dimsq}{\kapparz}\frac{\bruntrz^2}{4\Omzero^2}\five\text{(Eddington-Sweet time-scale)},\label{eq:tesdim}\\
    &\text{and}\nonumber\\
    \tnudim&\define  \frac{\rbcz\dimsq}{\nurz}\nonumber\\
    &\text{(viscous diffusion time-scale)}.\label{eq:tnufulldim}
\end{align}
\end{subequations}
%\FloatBarrier

Using these definitions, the control parameters appearing in Equations \eqref{eq:control} are
%\FloatBarrier
\begin{subequations}\label{eq:controltimeratio}
\begin{align}
    \pr &\define\frac{\tkappaczdim}{\tnuczdim},\label{eq:tratiopr}\\
    \prm&\define \frac{\tetaczdim}{\tnuczdim},\label{eq:tratioprm}\\
    \raf &\define \frac{\tnuczdim\tkappaczdim}{\tffdimsq},\label{eq:tratiora}\\
    \roc &\define \frac{\tomegadim}{\tffdim},\label{eq:tratioroc}\\
    \sigma &\define \sqrt{\frac{\tes\dimm}{\tnu\dimm}},\label{eq:tratiosigma}\\
    \ek &\define \frac{\tomegadim}{\tnuczdim},\label{eq:tratioek}\\
    \andd\bu &\define \left(\frac{\tomegadim}{\tbruntdim}\right)^2.\label{eq:tratiofr}
\end{align}
\end{subequations}
%\FloatBarrier

Equations \eqref{eq:controltimeratio} are valid for the nondimensional time-scales as well (i.e., after simply removing the asterisks from the right-hand-sides). Since we have chosen $(\twoOmzero)^{-1}$ as our unit of time, we define the nondimensional time-scales by:
%\FloatBarrier
\begin{subequations}\label{eq:timescales}
\begin{align}
\tbrunt &\define \bu^{-1/2}\\
\tomega&\define 1,\\
\tff&\define \ro_{\rm c}^{-1},\\
\tnucz &\define \ek^{-1},\\
\tkappacz&\define \pr\ek^{-1},\\
\tetacz &\define \prm\ek^{-1},\\
\tes &\define \left(\frac{\beta}{1-\beta}\right)^2\left(\frac{\nucz}{\nurz}\right) \sigma^2 \ek^{-1},\\
\andd \tnu &\define \left(\frac{\beta}{1-\beta}\right)^2\left(\frac{\nucz}{\nurz}\right)\ek^{-1}.
\end{align}
\end{subequations}
%\FloatBarrier

The nondimensional time-scales and the simulation run times are shown in Table \ref{tab:time-scales}. 

\begin{table*}
	\caption{Nondimensional time-scales for each simulation and the Sun. For the HD cases, we report the run time in units of $\tes$, while for the MHD cases, we report the run time in units of $\teta$. }
	\label{tab:time-scales}
	\centering
	\begin{tabular}{*{11}{l}}

    \multicolumn{11}{c}{\textbf{Nondimensional time-scales}} \\ % This acts as a title
\hline
        
Name & $\tbrunt$ & $\tomega$ & $\tff$ & $\tnucz$ & $\tkappacz$ & $\tetacz$ & $\tes$ & $\tnu$ & HD: $\trun/\tes$ & MHD: $\trun/\teta$ \\
\hline
Case 0 & 0.0131 & 1 & 2.50 & 1.87e3 & 1.87e3 & 7.50e3 & 3.35e8 & 5.74e4 & 1.74e-04 & 3.69 \\
\hline
Case 1 & 2.12 & 1 & 2.00 & 4.47e3 & 4.47e3 & 1.79e4 & 1.80e4 & 8.09e4 & 4.27 & 0.233 \\
Case 2 & 2.12 & 1 & 2.00 & 2.00e3 & 2.00e3 & 8.00e3 & 8.04e3 & 3.62e4 & 6.33 & 0.796 \\
Case 3 & 2.46 & 1 & 2.50 & 2.65e3 & 1.33e3 & 1.06e4 & 6.72e3 & 8.12e4 & 6.17 & 9.79 \\
Case 4 & 1.10 & 1 & 2.50 & 5.93e3 & 5.93e2 & 2.37e4 & 1.50e4 & 1.82e5 & 7.54 & 1.21 \\
Case 5 & 0.671 & 1 & 2.00 & 2.00e3 & 2.00e3 & 8.00e3 & 8.04e4 & 3.62e4 & 0.762 & 0.750 \\
Case 6 & 0.174 & 1 & 2.50 & 3.75e3 & 9.37e2 & 1.50e4 & 9.51e5 & 1.15e5 & 0.0459 & 2.68 \\
\hline
Sun  & 0.0067 & 1 & 2.4 & 3.3e13 & 4e7 & 7e9 & 2e13 & 1.3e15 & n/a & n/a \\
\hline    
	\end{tabular}
\end{table*}

\section{Dimensional Parameters for the Sun and Our Simulations}\label{ap:dim}
\subsection{Dimensional Properties of the Sun}
Table \ref{tab:sundim} shows the dimensional parameters of the solar interior inferred from Model S \citep{ChristensenDalsgaard1996}. Many parameters can be obtained directly from the Model S profiles, but others (especially the diffusivities) require derivation and are only accurate to about 10\% (e.g., \citealt{Spitzer1962}). 

The base of the solar CZ $\rbczsun$ is calculated according to Model S from the condition
\begin{align}
    \brunttilde\dimm(\rbczsun)=0,
\end{align}
which yields $\rbcz=\sn{4.96}{10}\ \cm=0.713\rsun$.

Estimating $\Delta s$ according to Equation \eqref{eq:deltas} is only appropriate for our simulations, for which convection is driven by a thermal boundary layer. For the Sun, $\Delta s$ is unknown. Instead, to estimate Rayleigh and Rossby numbers, we derive a convective free-fall velocity $u_{\rm ff}\dimm$ according to mixing-length theory:
\begin{align}\label{eq:uff}
u_{\rm ff}\dimm\define \left(\frac{\fluxnradcz}{\rhocz}\right)^{1/3}.
\end{align}
We estimate the solar convective length-scale to be
\begin{align}
    H\cz &\define \avcz{\hrhotilde\dimm}\\
    \where \hrhotilde\dimm &\define - \left(\frac{d\ln\rhotilde\dimm}{dr\dimm}\right)^{-1}
\end{align}
is the local density scale-height. We then estimate the convective free-fall time
\begin{align}
    \tffdim&\define \frac{H\cz}{u_{\rm ff}\dimm},
\end{align}
the solar flux-based Rayleigh number,
\begin{align}
    \ra_\odot \define\frac{\tnuczdim \tkappaczdim}{\tffdimsq}= \frac{\fluxnradcz^{2/3}H\cz^2}{\rhocz^{2/3}\nu\cz\dimm\kappa\cz\dimm},
\end{align}
and the solar convective Rossby number
\begin{align}
    \ro_\odot\define \frac{1}{2\Omega_\odot \tffdim} = \frac{\fluxnradcz^{1/3}}{2\Omega_\odot H\cz\rhocz^{1/3}},
\end{align}
whose values are given in Tables \ref{tab:inputnondim} and \ref{tab:sundim}.
% Because of the uncertainty in the Coulomb logarithm (Equation \eqref{eq:lambda}), we report all diffusivities to only two significant digits.

 To calculate the microscopic diffusivities, we use the formulae from \citet{Braginskii1958,Hazlehurst1959,Spitzer1962,Wendell1987} (as summarized nicely by \citealt{Skoutnev2025b}):
\begin{subequations}\label{eq:solar_diffusions}
    \begin{align}
        \nutilde\dimm &=\nutilde\dimm_{\rm ii} + \nutilde\dimm_{\rm rad},\nonumber\\
        \where\nonumber\\
        \nutilde\dimm_{\rm ii} &=\av{\frac{A^{1/2}}{Z^4\ln\Lambda}}\frac{2m_p^{1/2}(k_{\rm B}\tmptilde\dimm)^{5/2}}{5e^4\rhotilde\dimm}\\
        \andd\nonumber\\
        \nutilde\dimm_{\rm rad} &=\frac{4a\tmptilde^{*4}}{15c\tilde{\chi}\dimm\rhotilde\dimsq}\\
        \
        \kappatilde\dimm &= \frac{4ac\tmptilde^{*3}}{3\cptildedim{\chi}\dimm\rhotilde\dimsq}\\
        \
        \ 
        &\andd\nonumber\\
        \etatilde\dimm &= \av{\frac{Z\ln\Lambda}{\gamma_E}}\frac{\pi^{1/2} e^2m_e^{1/2}c^2}{16\sqrt{2}(k_{\rm B}\tmptilde\dimm)^{3/2}}
    \end{align}
\end{subequations}
where now the $\widetilde{(\cdots)}\dimm$ notation refers to dimensional profiles taken from Model S. Here, $m_p=\sn{1.7}{-24}\ \gram$ is the proton mass, $m_e= \sn{9.1}{-28}\ \gram$ is the electron mass, $k_{\rm B}=\sn{1.4}{-16}\ \erg\ \kelvin^{-1}$ is the Boltzmann constant, $e=\sn{4.8}{-10}\ \gram^{1/2}\ \cm^{3/2}\ \second^{-1}$ is the electron charge, $a=\sn{7.6}{-15}\ \erg\ \cm^{-3}\ \kelvin^{-4}$ is the radiation constant, $c=\sn{3.0}{10}\ \cm\ \second^{-1}$ is the speed of light, $\cptildedim$ is the specific heat at constant pressure, $\tilde{\chi}\dimm$ is the opacity (with units $\gram^{-1}\ \cm^2$), $Z$ is the atomic number of the ions, $A$ is the atomic weight of the ions, and $\gamma_E=O(1)$ is a correction factor to the magnetic diffusivity for electron-electron collisions (see \citealt{Spitzer1962}, p. 139 and use his Table 5.4 assuming solar abundances and $\gamma_E$ for metals of $\approx$0.85). The angular brackets refer to averages over three ionic species (H, He, and metals), which are assumed to have solar-like mass fractions $X_M\approx0.70$, $Y_M\approx0.28$, and $Z_M\approx0.02$, respectively. For the metals, we assume mean atomic weight $A_Z\approx15.5$ and mean atomic number $Z_Z\approx7.7$.

The Coulomb logarithm (which is different for each atomic species through the $Z$-dependence) is given by 
\begin{align}\label{eq:lnlambda}
\ln\tilde{\Lambda}\define\ln\left(\frac{d_{\rm max}}{d_{\rm min}}\right),
\end{align}
where 
\begin{subequations}\label{eq:dminmax}
\begin{align}
d_{\rm min}&\define\max\left(\frac{Ze^2}{k_{\rm B}\tmptilde\dimm},\frac{\hbar}{\sqrt{m_ek_{\rm B}\tmptilde\dimm}}\right)\\
\andd d_{\rm max} &\define \sqrt{\frac{k_{\rm B}\tmptilde\dimm}{4\pi \tilde{n}_e\dimm e^2}},
\end{align}
\end{subequations}
where $\tilde{n}_e\dimm$ is the electron number density. 

\begin{table*}
	\caption{Dimensional properties of the solar background state according to Model S \citep{ChristensenDalsgaard1996}. Quantities are volume-averaged over either the CZ (which here is taken to be the region $0.713\rsun\leq r\dimm\leq0.999\rsun$, i.e., excluding the very near-surface CZ where the diffusivities deviate significantly from their bulk values due to the extremely small-scale variation of $\rhotilde\dimm$ and $\tmptilde\dimm$) or the RZ (which here is taken to be the tachocline region $0.693\leq r\dimm\leq0.713\rsun$, i.e., assuming a width of the tachocline of $0.02\rsun$). }
	\label{tab:sundim}
	\centering
	\begin{tabular}{*{3}{l}}
    \multicolumn{3}{c}{\textbf{Dimensional properties of the solar interior}} \\ % This acts as a title
		\hline
Name & Symbol & Value\\
\hline
solar luminosity & $L_\odot$ & $\sn{3.85}{33}\ \erg\ \second^{-1}$\\
solar mass & $M_\odot$ & $\sn{1.99}{33}\ \gram$\\
solar radius & $R_\odot$ & $\sn{6.96}{10}\ \cm$\\
base of the solar CZ & $\rbczsun$ & $0.713\rsun = \sn{4.96}{10}\ \cm$\\
average RZ angular velocity & $\Omega_\odot$ & $\sn{2.70}{-6}\ \radsecond$\\
average nonradiative energy flux & $\fluxnradcz$ & $\sn{6.98}{10}\ \erg\ \cm^{-2}\ \second^{-1}$\\
average gravity & $g\cz$ & $368\ \meter\ \second^{-2}$\\
average density scale-height & $H\cz$ & $\sn{4.75}{9}\ \cm$\\
average density & $\rho\cz$ & $\sn{5.43}{-2}\ \gram\ \cm^{-3}$\\
average temperature & $T\cz$ & $\sn{8.57}{5}\ \kelvin$\\
average CZ pressure specific heat & $\cpcz$ & $\sn{3.92}{8}\ \unitentr$\\
convective free-fall velocity & $u_{\rm ff}\dimm$ & $109\ \meter\ \second^{-1}$\\
convective free-fall time & $t_{\rm ff}\dimm$ & $5.05\ \days$\\
average angular buoyancy frequency & $N\rz$ & $\sn{8.07}{-3}\ \radsecond$\\
average CZ kinematic viscosity & $\nu\cz$ & $4.1\ \stoke$\\
average CZ thermometric conductivity & $\kappa\cz$ & $\sn{3.2}{6}\ \stoke$\\
average CZ magnetic diffusivity & $\eta\cz$ & $\sn{1.9}{4}\ \stoke$\\
average tachocline kinematic viscosity & $\nu\rz$ & $10\ \stoke$\\
average tachocline thermometric conductivity & $\kappa\rz$ & $\sn{1.3}{7}\ \stoke$\\
average tachocline magnetic diffusivity & $\eta\rz$ & $\sn{1.4}{3}\ \stoke$\\
\hline

	\end{tabular}
\end{table*}

\subsection{Dimensional Analogs for Our Simulations}
For each nondimensional simulation we can compute an infinite number of dimensional analogs, the only requirement being that the analog have the same nondimensional parameters and background state. This is because there are far more dimensional parameters of the system than nondimensional ones. To pick a unique analog, typically the community has chosen to assign solar-like values to as many of the dimensional parameters as possible (for example, matching the values of Model S, as in Table \ref{tab:sundim}), while choosing diffusivities that are much higher than their solar counterparts. However, once the analog is chosen to have a given $\gravcz$ (i.e., a given stellar mass $M$), $\cpcz$, $\rhocz$, $\tmpcz$, and $H$ (i.e., a given stellar radius $R$), either the chosen luminosity $L$ (and hence $\fluxnradcz$) or the frame rotation rate $\Omzero$ usually needs to be non-solar-like as well. In our simulations, for which convection is driven by a thermal boundary layer, this is because the nondimensional parameter
\begin{align}
    \frac{\raf\ek^3}{\pr^2} = \frac{\fluxnradcz \gravcz}{8\Omzero^3\cp\rhocz\tmpcz H^2}
\end{align}
is independent of the diffusivities. Because $\raf\ek^3/\pr^2$ cannot be changed from the nondimensional simulation, $L$ (via $\fluxnradcz$) and $\Omzero$ cannot be chosen independently. 

% Because $\bu=(\bruntrz/\twoOmzero)^2$ cannot be changed either, $\bruntrz$ may also potentially be non-solar-like.
In this work, we arbitrarily choose $L$ to match the value from Model S and choose the $\Omzero$ to potentially be non-solar-like. We thus compute
\begin{align}
    \Omzero = \left(\frac{\pr^2}{\raf \ek^3} \frac{\fluxnradcz\gravcz}{8\cpcz\rhocz\tmpcz H^2}\right)^{1/3},
\end{align}
where $\fluxnradcz$, $\gravcz$, $\cp=\cpcz$, $\rhocz$, and $\tmpcz$ are taken from Table \ref{tab:sundim} and $H\define 0.228\rsun$.

Once $\Omzero$ is chosen, we choose
\begin{align}
    \nucz &= \ek (\twoOmzero H^2),\\
    \kappacz &= \frac{\nucz}{\pr},\\
    \etacz &= \frac{\nucz}{\prm},\\
    \andd \bruntrz &= \bu^{1/2} (\twoOmzero).
\end{align}

Table \ref{tab:simdim} shows the dimensional parameters for each simulation and the Sun, along with the CGS units for $[\vecu]$, $[\vecb]$, $[\rhoumer]$, $[\Delta L]$, and $[\rhotilde\vecu^2/2]$. Note that a torque density $\tau$ has the same units as $[\rhotilde\vecu^2/2]$ and $\int \av{\tau}dt$ has the same units as $\Delta L$. We reiterate that the units chosen in this way are not unique, the diffusivities are very far from solar/stellar values, and thus dimensional quantities extracted from our simulations (or any other simulations run to date) should be treated cautiously. 

\begin{table*}
	\caption{Dimensional parameters and associated units for each simulation, assuming all quantities are solar-like except the frame rotation rate and diffusivities. }
	\label{tab:simdim}
	\centering

    % === Part 1: The Header Row (Increased Spacing) ===

	\begin{tabular}{*{11}{c}}
    \multicolumn{11}{c}{\textbf{Redimensionalization scheme for the simulations}} \\ % This acts as a title
		\hline
        % Header Row Content (using \dfrac for the tall fractions)
        Name & $\dfrac{\Omzero}{\Omsun}$ & $\dfrac{\bruntrz}{\radsecond}$ & $\dfrac{\nucz}{\stoke}$ & $\dfrac{\kappacz}{\stoke}$ & $\dfrac{\etacz}{\stoke}$ & $\dfrac{[\vecu]}{\meter\ \second^{-1}}$ & $\dfrac{[\vecb]}{\gauss}$ & $\dfrac{[\rhoumer]}{\gram\ \cm^{-2}\ \second^{-1}}$ & $\dfrac{[\Delta L]}{\gram\ \cm^{-1}\ \second^{-1}}$ & $\dfrac{[\rhotilde\vecu^2/2]}{\erg}$\\
    \hline

Case 0 & 3.44 & 1.42e-03 & 2.50e12 & 2.50e12 & 6.26e11 & 2.95e3 & 2.44e5 & 1.60e4 & 2.55e14 & 4.73e9\\
\hline
Case 1 & 3.96 & 1.01e-05 & 1.21e12 & 1.21e12 & 3.02e11 & 3.40e3 & 2.81e5 & 1.85e4 & 2.94e14 & 6.28e9\\
Case 2 & 3.03 & 7.71e-06 & 2.07e12 & 2.07e12 & 5.17e11 & 2.60e3 & 2.15e5 & 1.41e4 & 2.24e14 & 3.67e9\\
Case 3 & 3.06 & 6.73e-06 & 1.58e12 & 3.16e12 & 3.94e11 & 2.63e3 & 2.17e5 & 1.43e4 & 2.27e14 & 3.76e9\\
Case 4 & 2.34 & 1.15e-05 & 5.40e11 & 5.40e12 & 1.35e11 & 2.01e3 & 1.66e5 & 1.09e4 & 1.74e14 & 2.20e9\\
Case 5 & 3.03 & 2.44e-05 & 2.07e12 & 2.07e12 & 5.17e11 & 2.60e3 & 2.15e5 & 1.41e4 & 2.24e14 & 3.67e9\\
Case 6 & 2.73 & 8.48e-05 & 9.94e11 & 3.98e12 & 2.49e11 & 2.34e3 & 1.94e5 & 1.27e4 & 2.02e14 & 2.98e9\\

\hline    
	\end{tabular}
\end{table*}

In Table \ref{tab:time-scales_dim}, we use the dimensional values from Table \ref{tab:simdim} to compute the appropriate dimensional time-scales for the simulations and the Sun. This can be done either by using the definitions in Equation \eqref{eq:timescalesdim} directly or more quickly by simply multiplying the values in Table \ref{tab:time-scales} by $(\twoOmzero)^{-1}$. From Table \ref{tab:time-scales_dim}, it is clear that the simulations have time-scales ``scrambled" from their ordering in the Sun and that because of our chosen redimensionalization, the diffusive time-scales in the simulations are much shorter than their solar counterparts. However, we reiterate the important point that the ordering $\tesdim\lesssim\tnudim$ (i.e., $\sigma\lesssim1$) holds for all the radiatively spreading cases (Cases 1--6). 
\begin{table*}
	\caption{Dimensional time-scales for each simulation and the Sun, using the redimensionalization scheme outlined in Table \ref{tab:simdim}. }
	\label{tab:time-scales_dim}
	\centering
	\begin{tabular}{*{11}{l}}
    \multicolumn{11}{c}{\textbf{Dimensional time-scales}} \\ % This acts as a title
		\hline
        
Name & $\tbruntdim$ & $\tomegadim$ & $\tffdim$ & $\tnuczdim$ & $\tkappaczdim$ & $\tetaczdim$ & $\tesdim$ & $\tnudim$ & HD: $\trundim$ & MHD: $\trundim$ \\
\hline
Case 0 & 11.8 min & 0.623 d & 1.56 d & 3.20 yr & 3.20 yr & 12.8 yr & 571000 yr & 98.0 yr & 99.2 yr & 47.2 yr \\
\hline
Case 1 & 1.15 d & 0.541 d & 1.08 d & 6.63 yr & 6.63 yr & 26.5 yr & 26.7 yr & 120 yr & 114 yr & 6.17 yr \\
Case 2 & 1.50 d & 0.708 d & 1.42 d & 3.88 yr & 3.88 yr & 15.5 yr & 15.6 yr & 70.1 yr & 98.7 yr & 12.3 yr \\
Case 3 & 1.72 d & 0.700 d & 1.75 d & 5.08 yr & 2.54 yr & 20.3 yr & 12.9 yr & 155 yr & 79.5 yr & 199 yr \\
Case 4 & 1.01 d & 0.915 d & 2.29 d & 14.8 yr & 1.48 yr & 59.4 yr & 37.7 yr & 455 yr & 284 yr & 71.8 yr \\
Case 5 & 0.475 d & 0.708 d & 1.42 d & 3.88 yr & 3.88 yr & 15.5 yr & 156 yr & 70.1 yr & 119 yr & 11.6 yr \\
Case 6 & 0.136 d & 0.785 d & 1.96 d & 8.06 yr & 2.01 yr & 32.2 yr & 2040 yr & 247 yr & 93.8 yr & 86.3 yr \\
\hline
Sun  & 21 min & 2.1 d & 5.1 d & 6e10 yr & 2e5 yr & 4e7 yr & 1.4e11 yr & 2e12 yr & n/a & n/a \\

\hline    
	\end{tabular}
\end{table*}

\section{Nondimensional Output Parameters}\label{ap:outputdim}
We characterize the degree of rotational constraint in each simulation by the realized mean and fluctuating Rossby numbers, 
\begin{equation}\label{eq:ro}
	\ro_{\rm mean} \define \avvolt{\lvert\av{\vecom}\rvert^2}^{1/2}\ \ \text{and} \ \ 	\ro_{\rm fluc} \define \avvolt{\lvert \vecom^\prime\rvert^2}^{1/2},
\end{equation}
respectively, where $\avvol{\cdots}$ refers to a volume-average over either the CZ or deep RZ. 

We characterize the level of turbulence by the realized Rayleigh number
\begin{align}
	\ra &\define \raf \Delta\entrover, \nonumber\\
	\where \Delta\entrover &\define \avspht{\entrhat}(\rbcz) -\avspht{\entrhat}(\rout), 
\end{align}
is the realized nondimensional entropy difference across the CZ, as well as the mean and fluctuating Reynolds numbers
\begin{equation}
	\re_{\rm mean} \define \frac{ \avvolt{\lvert\av{\vecu}\rvert^2}^{1/2}}{\ek\,\nu_{\rm v}/\nu_{\rm cz}} \ \ \text{and}\ \ \re_{\rm fluc} \define \frac{\avvolt{\lvert \vecu^\prime\rvert^2}^{1/2}}{\ek\,\nu_{\rm v}/\nu_{\rm cz}},
\end{equation}
respectively. %, where again $\avvol{\cdots}$ refers to a volume-average over either the CZ or deep RZ. 

The energetics of the system is characterized by the energy densities of the differential rotation
\begin{align}
	{\rm KE_{dr}}\define \frac{1}{2}\avvolt{\rhotilde \av{\uphi}^2},
\end{align}
meridional circulation
\begin{align}
	{\rm KE_{mc}}\define \frac{1}{2} \avvolt{\rhotilde \av{\umer}^2},
\end{align}
convection 
\begin{align}
	{\rm KE_{fluc}}\define \frac{1}{2} \avvolt{\rhotilde \vecu^{\prime2}},
\end{align}
and magnetic field 
\begin{align}\label{eq:me}
	{\rm ME}\define \frac{1}{2} \avvolt{\vecb^2}. 
\end{align}

Table \ref{tab:outputnd} shows these diagnostics for simulation suite, which were also discussed in Section \ref{sec:basic}.

\begin{table*}
	\caption{Output nondimensional numbers for the simulation suite, characterizing the rotational constraint, turbulence level, and partition of energy for the mean and fluctuating flows and fields. See Equations \eqref{eq:ro}--\eqref{eq:me} for definitions.}
	\label{tab:outputnd}
	\centering
	\begin{tabular}{*{10}{l}}

    \multicolumn{10}{c}{\textbf{Output parameters for the simulations}} \\ % This acts as a title
		\hline
		\multicolumn{10}{c}{CZ ($\rbcz$ to $\rout$)} \\
		\hline\hline
		Name & $\ro_{\rm mean}$ & $\ro_{\rm fluc}$ & $\Delta\entrover$ & $\re_{\rm mean}$ & $\re_{\rm fluc}$ &       $\rm KE_{dr}$ & $\rm KE_{\rm mc}$ & $\rm KE_{\rm fluc}$ & $\rm ME$\\
		\hline
        
Case H0 & 0.146 & 0.445 & 1.12 & 176 & 70.1 & 2.91e-03 & 1.11e-06 & 4.43e-04 &  - \\
\hline
Case H1 & 0.140 & 0.571 & 0.427 & 322 & 127 & 1.41e-03 & 8.67e-07 & 2.58e-04 &  - \\
Case H2 & 0.134 & 0.535 & 0.600 & 132 & 70.9 & 1.17e-03 & 1.54e-06 & 3.94e-04 &  - \\
Case H3 & 0.167 & 0.566 & 1.40 & 264 & 96.2 & 2.28e-03 & 1.34e-06 & 3.13e-04 &  - \\
Case H4 & 0.276 & 0.995 & 1.85 & 1120 & 320 & 8.82e-03 & 4.61e-06 & 7.20e-04 &  - \\
Case H5 & 0.130 & 0.730 & 1.23 & 115 & 81.6 & 6.07e-04 & 1.28e-06 & 3.57e-04 &  - \\
Case H6 & 0.226 & 0.712 & 1.64 & 560 & 172 & 6.44e-03 & 2.57e-06 & 5.60e-04 &  - \\
\hline
Case M0 & 0.0586 & 0.432 & 1.10 & 41.5 & 57.5 & 9.73e-05 & 4.96e-07 & 2.27e-04 & 1.15e-04\\
\hline
Case M1 & 0.0503 & 0.530 & 0.378 & 66.8 & 130 & 6.94e-05 & 6.28e-07 & 2.87e-04 & 1.77e-04\\
Case M2 & 0.0640 & 0.495 & 0.540 & 48.6 & 71.5 & 1.70e-04 & 1.07e-06 & 4.14e-04 & 2.15e-04\\
Case M3 & 0.0628 & 0.551 & 1.20 & 55.9 & 98.0 & 9.57e-05 & 8.54e-07 & 3.27e-04 & 2.63e-04\\
Case M4 & 0.0897 & 0.996 & 1.57 & 173 & 287 & 1.65e-04 & 2.08e-06 & 5.78e-04 & 4.69e-04\\
Case M5 & 0.0671 & 0.688 & 1.18 & 39.0 & 84.2 & 8.10e-05 & 1.14e-06 & 4.06e-04 & 2.26e-04\\
Case M6 & 0.0710 & 0.695 & 1.36 & 90.4 & 154 & 1.21e-04 & 1.13e-06 & 4.13e-04 & 3.57e-04\\

		\hline
		\multicolumn{10}{c}{deep RZ ($\rin$ to $r\drz$)} \\
		\hline\hline
		Name & $\ro_{\rm mean}$ & $\ro_{\rm fluc}$ & $\Delta\entrover$ & $\re_{\rm mean}$ & $\re_{\rm fluc}$ &       $\rm KE_{dr}$ & $\rm KE_{\rm mc}$ & $\rm KE_{\rm fluc}$ & $\rm ME$\\
        \hline
        
Case H0 & 0.0663 & 0.0356 & - & 221 & 27.0 & 4.13e-03 & 1.11e-11 & 3.75e-05 &  - \\
\hline
Case H1 & 0.0152 & 0.0538 & - & 59.3 & 18.3 & 4.73e-04 & 1.72e-07 & 2.81e-05 &  - \\
Case H2 & 0.0147 & 0.0509 & - & 23.1 & 9.60 & 3.42e-04 & 2.08e-07 & 3.77e-05 &  - \\
Case H3 & 0.0322 & 0.0548 & - & 157 & 39.0 & 1.23e-03 & 2.13e-07 & 4.80e-05 &  - \\
Case H4 & 0.0923 & 0.0889 & - & 814 & 147 & 7.56e-03 & 9.55e-07 & 1.51e-04 &  - \\
Case H5 & 0.00701 & 0.0328 & - & 11.8 & 5.91 & 1.15e-04 & 8.39e-08 & 1.64e-05 &  - \\
Case H6 & 0.0925 & 0.0599 & - & 701 & 99.4 & 1.14e-02 & 2.51e-08 & 1.52e-04 &  - \\
\hline
Case M0 & 0.00508 & 0.00840 & - & 9.52 & 3.98 & 7.78e-06 & 7.54e-12 & 7.74e-07 & 2.13e-05\\
\hline
Case M1 & 0.00883 & 0.0647 & - & 14.8 & 22.7 & 2.25e-05 & 1.20e-07 & 4.14e-05 & 4.99e-05\\
Case M2 & 0.00962 & 0.0531 & - & 8.51 & 10.6 & 3.72e-05 & 1.46e-07 & 4.47e-05 & 5.12e-05\\
Case M3 & 0.00742 & 0.0426 & - & 26.4 & 30.4 & 3.05e-05 & 1.32e-07 & 2.88e-05 & 8.78e-05\\
Case M4 & 0.00880 & 0.0595 & - & 72.6 & 82.1 & 5.02e-05 & 4.95e-07 & 4.10e-05 & 6.35e-05\\
Case M5 & 0.00640 & 0.0380 & - & 5.03 & 7.12 & 1.93e-05 & 7.02e-08 & 2.34e-05 & 4.18e-05\\
Case M6 & 0.00642 & 0.0234 & - & 27.2 & 18.4 & 1.75e-05 & 3.46e-08 & 4.52e-06 & 3.36e-05\\

		\hline
	\end{tabular}
\end{table*}

\section{Estimating $\sigma_{\MakeLowercase{c}}$: Circulation Amplitudes and Regime Boundaries}\label{ap:sigmac}
In this Appendix, we use the dynamics of the two types of spreading processes (viscous and radiative) to understand the achieved meridional circulation flow amplitudes in our simulated RZs. We also explain how we still achieve radiative spreading for $\sigma\approx3$, i.e., why $\sigma_c>1$.

Both viscous and radiative spreading can be described by the maintenance of a particular set of dynamical balances holding in the HD fluid equations. These balances are thought to be:

\noindent (1) A time-dependent angular momentum equation, with transport by viscosity and/or meridional circulation primarily driving the evolution,
\begin{align}\label{eq:torque_burrowing}
\rhotilde\lambda^2\pderiv{\Omega}{t}=&-\rhoumer\dotgrad(\lambda^2\Omega) +\ek\Div(\nutilde\lambda^2\nabla\Omega),
\end{align}
(2) A time-independent, anelastic continuity equation for the meridional flow, 
\begin{align}\label{eq:contmer}
\Div\rhoumer\equiv0,
\end{align}
(3) A time-independent thermal wind balance in the meridional-circulation equation (see \citealt{Matilsky2023} for a comprehensive discussion), 
\begin{align}\label{eq:tw}
    \pderiv{\Omega^2}{z}=\rocsq \frac{\gravtilde}{r^2\sin\theta}\pderiv{\entrhat}{\theta},
\end{align}
and

\noindent (4) A time-independent advective--diffusive balance in the heat equation.
\begin{align}\label{eq:advdiff}
\av{\urad}=\frac{\rocsq\ek}{\sigma^2} \frac{\gravtilde}{\rhotilde\tmptilde\brunttildesq}\Div(\rhotilde \tmptilde \kappatilde \nabla \av{\entrhat}). 
\end{align}

\subsection{Estimating Meridional Circulation Amplitude During Spreading}\label{sec:radspreadsigmadyn}
%We can estimate the relative magnitudes of the viscous and radiative spreading processes (i.e, $\sigmadyn(t)$) by using Equations \eqref{eq:torque_burrowing}, \eqref{eq:contmer}, \eqref{eq:tw}, and \eqref{eq:advdiff} from the preceding section.   
% $|\nabla\Omega|\sim\dOm/\Delta$

%\com{NB: This section is all to justify $\sigma_c > 1$.  As an explanation, this section is pretty dense, but I wonder if this is a bit we could move to an appendix or something if was want to cut length} \com{[LM: This is now the final appendix!]}

In the estimates that follow, we consider spreading dynamics for a relatively thin shear layer in our anelastic spherical RZ, with time-dependent characteristic thickness $\Delta(t)$ and shear amplitude $\dOm(t)$.%, and fixed location $\rbcz$.

As long as $\Delta(t)\lesssim \hrho$ (certainly true for all our simulations and the solar tachocline), we can use Equation \eqref{eq:contmer} to estimate 
\begin{align}\label{eq:contmerest}
	\frac{\abs{\avurad}\est(t)}{\abs{\avutheta}\est(t)}= \frac{\Delta(t)}{\rbcz},
\end{align}
where the ``est" subscripts denote an estimated quantity as a function of $t$.

We assume weak differential rotation (or $2\Omega\approx1$) and $\dOm\ll\Omega$. We also assume significant departures from the Taylor-Proudman state $\pderivline{\Omega}{z}\equiv0$, as is true for most of our simulations and for the solar tachocline. We thus have $|\pderivline{\Omega^2}{z}| \sim \dOm(t)/\Delta(t)$. Equation \eqref{eq:tw} then gives the estimate
\begin{equation}\label{eq:stellartwest}
	\abs{\aventrhat}\est(t) = \frac{\rbcz^2\dOm(t)}{\rocsq\Delta(t)}, 
\end{equation}
where we assume all nondimensional background-state quantities are $O(1)$. 

Finally, Equation \eqref{eq:advdiff} gives the estimate
\begin{equation}\label{eq:advdiffest}
	\abs{\avurad}\est(t) = \frac{\ek\rocsq}{\sigma^2}\frac{\abs{\aventrhat}\est(t)}{\Delta(t)^2}. 
\end{equation}

Combining the estimates \eqref{eq:contmerest}, \eqref{eq:stellartwest} and \eqref{eq:advdiffest}, yields the following estimate for the amplitude of a burrowing circulation that simultaneously satisfies thermal wind balance and advective--diffusive balance of heat: 
\begin{subequations}\label{eq:mcest}
	\begin{align}
		\abs{\avurad}\est(t)&=\frac{\ekt}{\bu}\frac{\rbcz^2\dOm(t)}{\Delta(t)^3}\\
		\andd \abs{\avutheta}\est(t) &=\frac{\ekt}{\bu}\frac{\rbcz^3\dOm(t)}{\Delta(t)^4}, \label{eq:avuthetaest}
	\end{align}
\end{subequations}
where 
\begin{align}
    \ekt\define \frac{\ek}{\pr} = \frac{\kappacz}{\twoOmzero H^2}
\end{align}
is the thermal Ekman number. 

\subsection{Estimating Torques and $\sigma_c$}
For $\dOm\ll\Omega$, Equation \eqref{eq:amom} yields $\amom\approx \lambda^2/2$. Thus, the meridional circulation torque is typically dominated by the advection of the background angular momentum:
\begin{equation}\label{eq:taumcapprox}
\taumc \approx \rhotilde\lambda\av{\ulambda}.
\end{equation}

Neglecting geometric factors like $\cos\theta$ and $\sin\theta$, we estimate $\avulambda\sim\avutheta$, and Equation \eqref{eq:taumcapprox} then yields the estimate
\begin{equation}\label{eq:taumcest}
	\abs{\taumc}\est(t) = \rbcz \abs{\av{\utheta}}\est(t).
\end{equation}

From Equation \eqref{eq:tauv}, we  estimate
\begin{equation}\label{eq:tauvest}
	\abs{\tauv}\est(t)=\ek\rhotilde\nutilde r^2\frac{\dOm(t)}{\Delta(t)^2}.
\end{equation}

Referring to the torque estimates \eqref{eq:taumcest} and \eqref{eq:tauvest}, we estimate
\begin{subequations}\label{eq:bothestimates}
	\begin{align}
		\abs{\taumc}\est(t)&=\frac{\ek}{\sigma^2}\frac{\dOm(t)}{[\Delta(t)/\rbcz]^4}\\
		\andd \abs{\tauv}\est(t)&= \ek \frac{\dOm(t)}{[\Delta(t)/\rbcz]^2}.
	\end{align}
\end{subequations}

We note that with some amount of manipulation, the estimates \eqref{eq:bothestimates} can be be obtained from Equation (4.9) in \citetalias{Spiegel1992}. 

We thus estimate $\sigmadyn(t)$: 
\begin{align}\label{eq:sigmadynest}
	\sigmadynest(t) &\define \sqrt{\frac{\abs{\tauv}\est(t)}{\abs{\taumc}\est(t)}}  = \frac{\Delta(t)}{\rbcz}\sigma.
\end{align}

The key implication of Equation \eqref{eq:sigmadynest} is that while the shear layer is still thin ($\Delta(t)\ll \rbcz$), even for $\sigma$ substantially higher than unity, the spreading dynamics are still expected to be dominated by circulation burrowing. This is the fundamental reason why we still achieve radiative spreading for Case 6, even though $\sigma\approx 3$. 

We can estimate whether or not the system will be in the radiative spreading regime (i.e., estimate 
$\sigma_c$) if we know the value of 
$\Delta(\thalf)$. This is also an emergent property of the system, but from Table \ref{tab:thalf}, the values of $\Delta(\thalf)$ appear to all be roughly near their average value of $\approx$0.3 for both the HD and MHD cases. We can thus estimate a very rough  universal value of $\sigma_c$ for our simulation suite of by plugging $\Delta(\thalf)\approx0.3$ into Equation \eqref{eq:sigmadynest}. This yields
\begin{align}\label{eq:sigmac}
\sigma_c \define \frac{\rbcz}{\Delta(\thalf)}\approx 10, 
\end{align}
which, although somewhat crudely estimated, is substantially larger than unity.

% bibliography
%\newpage
%\bibliography{library, proceedings, books}
%\bibliography{library_jstyle, proceedings, books}
%\bibliographystyle{aasjournalv7}

\end{document}